\documentclass[10pt,twocolumn,secnumarabic,prd,nofootinbib]{revtex4-1}

\usepackage[utf8]{inputenc} 

\usepackage{amsmath} 
\usepackage{amssymb} 

\usepackage[T1]{fontenc}
\usepackage{textcomp}
\usepackage{txfonts}
\usepackage{bm}

\usepackage{fancyhdr}
\usepackage{graphicx}
\usepackage[linktocpage=true,colorlinks=true,linkcolor=blue,citecolor=blue,urlcolor=blue]{hyperref}

\begin{document}

\def\arcsec{$^{\prime\prime}$}
\title
{The expansion of the universe observed with supernovae}
\rhead{{\small The expansion of the universe observed with SNe}}

\author{Pierre Astier}

\affiliation{LPNHE, CNRS/IN2P3, Universit\'e Pierre et Marie Curie Paris 6, Universit\'e Paris Diderot Paris 7, 4 place Jussieu, 75005 Paris, France}
\begin{abstract}
  Over the last 20 years, supernovae have become a key tool to
  constrain the expansion history of the Universe through the
  construction of Hubble diagrams, using luminosity distances to
  supernovae belonging to the ``Ia'' subtype. This technique was key
  for the discovery that the expansion of the Universe is now
  accelerating. We review the principle and difficulties of the
  measurements, the classification and diversity of supernovae, and
  the physics of the explosion. We discuss the systematic
  uncertainties affecting the cosmological conclusions with some
  emphasis on photometric calibration. We describe the major supernova
  cosmology surveys, the presented analyses and their conclusions,
  together with the present status of the field. We conclude on the
  expectations for the near future. 
\end{abstract}

\maketitle
\tableofcontents
\section{Introduction}
\thispagestyle{empty} 
   Kinematics of test masses
probe the sources of gravitational field, both in a Newtonian framework
and in general relativity (GR). In a Universe following the 
cosmological principle, namely homogeneous and isotropic,
the motion of test masses (e.g. galaxies) constrains
the mean density of the Universe and its evolution with time. 

In 1927, Lemaître derived from Einstein's general relativity equations
that the velocity of recession is proportional to distance
to the first order, and also extracted from the measurements of
distances to galaxies (deduced from their apparent brightness) and 
their recession velocity (from their spectral shift) that this property
might be realized in nature \cite{Lemaitre27}\footnote{Note that the English translation
\cite{Lemaitre27-translation}, probably carried out by Lemaître himself
in 1931, misses the key paragraphs that propose the observational 
evidence for expansion.}. In 1929, Hubble presented distances to a 
sample of galaxies extending further out and also proposed 
a linear relation between distance and recession velocity \cite{Hubble29}.
This was the 
discovery of the expansion of the Universe and the Hubble
law states that distance is proportional to recession velocity,
at least at a small velocity, and ignoring ``peculiar motions'' due to spatial
density variations. The proportionality coefficient is now 
called the ``Hubble constant'' $H_0$. Although its
value has been revised by almost
an order of magnitude since Hubble's initial determination, the discovery
of the proportionality itself remains one of the major early steps of
observational cosmology. In modern parlance, the fractional 
spectral shift towards the red, the ``redshift'' $z$ has replaced the 
recession
velocity: $z=\lambda_\mathrm{reception}/\lambda_\mathrm{emission}-1$.
Because it is a direct observable, redshift is used by cosmologists 
to index cosmic time. In the current standard model of cosmology,
$z=0.1$ corresponds to $\sim$10\% of the current age of the 
Universe (about 13.7 Gy), $z=0.5$ to 40\%, $z=1$ to 55\%. The first 
stars were probably formed around $z=10$ (0.5 Gy after the Big Bang).

The Hubble constant $H_0$ has the dimension of the inverse of a time, 
usually expressed in units of km/s/Mpc. It defines the
expansion rate today, and sets the time and distance scales of the
expansion. Without diminishing the importance of the Hubble constant
determination, one should note that its value does not convey
sufficient information to define the current average energy content of
our Universe: GR relates the expansion rate at a given
epoch with energy density {\it and} spatial curvature at the same
epoch. The Hubble constant defines the average density of the Universe
only if the latter is flat (i.e. has zero curvature). The 
density for which the curvature changes sign is called the critical density
and depends only on the Hubble constant and Newton's constant.

In order to constrain gravitation sources driving the expansion, one
has to measure the expansion beyond first order in recession velocity (or redshift).  Namely, as
one studies the velocity {\it variations} with time of free
falling bodies to measure the gravitational field of the Earth, one
should probe how the expansion rate evolves with time in order to
characterize the mean Universe content. In the GR
framework, the expansion rate {\it evolution} provides a handle on the
fractions of various Universe components, including curvature.
Practically, probing the distance-velocity relation beyond first order
makes it possible to constrain the evolution of the energy
density in the Universe, which in turn constrains the proportions of
various components in the admixture, after assuming how each of the
involved components dilutes with expansion. We will discuss in \S \ref{sec:standard-candles} why determining the proportions of the various components
does not require measuring the Hubble constant $H_0$.

How practical is it to measure the distance-redshift relation?
Measuring a redshift might be considered as relatively easy, at least
if the target is bright enough to allow one to acquire a spectrum:
redshifts of galaxies are routinely measured to three decimal places
from spectroscopy in the visible.  But measuring cosmological
distances is not as straightforward. Hubble derived distances to
galaxies from assumptions about intrinsic luminosities of stars. The
overall intrinsic luminosity scale turned out to be wrong and yielded
a Hubble constant about 7 times larger when compared with modern
determinations.  But the real weakness of galaxies as distance
indicators is in fact more serious (e.g. \cite{OstrikerTremaine75}):
these are evolving too rapidly with time to provide reliable
distances, even if one concentrates on relative distances. Supernovae
were then introduced and used as a more reliable distance indicator
\cite{Kowal68,KK1974}, and eventually proposed to constrain the
evolution of expansion rate \cite{Wagoner77}.

Walter Baade and Fritz Zwicky introduced, in 1934, the term
``supernova'' to describe extremely bright explosions of stars that
appear and fade on timescales of months to years. Supernovae are
classified today into a handful of types and subtypes. The subtype of
type Ia supernovae (SNe~Ia) displays reproducible luminosities: 
at low redshift, the apparent brightness scales as the
inverse square of the redshift (e.g. \cite{Kowal68,Hamuy96c}). By extension,
plots of apparent brightness of similar objects versus redshift are
often called Hubble diagrams.

Precision supernova photometry started in the 1990s with the advent of
CCD imagers mounted on telescopes. The Calan-Tololo survey delivered,
in 1996, precise and homogeneous CCD-based photometry in several bands
for 29 SNe~Ia events \cite{Hamuy96b}, and derived distances
to those \cite{Hamuy96c}. Although the Calan-Tololo survey should be regarded
as the keystone of supernova cosmology, redshifts were limited to
$z \lesssim 0.1$, too low to constrain the
distance-redshift relation to second order. At about the same time,
two teams targeted the detection of SNe~Ia at significantly higher 
redshifts, also using CCD imagers. After a slow initial period,
both teams eventually succeeded in finding and measuring redshifts
and distances to about 50 SNe~Ia events in total, at $\langle z
\rangle \sim 0.5$. Both teams compared their own distant sample to
the Calan-Tololo nearby events (\S \ref{sec:scp-hi-z} \&
\cite{Riess98b,Perlmutter99}) and reached, in 1998, the same striking
conclusion: the distant events appear too faint compared with nearby
ones, in any matter-dominated Universe.  In terms of kinematics, this
means that the expansion is accelerating at late time, which is
impossible under matter domination (where matter is defined as
``tracing the expansion''). Cosmologists named the
conjectured fluid that sources the acceleration ``dark energy'', 
and it represents about 3/4 of the total energy density now, so that our
  Universe is no longer matter-dominated.  The discovery of accelerated
  expansion (or dark energy) constitutes one of the major recent discoveries in
  physics: it causes persistent perplexity (for reasons summarized in \S
  \ref{sec:dark-energy-models}), and still sources an enormous flow of scientific
  literature. The discoverers of accelerated expansion were awarded the 2011
Nobel Price in Physics. Dark energy studies constitute the main scientific driver for
large astronomical projects such as LSST and Euclid (\S \ref{sec:prospects}).

There had been earlier hints that our Universe might not follow the
flat matter-dominated scenario. Different indications of a low matter
density (i.e. well below the theoretically favoured critical density)
were presented in the early 1990s, mostly based on galaxy clustering
(e.g. \cite{Efstathiou90}) or primordial nucleosynthesis and galaxy
clusters (e.g. \cite{White93}).  The arguments in favour of a low
matter density turned out to be correct, but this does not directly
imply an accelerated expansion. In the 1990s, settling between a
(observationally favoured) low-density Universe and a (theoretically
favoured) critical Universe constituted a key question, and the 
kinematics of expansion seemed an obvious decisive observational test,
if possible. The enterprises aiming at producing Hubble diagrams of
SNe~Ia were conceived within this framework.

Within the cosmological principle and GR, acceleration
of the expansion happens when a fluid that does not expand (or expands
slowly) starts to dominate the energy density budget. The simplest
cosmological model that describes the observations consists in a flat
Universe consisting, today, of about 1/4 of matter (diluting with
expansion and time) and 3/4 of ``dark energy'', insensitive
to expansion and of constant density.  Their densities were equal at
$z\sim 0.45$, and matter dominated earlier, while the initial 
domination of radiation energy over matter lasted until $z\simeq 3300$.  
A possible incarnation of dark energy of static density is the
cosmological constant $\Lambda$ introduced by Einstein in the GR
field equations. It plays, in the cosmological framework, exactly 
the same role as a uniform fluid with static
density. The ``$\Lambda$CDM'' cosmological model consists (at late time) in this admixture
of a fluid of static density $\Lambda$ and matter which has to be dark and cold,
hence coined cold dark matter or CDM.

Cosmologists have considered a broader range of dark energy models and
describe them in terms of the ``equation of state'' of dark energy,
namely how the dark energy density varies with time and redshift (\S
\ref{sec:Friedman}). Describing dark energy using the equation of
state concept should not be taken too literally: this constitutes a
phenomenological description which does not require the assumption
that acceleration is indeed due to some fluid. This description of the
phenomenon applies to actual fluids, to the cosmological constant, and
it can even be rephrased in terms of alterations to GR
if one ignores density perturbations.

The discoverers of the accelerated expansion from supernovae readily
considered alternatives to the cosmological constant paradigm, and
found their observations to be compatible with it(\S
\ref{sec:scp-hi-z}). Since then, the result has been consolidated
mainly on two fronts: firstly, second-generation supernova surveys have been
conducted and have confirmed the early results with more events and
better measured distances extending to higher redshifts. Secondly, we now
have strong evidence for dark energy from other cosmological probes
although with less sensitivity (e.g. \S 4.1 in
\cite{Spergel07}). To date, large-scale cosmological observations have been 
described well by the $\Lambda$CDM model (e.g. \cite{Komatsu11}), sometimes
called the concordance model.

Constraining cosmology from distances to supernovae relies on a strong
hypothesis: SNe~Ia are good ``standard candles''. The tight scatter
of the Hubble diagram of SNe~Ia indicates that the events are
reproducible over a small redshift interval, but this does not prove
that distant and nearby events are identical.  They could, indeed, be of
different nature because distant events happen earlier in star
formation history than nearby ones.  This whole issue is referred to
as ``evolution'' and can be addressed observationally: the standard
candle hypothesis can be falsified, and is reinforced by every failed
attempt to do so. We will, however, see that the picture we have of the
SN~Ia explosion mechanism makes it a fundamental ``self-calibrated''
process, weakly sensitive to environmental influences.

Photometric calibration has recently become a serious issue for
supernova cosmology. Although the core of the method consists in
comparing fluxes, these fluxes have to be measured in different
observer bands because of redshift: nearby supernovae are usually
measured in blue and green bands, and the same restframe region lies
around 1~$\mu m$ for supernovae at $z\sim 1$. To take full
advantage of current supernova statistics, these flux ratios should
be calibrated to better than
1\%, which hits fundamental limitations of current astronomical
photometric calibration techniques.  Today, supernova cosmology is probably
the most demanding field of astronomy in terms of photometric
calibration across the visible and near infrared range.

The plan of this review goes as follows: \S \ref{sec:cosmo-framework}
covers the minimal necessary cosmological background, how distances
constrain cosmological parameters, and a summary of the leading
hypotheses for dark energy; \S \ref{sec:obs-tech} introduces the minimal
background in astronomy techniques required for this review, for readers
unfamiliar with the field; \S \ref{sec:pioneers} covers the
observational aspects of supernova taxonomy; \S \ref{sec:sneIa}
concentrates on the ``Ia'' subclass delivering almost all cosmological
constraints, and \S \ref{sec:explosions} covers the pre-explosion and
explosion issues. We then enter the observational and reduction
techniques (\S \ref{sec:obs-reduc}), and devote the next section (\S
\ref{sec:fitters}) to light curve fitting, almost always involved in
distance estimation. \S \ref{sec:photom-calib} covers photometric
calibration, the basement of supernova cosmology, and studies of other
astrophysical systematic uncertainties are reviewed in \S \ref{sec:systematics},
including gravitational lensing (\S \ref{sec:lensing}).
\S \ref{sec:surveys} covers
supernova surveys and their cosmological conclusions;
\S \ref{sec:prospects} briefly discusses
the prospects of the field. We summarize and conclude in \S 
\ref{sec:conclusion}.

For interested readers, we provide here an incomplete set of reviews
with similar or highly related scopes: \cite{GoobarLeibundgut11}
parallels this one; \cite{Hillebrandt00} concentrates on explosion
models; \cite{HowellReview11} focuses on the physics of SNe~Ia in
relation to their cosmological use; \cite{FreedmanH010} discusses
the determination of the Hubble constant which we will not discuss
here; \cite{FriemanRA&A} presents theoretical and observational
aspects of dark energy and the accelerating Universe, while 
\cite{Weinberg12} studies dark energy probes in great detail.

\section{Cosmological framework\label{sec:cosmo-framework}}

\subsection{The cosmological principle and Friedman equation(s) \label{sec:Friedman}}

The cosmological principle states that the Universe is isotropic and
homogeneous. Although we have ample evidence that this is wrong on
small distance scales, the quasi-isotropy of the cosmic microwave
background (CMB) \cite{Smoot91}, and the homogeneity of the distribution of
galaxies \cite{Hogg05} support this framework for large scales. Within General
Relativity, this principle is encoded into symmetries of the metric:
$$
ds^2 = dt^2-R^2(t)\left (\frac{dr^2}{1-kr^2}
              +r^2(d\theta^2+\sin^2\theta d\phi^2) \right )
$$
usually called the Friedman-Lemaitre-Robertson-Walker (FLRW) metric. 
$R(t)$ is called
the scale factor, and $k = -1$, $0$ or $1$, is the sign of spatial curvature.
We set ``$c=1$'' in what follows.
Objects with constant coordinates $(r,\theta,\phi)$ are called comoving.
Cosmologists attach comoving coordinates 
to the matter fluid, so that $R(t)$ scales with the 
separation of matter tracers (typically galaxies). 
In the FLRW framework, it is easy to show that photons (following paths
with $ds^2=0$), emitted by comoving sources and detected by comoving observers,
see their wavelength scale with $R(t)$:
$$
\frac{\lambda_\mathrm{reception}}{\lambda_\mathrm{emission}} = 
\frac{R(t_\mathrm{reception})}{R(t_\mathrm{emission})} \equiv 1+z
$$
where $z$ is called the redshift of the emitting source.
GR postulates relations between the metric and the 
gravitation sources. For the FLRW metric, these are two differential
equations for $R(t)$, the Friedman equations \cite{Friedmann24}:
\begin{align}
H^2(z) \equiv \left( \frac{\dot R}{R}  \right)^2 &= 
    \frac{8 \pi G}{3} \rho - \frac{k}{R^2(t)} + \frac{\Lambda}{3} \label{eq:friedmann_1}\\
\hskip 22mm \frac{\ddot R}{R} &=  - \frac{4\pi G}{3} (\rho+3p) + \frac{\Lambda}{3} \label{eq:friedmann_2}
\end{align}
where $\Lambda$ is the cosmological constant, $\rho$ stands for the proper 
energy density and $p$ for the pressure. $H(z)$ is the expansion rate and $H_0\equiv H(z=0)$.
The energy conservation equation:   
\begin{equation}
\label{eq:energy-conservation}
\frac{d}{dt}(\rho R^3) = -3 p R^2 \dot{R}
\end{equation}
relates pressure to density evolution and also applies separately
to the various fluids in the Universe. For non-relativistic matter, $\rho_M R^3$
is constant and hence $p_M=0$. A fluid with static density ($\dot \rho = 0$)
has $p = -\rho$. The cosmological constant $\Lambda$ plays exactly the same
role as a fluid with static density $\rho_{\Lambda}=-p_\Lambda = \Lambda/8\pi G$,
and can be summed into the $\rho$ and $p$ quantities in the Friedman
equations.
Relation (\ref{eq:energy-conservation}) can be obtained by eliminating 
$\ddot R$ between equations \ref{eq:friedmann_1} and \ref{eq:friedmann_2}.

Fluids can be characterized by a relation between
$p$ and $\rho$. The equation of state of each fluid $w_X$ is defined
by $p_X=w_X\rho_X$, and equation \ref{eq:energy-conservation} implies $\rho_X(t) \propto R(t)^{-3(1+w_X)}$. Given
the densities at one epoch (e.g. now) and the equations of state of
the fluids of the Universe, one can solve the first Friedman
equation (equation \ref{eq:friedmann_1}) for $R(t)$. 
The critical density is defined as the energy density 
for which $k=0$:
$$
\rho_c = \frac{3 H_0^2}{8\pi G}
$$
where $H_0 = (\dot R/R)_{now}$ is the Hubble constant. 
Densities are usually 
parametrized through their value ``today'', in units 
of the current critical density:
$$
\Omega_M \equiv \frac{\rho_M}{\rho_c} = \frac{8 \pi G \rho_M}{3 H_0^2}, \hspace{5mm}
   \Omega_\Lambda \equiv \frac{\Lambda}{3 H_0^2}, \hspace{5mm} \Omega_k \equiv - \frac{k}{R_0^2 H_0^2} 
$$
and the first Friedman equation implies $1=\Omega_M+\Omega_\Lambda+\Omega_k$.
This assumes that our Universe is composed only of matter and $\Lambda$.
One might consider other fluids such as an implementation of dark 
energy different from the cosmological constant (see \S \ref{sec:dark-energy-models}) and radiation density.

\subsection{Cosmological distances and dark energy}
\label{sec:cosmo-distances}
Cosmological distances should be defined from observables.
For example, one defines the luminosity distance through the energy flux $f$
of a source of proper energy luminosity $\mathcal{L}$ through 
$d_L^2 = \mathcal{L}/4 \pi f $.
For a source observed at a redshift $z$, it reads \cite{CPT92}:
\begin{align}
d_L(z) = &\frac{(1+z)}{H_0 \sqrt{|\Omega_k|}}Sin \left\{  |\Omega_k|^{1/2}\int_0^z\frac{ dz'}{H(z')/H_0} \right\} \label{eq:r(z)} \\
  \frac{H^2(z)}{ H^2_0} = & \left[  \Omega_M(1+z')^3+\Omega_\Lambda+ 
   \Omega_k(1+z')^2 + \Omega_R (1+z)^4 \right] \label{eq:H(z)} 
\end{align}
where $Sin(x)=sin(x),x,sinh(x)$ for $k=1,0,-1$; note that the expression is 
continuous in $\Omega_k=0$. In $z=0$, the integrand is 1 and the Taylor expansion reads
$d_L(z)=z/H_0+O(z^2)$, so that densities are not constrained
by distances at a small redshift. Distances to higher redshifts become sensitive
to densities, although in a somehow degenerate way. For sake of 
completeness, the expression for $H^2(z)$ (\ref{eq:H(z)}) contains a radiation term $\Omega_R$,
which now represents less that $10^{-4}$ of the energy budget
and is usually neglected. It represents about 10\% of the total at $z\sim 1000$.

One can generalize equation~\ref{eq:r(z)}  to alternatives to $\Lambda$ by replacing
the $\Omega_{\Lambda}$ term by the (reduced) density of the considered fluid.
For a constant equation of state, $\Omega_\Lambda \rightarrow \Omega_X(1+z)^{3(1+w)}$, and for a varying equation of state w(z), $\Omega_\Lambda \rightarrow \Omega_X \ {\rm exp} [ 3 \int_0^z\frac{1+w(z')}{1+z'} dz' ]$.

The discovery of accelerated expansion relied on supernovae at $z\simeq 0.5$
being observed to be fainter (i.e. at a larger distance) than expected from their low
redshift counterparts in a matter-dominated Universe. Inspecting
equation~\ref{eq:r(z)}, one notes that for a flat Universe ($\Omega_k=0$,
$\Omega_\Lambda=1-\Omega_M$), a positive $\Omega_\Lambda$ indeed increases
distances (at given $z$) with respect to $\Omega_\Lambda=0$. The argument also
holds for non-flat universes.

The angular distance $d_A$ is defined via the apparent angular size
$\theta$ of an object of comoving physical size D : $d_A \equiv
D/\theta$. Because photons follow null-geodesics of the metric
\cite{Etherington33}, we have $d_L = (1+z)^2d_A$ and hence $d_L$ and $d_A$
convey the same cosmological information. $d_L$ measurements rely on
``standard candles'' while $d_A$ measurements rely on ``standard
rulers''.

\subsection{Standard candles and Hubble diagram \label{sec:standard-candles}}
Standard candles are defined as objects of reproducible luminosity and
enable one to measure distances through their observed energy flux: $f
= \mathcal{L}/4 \pi d_L^2(z)$, where $\mathcal{L}$ stands for the
proper intrinsic luminosity (energy per unit source proper time).
Practically, we have to measure fluxes within bounded spectral
bands, and in order to compare objects at different redshifts we will do
our best to compare the same {\it restframe} bands which implies
measuring high-redshift objects through redder (observer) bands than
nearby objects. In supernova parlance, a Hubble diagram reports the
apparent flux (or magnitude) of standard candles as a function of
redshift. 

If one measures a Hubble diagram, we note that all fluxes scale with
the factor ${\mathcal L}H_0^2$, which conveys no information about
cosmological densities. In the cosmological analysis, this overall
factor will be estimated together with the other cosmological
parameters: it enters as a nuisance parameter in the analysis. Since
$H_0 d_L(z)$ does not depend on $H_0$, the Hubble constant enters
the problem only through this global flux scale. If the luminosity of the
standard candle ${\mathcal L}$ can be estimated without using the
flux-redshift relation, measuring ${\mathcal L}H_0^2$ from the latter
constrains $H_0$. We hence note that objects with an unknown
  luminosity do not help determining $H_0$ from a redshift-flux
  relation, and the initial determinations of the Hubble constant were
  off because a wrong luminosity was assumed.  We also note
  that determining $H_0$ entirely decouples from measuring the reduced
  densities from a Hubble diagram, and we refer to
\cite{FreedmanH010,Riess11} (and references therein) for a detailed
description of the role of SNe~Ia to measure $H_0$. We just
  mention here that, using the standard candle property, SNe Ia are
  used to extrapolate distances from local events to distant ones, and
  constitute the last step of the ``cosmic distance ladder'' used to
  establish the value of $H_0$ and hence the cosmic distance
  scale.

\subsection{What can we probe with distance data? \label{sec:distance-data}}

A set of distances as a function of redshift probes a weighted sum of
densities (equation \ref{eq:r(z)}), with weights evolving with redshift.
Since distance measurements deliver a single number at every redshift,
a redshift span is mandatory to break degeneracies of cosmological 
parameters \cite{Goobar95}.
As discussed in the above section, supernovae impose (at least) one extra
nuisance parameter, which illustrates why measuring a single cosmological
parameter like $\Omega_M$ in a matter-dominated Universe requires a finite
redshift interval. 

Although distances over a finite redshift interval formally break the
degeneracy of the sum, supernova surveys still deliver elongated
constraints in the $(\Omega_M, \Omega_\Lambda)$ or ($\Omega_M,w$)
planes (see e.g. \cite{Riess98b,Perlmutter99}), even for surveys with
the largest redshift span (e.g. figures 8 and 9 in \cite{Riess04}).

Accelerated expansion is indeed supported by supernovae alone,
  since the first measurements (discussed in \S \ref{sec:scp-hi-z}) and even
  more today (see figure \ref{fig:snls3-omol}), but with some caveats,
  (e.g. \cite{Cattoen07}). However, distances to supernovae are most
usually merged with other cosmological probes in order to produce the
tightest cosmological constraints and to decouple parameters. Standard
examples include anisotropies of the cosmic microwave background
(see e.g. \cite{Durrer01,Komatsu11}), and baryon acoustic oscillations 
(BAOs, see e.g. \cite{SeoEisenstein03,Blake11BAO})  measured in the
correlation function of galaxies.
BAO mostly deliver an (angular) distance constraint, with the same
degeneracies as luminosity distances, but the size of the measured
``standard rod'' also depends on cosmological parameters (mainly
$\Omega_M$ in the concordance model, see e.g. \cite{Eisenstein05}), and
the constraints hence exhibit different degeneracies than mere
distances. As shown in \cite{Blake11BAO}, BAO now deliver distance-redshift
relations similar to the ones obtained from SNe, but not yet as precise.
CMB anisotropies probe the same geometrical feature as BAO,
but at a much higher redshift ($z \simeq 1089$) which makes CMB
constraints even more complementary to distances to supernovae
typically at $z \lesssim 1\ {\rm to}\ 2$.

Even if supernovae on their own constrain only combinations of
cosmological parameters, they are currently key to constrain
dark energy: ignoring supernova distances degrades dark energy
constraints by factors of $\sim 2$ or more (see e.g. table 4 and fig 12 of
\cite{Komatsu11}).

\subsection{A few possibilities for dark energy \label{sec:dark-energy-models}}
Right after the discovery of the accelerated expansion,
modellers produced an enormous corpus of literature
devoted to implementations of dark energy. Those can be 
classified into three categories: the mere cosmological constant,
vacuum energy, and alterations to GR.
All these implementations involve new physics. A fourth
class of proposals for dark energy, free of new physics, consists in
noting that the effective Friedman equation in a non-homogeneous
matter-dominated Universe plausibly contains a term somehow mimicking 
dark energy, increasing with density fluctuations. We now briefly
describe these possibilities and refer to \cite{PeeblesRatra03} 
for more details. From an observational point of view,
the current tightest constraints (\S \ref{sec:surveys}, \ref{sec:ESSENCE-SNLS-SDSS})
are compatible with a static dark energy density, which 
can be incarnated by the cosmological constant.

\subsubsection{The cosmological constant}
In the original papers that introduce GR, the
cosmological constant was absent. It was later introduced by Einstein
himself in order to allow for a static Universe (see \cite{Carrol01},
and \cite{Weinberg89} for the historical aspects). Although the
addition became useless after the discovery of expansion, it is very
natural to allow for such a term in the Einstein equations. One even
expects such a term because it is algebraically indistinguishable from
the contributions as gravitational sources of all ``zero point
energies'' (also called vacuum energy) expected in theories of
particle physics.  ``The cosmological constant problem'' (see
\cite{Weinberg89} and references therein), is that the expected energy
density from zero point energies of particle physics theories is
``about'' 120 orders of magnitude larger than what astronomical
observations indicate\footnote{This $10^{120}$ ratio of energy
  densities comes from summing the modes of quantum oscillators up to
  the Planck scale.  This is arguable, but setting the cutoff to the
  electroweak scale ``only'' reduces the discrepancy to $\sim
  10^{60}$. In order to reconcile the zero point energies with
  cosmological densities, the cutoff should lie in the meV
  regime \cite{Weinberg89}.}.  Several
classes of solutions together with their weaknesses are discussed in
\cite{Weinberg89}.  For example, super-symmetry would solve the
problem because fermions and bosons then contribute equally but with
opposite signs to vacuum energy, but exact super-symmetry also cancels
the anomalous magnetic moment of the electron, which is now measured
to be non-zero to nine decimal places \cite{Odom06}.

So in practice, one is forced to assume that the contribution of
vacuum energy to gravitation sources is zero or extremely small, for
some yet unknown reason. This does not forbid the introduction of a
fundamental cosmological constant directly in the theory, and the
observational consequence is that the associated (dark) energy density
does not vary with time: it has an equation of state $w=-1$. In addition to
postulating one more fundamental constant, this incarnation of dark
energy raises the ``coincidence problem''
(e.g. \cite{PeeblesRatra03}): how come that the matter density
varying as $(1+z)^3$ and dark energy of constant
density now take similar values?

\subsubsection{Quintessence}
The word ``quintessence'' was introduced in \cite{Caldwell98} to name a
(fifth) component (after baryons, dark matter, neutrinos and
radiation) in the Universe, proposed there as a scalar field, with a
self-interaction parametrized with its potential energy. In most 
implementations, it 
interacts with other fluids only through gravitation. The dynamics of
these models depend on the initial conditions and the shape of the
potential, for which we have little theoretical guidance. In order to
be less arbitrary than the cosmological constant paradigm, these models
try to avoid the ``coincidence problem'' by proposing a potential in
which the scalar field density tracks the matter density, being
subdominant at early times and becoming slightly dominant at late
times, and where the dynamics exhibit an attractor, making the system
insensitive to initial conditions.  Power-law and exponential
potentials
(e.g. \cite{Wetterich88,RatraPeebles88,FerreiraJoyce98,Caldwell98,Wetterich08})
deliver dynamics roughly compatible with the observations. Such
potentials not only emerge in effective field theories (e.g. from
extra dimensions or standard model extensions like SUGRA), but also
provide the small numerical factors required to generate the very low
observed density \cite{BraxMartin99}. The equation of state of a
scalar field is confined to [-1,1]: at $w=-1$ the potential energy
dominates, while the kinetic energy dominates at $w=+1$. In order to
reproduce the observational constraints (i.e. $w\simeq -1$),
quintessence dark energy models should be essentially static at late
times. ``Natural'' quintessence models commonly exhibit late equations
of state around $w > \sim -0.8$ (e.g. \cite{BraxMartin99}) which falls
about 3 $\sigma$ away from the latest measurements. 
Without fine tuning, the effective mass of the quintessence field is
of the order of $H_0$ \cite{Linder06} (i.e. $10^{-33} eV$), and hence
its spatial inhomogeneities only develop on scales comparable to the
horizon today. Well-motivated quintessence models should ``protect'' such a
small mass from quantum corrections by some symmetry (as proposed in 
e.g. \cite{Wetterich08}).

As measurements increasingly favour a static dark energy density, the
quintessence scenario loses its appeal. In contrast, if we
detected an evolution of the dark energy density,
quintessence would likely become a very serious candidate.

\subsubsection{Alterations to gravity}

Since dark energy is required when assuming that GR
holds on large scales, one may question the theory itself. In the
GR framework, the evolution of structure formation is sensitive to the
expansion history and hence constitutes an alternative probe of the
Universe content.  Precision measurements of the growth of structure
in the past constitute a key test of the dark energy paradigm in the
GR framework and a strong justification of dark energy
space missions (see e.g.  \cite{Guzzo08,EuclidRB}). Correlations of
the cosmic shear (e.g. \cite{Linder03}) are a dark energy probe
sensitive to both structure formation and expansion history.

Before altering GR, one should note that it successfully passed
demanding tests in the solar system (e.g. \cite{Jaekel11}) and that
alterations to GR should only affect cosmological scales\footnote{ The
  ``Pioneer anomaly'' \cite{Anderson02,Levy09} is the only evidence a
  GR ``failure'' on small scales. Improved models of radiative
  transfers on the spacecraft have recently been presented which may
  account for the whole effect \cite{Francisco11}.}. Embedding our 4D
space-time into a 5D Minkowski space provides some room for
alterations to standard GR (\cite{DGP00}, and \cite{Deffayet02} for a
simplified introduction). Although one could originally fit the data
without dark energy in this framework \cite{Deffayet02}, improved data
sets now rule out some specific 5D models \cite{Fairbairn06,Fang08}.

The standard 4D GR action integrates $\mathcal{R}$ (the scalar curvature, 
not the scale factor) and 
alterations to this action are called ``$f(\mathcal{R})$'' theories
and have been considered as alternatives to dark energy. Quoting
\cite{Amendola07a}, ``All $f(\mathcal{R})$  modified gravity theories are
conformally identical to models of quintessence in which matter is
coupled to dark energy with a strong coupling'', while in the simplest
effective dark energy models (such as $\Lambda$), matter and dark
energy are coupled only gravitationally. One should hence expect a
significantly different phenomenology: comparing predictions to data,
several works \cite{Bean07,Amendola07a,Amendola07b} exclude a wide
class of $f(\mathcal{R})$ models and disfavour those as an alternative to dark
energy. 

\subsubsection{Back-reaction}

The standard derivation of Friedman equations assumes a homogeneous
Universe which is obviously a simplification. In \cite{Buchert00}, it
is shown that in an inhomogeneous Universe, a spatially averaged
Friedman equation may convey a term causing acceleration, only due to
matter inhomogeneities. This ``back-reaction'' mechanism relies on
the non-linear structure of GR, and at variance with
all other dark energy incarnations does not involve new physics. In
this scheme, dark energy is a ``mirage'' from structure formation and
the ``coincidence problem'' vanishes.

Since one should question whether alterations of the average Friedman
equation indeed alter the average distance-redshift relation in the
same way as for homogeneous models, explicit toy models have been
built.  By ``punching'' low-density spherical voids in a homogeneous
matter background, with a continuous density, one obtains a
Swiss-cheese Universe, for which analytic solutions have been derived
(see e.g. \cite{Marra08} and references therein). In a periodic
lattice of spherical voids, distances differ significantly from a
homogeneous dark energy free Universe, but do not mimic the
concordance model distance-redshift relation (figure 5 of \cite{Marra08}),
even with voids as large as 350 Mpc. Randomizing void locations makes
the disagreement even worse \cite{Vanderveld08}.

Another stream of back-reaction attempts develops second-order
perturbation theory in order to evaluate the contribution of
perturbations to source terms of Einstein's equations (see
e.g. \cite{Mukhanov97, Nambu02}). Although one finds contributions
that would contribute as some effective dark energy in the Friedman
equations, their physical meaning and gauge invariance have been
questioned \cite{Ishibashi06}.  It has also been argued that
corrections to large-scale distances from density contrasts are very
small (e.g. \cite{Behrend08}).

Because back-reaction is technically very involved, whether it constitutes
an alternative to dark energy is debated: compare, e.g., \cite{Buchert08}
and \cite{Ishibashi06}.  One should, however, remark that (perhaps for
excellent reasons, \cite{Buchert08}) the back-reaction mechanism
does not deliver yet the $\Lambda$CDM phenomenology without dark energy.

\section{Astronomical observation techniques \label{sec:obs-tech}}

This short section, for readers unfamiliar with astronomy, 
introduces the minimal concepts
and parlance of astronomical observation at optical wavelengths, and
we refer to \cite{Lena98} for a textbook devoted to observation techniques.

\begin{figure} [h]
\begin{center}
\includegraphics[width=1.03\linewidth]{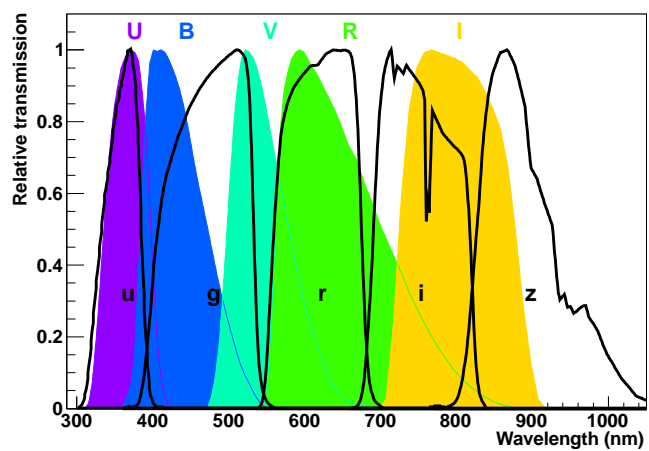}
\caption{Shape of commonly used photometric bands in the visible,
arbitrarily normalised.
$U$, $B$, $V$, $R$, $I$ refer to standard bands from \cite{Bessel90}, and $u$, $g$, $r$, $i$, $z$ to the SDSS bands from \cite{Fukugita96}. The latter integrate representative 
atmospheric absorptions experienced from the ground.
\label{fig:bands}}
\end{center}
\end{figure}

Astronomical observations at optical wavelengths consist mostly in
imaging and spectroscopy. Imaging is amost always carried out through
filters, in particular for measuring distances to SNe~Ia.  There are
two sets of common filters in the visible, the UBVRI ``Standard
filters'' \cite{Bessel90} and the ugriz SDSS filters \cite{Fukugita96},
displayed in figure \ref{fig:bands}. Practical implementations of
filters are always slightly different from the archetypes. In the
visible range, light sensors are now essentially always silicon CCDs,
which have a red cutoff at around 1.1 $\mu$m and deliver pixelized
images that can now reach millions of pixels per device.  Before the
1990s, images were mostly acquired with photographic plates (in
particular wide-field images), and precision photometry used vacuum
tubes, typically photo-multipliers. CCDs became common during the 1990s
and wide-field imaging is now carried out using mosaics of CCDs, with
tens to thousands of individual devices. Typical imaging observations
consist in several minutes of integration (where the integration time
is defined by a mechanical shutter) followed by about 1 min of
read out. Typical readout cadences are O(100 000) pixels per second,
in order to limit the contribution of electronic noise to $\sim$5
electrons per pixel. The dominant noise affecting ground-based CCD
images is most usually the shot noise of the sky glow, except for the
brightest sources where the shot noise of the object itself
contributes. Image quality (IQ), which directly determines the
photometric accuracy for point-like objects such as stars (and supernovae),
is usually described through the size of these point-like sources, and
the best ground-based sites routinely reach IQ below 1\arcsec~
FWHM in the visible. This figure is almost one order of magnitude better in space.
Astronomers usually report flux measurements as magnitudes defined as
$m=-2.5 log_{10} (\mathrm{flux})+C$, where $C$ is defined to attribute
a certain magnitude to some reference object, either real (such as
Vega) or some spectrum such as AB magnitudes (see \cite{Oke83}).  How
these magnitude systems are related to physical fluxes is described
in \S \ref{sec:photom-calib}.  ``Colour'' refers to magnitude
combinations insensitive to overall brightness, such as $B-V$. For
quasi-static stars, colours are a rough proxy for temperature.

For spectroscopy, one relies on a dispersing element, commonly a
grating. To select the light from the object of interest, the most
common technique consists in positioning a narrow slit in the image
plane and then dispersing the light in the direction perpendicular to
the slit. The slit can be replaced by fibre
optics \cite{Gunn-SDSS-telescope06} or systems of
lenslets \cite{Sauron00}. Multi-object spectroscopy (MOS) consists in
simultaneously acquiring spectra, either using fiber optics positioned
on the targets in the telescope focal plane, or machined multi-slit
masks. Currently, high-multiplex MOS instruments typically acquire
O(1000) spectra at a time. In space, one can perform slit-less
spectroscopy by placing a disperser in front of an imager, and still
separate the object light from the sky light, thanks to the low sky
brightness. The spectral resolution $\delta \lambda$ is commonly
expressed through $R \equiv \lambda/\delta \lambda$. As imaging,
spectroscopy in the visible also uses CCDs. 
Synthetic photometry from slit spectra is generally inaccurate, because
the fraction of light missing the slit is both unknown and 
wavelength dependent. Some spectrograph designs 
explicitly tackle this limitation (e.g. \cite{Sauron00,SnifsSPIE04}).

Observations in the near infrared (NIR) now rely on low band-gap
pixelized semiconductor devices (e.g. \cite{Hodapp96}), coupled to
integrated readout electronics. Observing around and above 1 $\mu$m
from the ground is difficult, because the atmosphere glow rises very
rapidly with wavelength, and also features numerous absorption
lines. Both atmospheric glow and absorption are variable on timescales 
of minutes to hours.

Observations from space benefit from a much improved IQ.
Compared to ground, the sky background is reduced, with a dramatic
improvement in the NIR \cite{Leinert98}. On the other hand,
ionizing radiation hitting the sensors affects the acquired data via
local charge deposition, and also degrades the performance with
integrated dose.

\section{Pioneering works on supernovae and the path to current classification \label{sec:pioneers}}
Already in the late 20's, Baade and Zwicky were searching transients
by visually comparing photographic plate pairs exposed for about 30 min, 
about one month apart, on the Palomar 18 inch telescope, in order to identify
transient events.
In 1934, they proposed the name ``supernovae'' \cite{BaadeZwicky34a}
to denote a new class of extragalactic  ``novae'' typically $10^4$ times
brighter than regular novae. They declared these two event classes 
as totally different phenomena and proposed a supernova rate of about 
one event per galaxy and per several centuries. They identified
that nuclear explosions are necessary to generate the enormous amount
of energy they observed. They were able to acquire spectra of their events,
which exhibit the broad features characteristic of fast moving ejecta,
and which do not show evidence of hydrogen lines. 
In 1941, a second class of supernovae was proposed \cite{Minkowski41}: 
these ``type II supernovae'' 
exhibit hydrogen lines and are typically fainter than the 
original ``type I'' supernovae.
The classification of type I supernovae was refined 
in 1985 \cite{WheelerLevreault85}, when a subclass of peculiar 
type I were isolated: those are typically fainter than regular type I
and miss the Si II broad absorption at $\sim$610~nm. Type Ia supernovae
were then defined as events that display Silicon and no hydrogen
whereas  type Ib and and Ic do not display silicon nor hydrogen. 
Type Ib display a strong helium line, and type Ic do not display it.

The current taxanomy of supernovae was sketched in 1993
\cite{DaSilva93}, and its current status and history are detailed in
\cite{Filippenko97}.  From a physical point of view, type Ia
supernovae are thermonuclear explosions, while all the other types are
core-collapse explosions. We refer to \cite{Wheeler03,Woosley86}
for introductory and more advanced material on theory of supernova explosions.

Core-collapse progenitors are massive stars (typically above 8
$M_{\odot}$ \cite{Smartt09}) which run ``out of fuel'' when the chain
of successive fusions stops on Fe nuclei: further fusions are
endothermic.  The gravitational pressure can then no longer be
balanced by heat radiation, and the core collapses
(e.g. \cite{Woosley05} and references therein). Iron nuclei are
photodisintegrated and protons are turned to neutrons by electron
capture and neutrino emission, producing a neutron star, on which the
outer shells fall and bounce.
The spectral characteristics depend mainly on the composition of the outer 
shells, where Ib and Ic have been stripped of H and/or He, and type II
have retained their hydrogen. Since the progenitors are massive, they
experience a rapid evolution (lifetimes of $10^{6}-10^{7}$ years), so 
that the core-collapse SN rate traces the star formation rate.
Core-collapse supernovae arise from a variety of progenitors and
exhibit a broad distribution of luminosity, even within each subclass
\cite{Richardson02,Li10II}.
As already noted, these events are on average fainter than SNe Ia \cite{Li10II},
and the population of the faint tail is still poorly known.

\section{Observational characteristics and diversity of SNe Ia\label{sec:sneIa}}

\subsection{Light curves and colours}

SNe~Ia emit most of their energy in the visible, and the
remaining part is essentially confined to near UV and near IR. During the
bright phases, the spectrum peaks around 400~nm (see Fig
\ref{fig:sn2003du}). Very rapidly, observers noted that SNe Ia (which
were still called SNe I) exhibit reproducible light curves
(e.g. \cite{Minkowski64}).

In the $B$-band (blue light, see figure \ref{fig:bands}), the light curve
rises in about 19 (restframe) days on average
(e.g. \cite{ConleyRiseTime06}). About 35 days past maximum the
light curve follows a quasi-exponential decay (see
figure \ref{fig:sn2006D}). If one sums the emitted energy, thus building
the bolometric light curve, the observed decay rate beyond $\sim$ 50
days after maximum is related to the $^{56}Co$ half-life: the $\beta$
and $\gamma$ decay chain $^{56}Ni \rightarrow ^{56}Co \rightarrow
^{56}Fe$ powers the light curve
(e.g. \cite{Colgate69,Colgate80,RuizLapuente98,Stritzinger06}, and \S
\ref{sec:explosions}).

In the $R$-band ($\sim$650~nm), the light curves exhibit a shoulder 20
to 30 days after maximum, which transforms into a second maximum
towards redder bands (up to $\sim$2${\rm \mu m}$), taking place 20-30
days after the first one (see Fig \ref{fig:sn2006D}); the feature is more prominent for bright
events, faint events might miss it.

The $B-V$ colour (blue versus green, where $V$ stands for
``visible'', see also figure \ref{fig:bands}) reddens during the first 50 days, indicating that the
emitting material is cooling. At about 40 days after maximum
(i.e. $\sim$ 60 days after explosion) the B-V colour changes slope and
then becomes bluer and bluer (e.g. figure 4 of (\cite{BranchRev92},
figure 1 in \cite{Riess96a}). Other colours have different behaviour.

\subsection{Spectra}

\begin{figure*}
\begin{minipage}[t]{0.48\textwidth}
\begin{center}
\includegraphics[width=\linewidth]{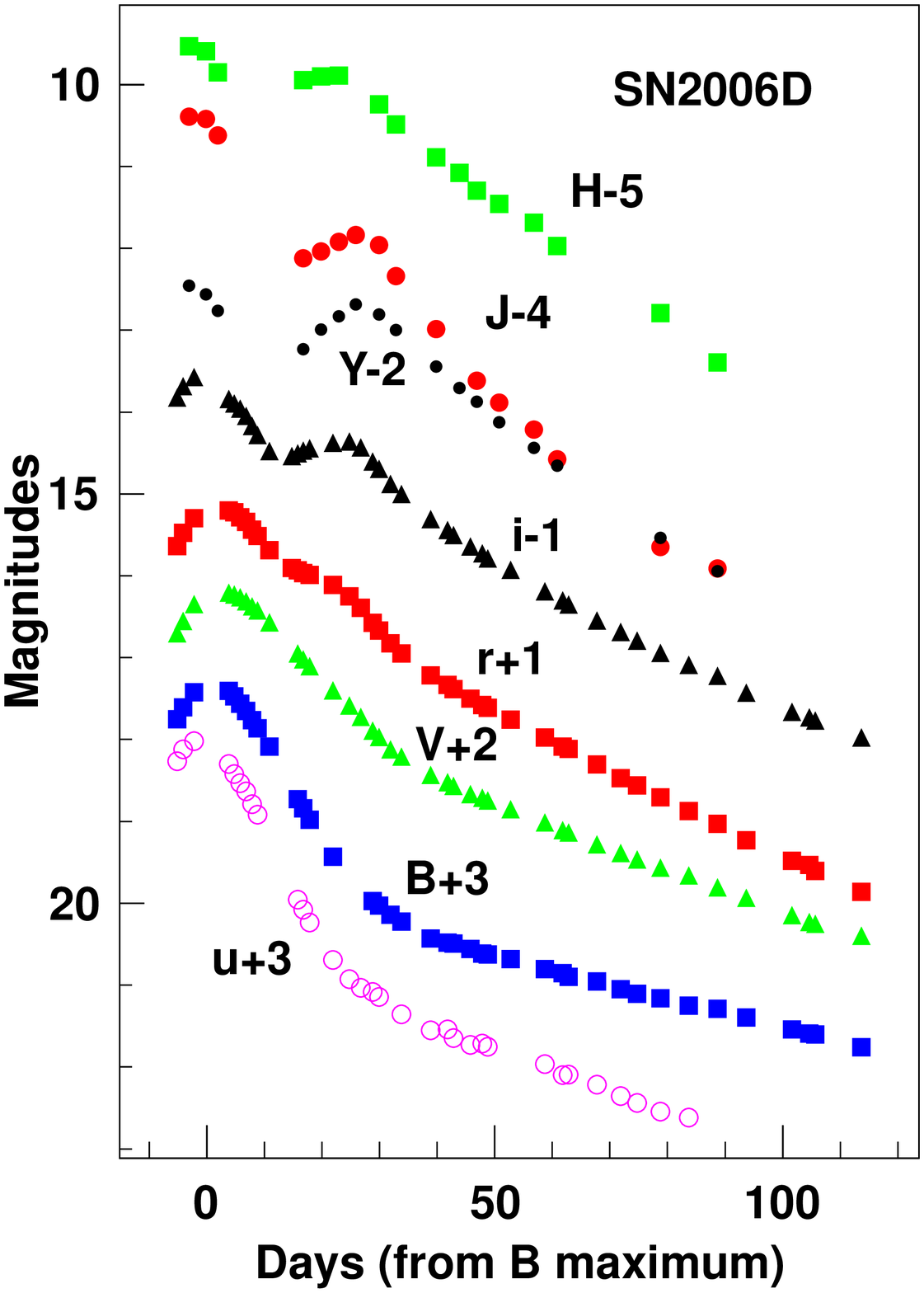}
\caption{Light curves of SN2006D (\cite{Contreras10}) in bands ranging from
the near UV (bottom) to the NIR (top). Bands labelled by upper case letters
refer to Vega magnitudes, lower case letters to AB magnitudes.\label{fig:sn2006D}}
\end{center}
\end{minipage}
\hfill
\begin{minipage}[t]{0.48\textwidth}
\begin{center}
\includegraphics[width=\linewidth]{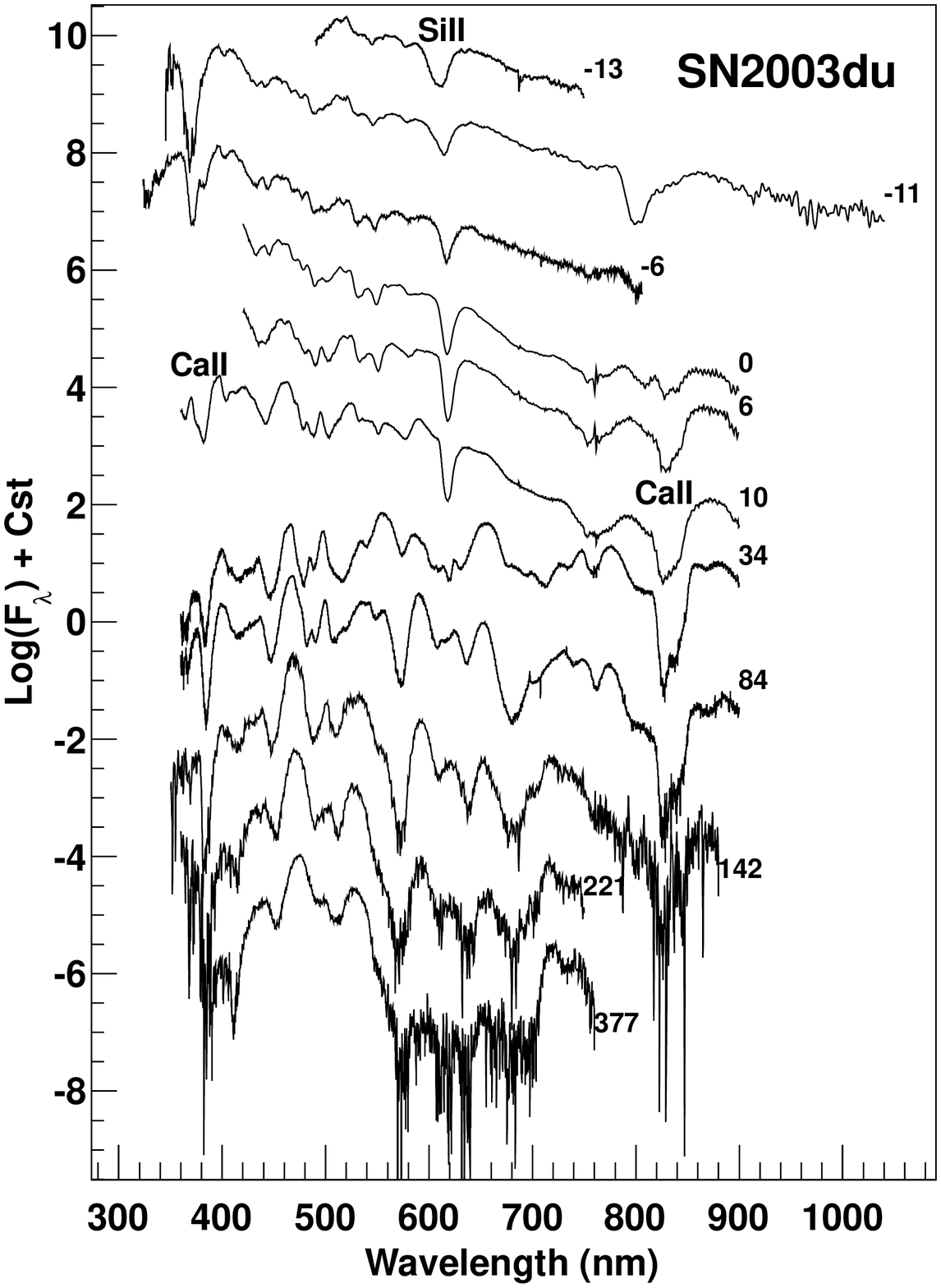}
\caption{Spectral sequence of SN2003du (\cite{Stanishev-SN2003du-07}).
Spectra are labelled by their phase w.r.t B maximum. Absorption features
are attributed to Si and Ca elements. Note that the absorption features
of the first spectra shift to the red with time.\label{fig:sn2003du}}
\end{center}
\end{minipage}
\end{figure*}

SNe~Ia spectra captured early or around maximum are characterized by
broad absorption lines of ``intermediate-mass'' elements, mainly Si
and Ca, shifted to the blue with respect to the SN host galaxy
redshift, because the absorption takes place in layers ejected at high
velocity towards the observer. The velocity of the absorbing region is
measured from the blue-shift of absorption features in the spectrum
and decreases with time (e.g. \cite{Branch88,Benetti05}, and
figure \ref{fig:sn2003du}). The light-emitting region (called
photosphere) is characterized by a material density below which
photons can escape.  As the supernova expands, its density decreases
and the light-emitting region proceeds towards the core of the 
star that just exploded (its nature will be discussed at \S \ref{sec:explosions}). The supernova eventually enters a nebular phase where emission
features dominate.

\subsection{Variability \label{sec:variability}}
SNe Ia exhibit a variability of the width of their light curve, even
after accounting for time dilation in order to compare restframe
widths.  This variability was initially noticed through a correlation
between the decay slope of the B ($\sim$450~nm) light curve and its
peak brightness (\cite{Pskovskii77,Pskovskii84}). This correlation was
refined in 1993 from a small sample of well-measured events
\cite{Phillips93}. The ``brighter-slower'', or Phillips relation
quantifies that slower decliners are also intrinsically brighter,
which is of paramount importance when measuring distances: all
supernova cosmology analyses implement it in some way. Still in the
$B$-band, it has been proposed to describe this variability of
light curves by stretching the time axis of a single light curve
template \cite{Perlmutter97,Goldhaber01}, and the rise and fall 
timescales seem to vary together for \cite{Goldhaber01} and
\cite{ConleyRiseTime06}, while \cite{Hayden10} finds them essentially
independent.

The distribution of decline rate (or stretch factor) has an rms
variation of $\sim$10\% \cite{Guy10}, exhibits a long tail of
hard-to-catch rapidly declining and faint events
\cite{GonzalesGaitan11}, as well as some bright and slow events.
Cosmological analyses have to exclude these extreme events, because
the empirical light curve shape parametrizations behave poorly for
those \cite{Guy05,ConleySifto08,Guy10}, and also because the
brighter-slower relation changes \cite{Phillips99,GonzalesGaitan11}.
Fast decliners are on average redder than slow decliners
\cite{Riess96b, Phillips99, Nobili08}.

We will discuss parametrizations of this light curve shape
variability when we come to light curve models in \S
\ref{sec:lightcurve-models} but all turn out to
be essentially equivalent, and all extract the relation of 
their parameter to luminosity from data.

SNe~Ia also exhibit variability of their colours (measured, e.g., at
maximum) even at a fixed decline rate. For example, the distribution of the restframe B-V colour
has an rms of about 0.1 mag. (see e.g. \cite{Guy10}) and a tail of highly
reddened events (e.g. figure 5 in \cite{Blondin09}). Different colours
of the same event seem to be well correlated: $B-V$ and $U-B$ are
related by a linear relation (see e.g. figure 7 in \cite{Guy10}), as
expected if colour variability is attributed to varying amount of
absorption along the line of sight. More precisely, colour relations
of SNe (such as $U-B$ versus $B-V$) slightly depend on the decline rate,
which is easily accounted for (e.g. \cite{ConleySifto08}). As one considers
bands bluer than $B$, the relation between 
different colours becomes poorer and poorer (e.g. figure 7 
in \cite{Guy07}, \cite{Ellis08, Balland09}), a tendency
confirmed by space-based UV photometry down to 180~nm 
\cite{Milne10}. As for extinction, one observes
that bluer supernovae are brighter, and that different colours do 
not vary independently: one colour seems enough to
parametrize this colour variability. The 
source of colour variability unrelated to the decline rate is still debated: it could be 
intrinsic to supernovae (e.g. \cite{Foley11}), it could be due to extinction 
by dust in the host galaxy, or more likely a mixture of both.
There is evidence of narrow Na absorption lines, indicative of dust,
particularly for highly reddened events (e.g. \cite{Blondin09} and 
references therein), which are usually excluded from cosmological analyses.
One might, however, note that the quantitative agreement of SNe colour variations
with known dust properties is poor (see
e.g. \cite{Tripp98,Guy05,ConleySifto08, Nobili08}), with one notable
exception \cite{Chotard11}.
We will discuss relevant
parametrizations in \S \ref{sec:brighter-bluer} \& \ref{sec:lightcurve-models} .

The brighter-slower relation
becomes shallower when going to redder wavelengths \cite{Hamuy96a},
and eventually vanishes at NIR wavelengths 
\cite{Krisciunas04,WoodVasey08}. The brighter-bluer relation also
becomes shallower, and ignoring both of them delivers accurate
distances in NIR bands in particular in the H-band ([1.5,1.8]$\mu$m) \cite
{WoodVasey08} (but see also \cite{Kattner12} and references therein).
These findings are so far limited to nearby events, because these 
NIR restframe bands become increasingly difficult to observe
as redshift increases.

SNe~Ia variability is often described as a one-parameter family, with
extinction by dust in the host galaxy causing some extra colour
variation. In practice, and whatever the source of colour variations,
cosmological works mostly consider these events as a two-parameter family,
where one parameter indexes light curve shape and the other some
colour, or colour excess.

\subsection{Correlations with host galaxy properties}

The physics in galaxies is broadly indexed by the mean age of their
stellar population. Early-type galaxies mainly contain evolved stars
and form little or no new stars, while late-type galaxies have varying
amounts of star formation going on. One might broadly rank galaxies
from big, red, passive and elliptical to small, blue, star-forming and
spiral or irregular. SNe~Ia are observed in all galaxy types, while
core-collapse supernovae are observed only in star-forming hosts (see,
however, \cite{Kawabata10}) because their massive progenitors have a
very short lifetime.

The SNe~Ia rate can be split across the galaxy types (see
\cite{Mannucci05,Sullivan06} and references therein), and one finds
that star-forming galaxies produce more than 10 times more SNe~Ia (per
unit stellar mass) than passive galaxies. Fast decliners happen
preferably in old stellar environments, or early-type hosts (figure 4. of
\cite{Hamuy96b}, e.g. \cite{Sullivan06, Lampeitl10,Brandt10}), and the cause
remains unclear. This probably explains the early finding that SNe~Ia
are brighter in late-type galaxies \cite{Pskovskii67}, and the
observation that the average decline rate decreases with redshift
\cite{Howell07}. Colours have been compared across environments and
found to be compatible \cite{Sullivan10, Lampeitl10}, which is somehow
surprising because if extinction by dust contributes significantly to
colour variations, one expects less extinction in passive galaxies
because they have less dust than active ones
(e.g. \cite{Tran01,Temi05}). Even more surprisingly, the largest 
homogeneous SNe~Ia sample to date recently indicated \cite{Smith12}
that supernovae in passive hosts are slightly but significantly redder
than those in star-forming hosts. A similar trend was 
reported earlier in \S 4.2.2 of \cite{Hicken09}. Some other
aspects related to host galaxies are discussed in \S\ref{sec:segregation}.

\section{The paradigm of SNe~Ia explosions\label{sec:explosions}}

\subsection{Progenitors and initial state at ignition}
The current model for type Ia supernovae explosions assumes that
a carbon-oxygen (C-O) white dwarf (WD) accretes mass from a companion
until it reaches the Chandrasekhar mass ($M_{Chan}\equiv 1.38 M_\odot$), at which 
the pressure from the degenerate electron gas no longer balances
the gravitational pressure \cite{Chandrasekhar31}. The temperature can then increase
enough to trigger carbon fusions \cite{Hoyle60}, and the nuclear flame can 
``burn'' the whole star material within a few seconds. A large consensus
has built up around this paradigm \cite{Livio00}, because it is 
supported by several observational facts.
\begin{itemize}
\item Fusions of C and O nuclei produce Si, S and Ca, which are observed
in the spectra (see figure \ref{fig:sn2003du}).
\item $^{56}Ni$ is the expected end of a fusion chain starting with C
  and O since further fusions would not release energy.  The $^{56}Ni
  (T_{1/2}=6.1 d) \rightarrow ^{56}Co (T_{1/2}=77d )\rightarrow
  ^{56}Fe$ seems to power the light curve, both because of its decay
  time
(e.g. \cite{Colgate69,Colgate80,RuizLapuente98,Stritzinger06}), and because of the evolution of line strength
  in the late spectra \cite{Kuchner94}.
\item The reproducibility of luminosity (e.g. \cite{Hamuy96a}) is a
  logical consequence of the explosion occurring at a threshold
  depending only on fundamental constants.
\item The lack of H and He in the spectrum is just due to their
  absence in the initial state.
\item The fact that SNe Ia are the only type of SNe observed in old
  stellar environments such as elliptical galaxies suggests a low-mass
  (hence long-lived) evolved progenitor.  White dwarfs form at the end
  of the evolution of stars of less than 8 $M_\odot$, which eject most
  of their outer layers, leaving a
  $\sim 1 M_\odot$ C-O compact star, too cool to host nuclear
  reactions, and slowly radiating its energy, possibly for billions of
  years.
\end{itemize}

This scenario is strongly supported by a quantitative study of the
outcome of nuclear reactions \cite{Nomoto84III,Hillebrandt00}. It, however,
faces a major issue: the path to the initial state is still
ambiguous (\cite{Livio00,HowellReview11} for detailed discussions of
the various issues). Since the Chandrasekhar mass has to be reached
from below, a mechanism for the WD to gain mass should be
proposed. Binary systems (i.e. bound systems of two stars) provide two
different paths, the so-called single-degenerate (one WD and some
other star) and double-degenerate (two WDs) scenarios. A recent survey
of WDs in Sun's neighbourhood \cite{Holberg08} reports that 25\% of
those are in binary systems, and 6\% in double-degenerate
systems. Among known low-mass WD binaries, more than 10
systems will merge in less than a Hubble time \cite{Kilic11}.

The single-degenerate scenario consists in a low-mass WD
accreting material from a higher mass companion star until it reaches
the Chandrasekhar mass (see e.g. \cite{Whelan73}). This requires that
the mass donor is a low-surface-gravity and non-degenerate star, and
late main-sequence stars or red giants are the most likely
candidates. H and He are the more likely elements to be accreted.
However, plausible accretion rates face serious issues (see p 206 of
\cite{Hillebrandt00} and reference therein): slow accretion rates do
not compensate for the mass loss due to novae eruptions, or helium
flashes; in contrast, at high accretion rates, the steady burning
of H and/or He should make these progenitors so bright that they
should be easily detectable. The low observed X-ray flux from
elliptical galaxies might forbid \cite{Gilfanov10}, or
not \cite{Hachisu10}, the single-degenerate scenario, at least in
elliptical galaxies.  However, the arguments relying on the luminosity
induced by material accretion may not apply if the progenitor
binary systems are enshrouded during their accretion phase.  We should
stress here that the binary system (and its environment) is
complicated enough to take the outcome of models with caution. For
example, the role of H and/or He flashes is also controversial
\cite{Cassisi98,Kato99}. The single-degenerate scenario is
  more and more challenged by the donor star escaping observations: none
  was found in pre-explosion archival images \cite{Maoz08,Li11}, nor
  within SN~Ia remnants (e.g. \cite{Schaefer12} and
  references therein); the expected UV and X-rays emission from the
  explosion material colliding with it was not seen \cite{Brown11,Brown12}.
  We cannot entirely rule out the single-degenerate scenario, but its
  viability will be questioned if the donor star remains elusive. Note
  that WD are generally far too faint to be detected in
  pre-explosion archival images.

The double-degenerate scenario assumes an initial binary system of two
WDs \cite{Iben84,Webbink84}, losing orbital angular momentum
by means of gravitational radiation. Stellar population synthesis
calculations predict the required statistics for mergers
\cite{Livio00}.  While this scenario naturally explains the absence of hydrogen
in the spectrum, there is no built-in mechanism that leads to a
reproducible state at ignition, a simple explanation for the
empirical reproducibility of SNe~Ia. On the other hand, this initial
state naturally provides a mechanism to explain extremely luminous events
such as SNLS-03D3bb, which seems to require a super-Chandrasekhar
initial state to produce enough $^{56}Ni$ to explain its apparent
luminosity \cite{Howell06}. However, the single-degenerate initial state
might also generate supra-Chandrasekhar amounts of $^{56}Ni$ \cite{Hachisu12}.
 A threat to this double-degenerate
initial state is that WD mergers lead to off-centre ignition
causing gravitational collapse rather than the expected thermonuclear
disruption (e.g. \cite{Saio85,Woosley86,Saio98}), but a recent
simulation actually obtains an under-luminous SNe~Ia from two 0.9
$M_\odot$ WDs \cite{Pakmor10}, but requires a narrow mass difference of the
involved stars.

\subsection{Explosion models}
Assuming the initial state can indeed be obtained, modelling the 
outcome of the explosion is at least a two-step process: one first has
to go through the nuclear fusion reactions, and second to evaluate
the radiation transfers in the expanding ejecta. Both aspects are 
physically and computationally extremely challenging \cite{Hillebrandt00}.

The production of elements through successive fusions during the
thermonuclear runaway depends on the reaction conditions, mainly
density and temperature: the reaction rates vary extremely
rapidly with temperature, and the production of iron-peak elements\footnote{``Iron peak'' refers to a local maximum, around Fe, in the abundance of elements as a function of atomic number. The expression refers here to Fe,Co and Ni.}
favoured under high density. There are roughly two regimes for the
propagation of the flame: deflagration (subsonic) and detonation
(supersonic), where deflagration allows the unburnt
material to expand before reacting.  Detonation models (see
e.g. \cite{Arnett79}) very efficiently convert the input material into $^{56}Ni$,
and then under-produce intermediate-mass elements (Si, Ca, S,...)
which constitute the prominent features of the spectrum around maximum
light. The famous W7 explosion model \cite{Nomoto84} produces intermediate-mass
elements via deflagration, and is a fair match to the observed light
curve and spectrum at maximum.  So, a deflagration phase is
required in order to reduce the density of the stellar material prior
to nuclear reactions in a $M_{Chan}$ WD
\cite{Khokhlov91,Hillebrandt00}, in order to avoid the nearly
exclusive production of $^{56}Ni$ \cite{Nomoto76, Nomoto84,
  Woosley86}.

However, pure deflagration models face some discrepancies with
observations \cite{Nomoto84,Khokhlov91}: overproduction of $^{54}Fe$
w.r.t to solar abundances (sometimes referred to as over-production of
neutron-rich elements), some indication of a brighter-faster relation
(while we observe a brighter-slower relation), and too narrow
an interval of expansion velocities of intermediate-mass elements.  To
cure these issues, \cite{Khokhlov91} proposes that after an initial
episode of deflagration, the flame transitions to detonation
(deflagration-detonation transition (DDT)), as observed in laboratory
chemical combustion experiments (e.g. \cite{Smirnov95}).  Through the
alliance of both combustion modes, intermediate-mass elements are
produced over a large range of ejection velocities. DDT simulations
hence improve the match to observed light curves and spectra
\cite{Hoeflich96}, and provide some hints that the density at
transition might source the explosion strength variability
\cite{Hoeflich96,Hoeflich96b}.

\subsection{Radiative transfer}
The explosion determines the chemical composition of the ejecta and
velocities of the various layers, mostly driven by the energy released
in the nuclear fusions. From this intermediate state, radiative
transfer calculations generate synthetic light curves and spectra,
which are mostly powered by the ${56}Ni$ $\beta$ and $\gamma$ decay chain
(e.g. \cite{Colgate69,Colgate80,RuizLapuente98,Stritzinger06}) as
already discussed. Broadly speaking, the supernova emits less energy
than it produces until the maximum, when both quantities are equal 
\cite{Arnett79}. Arnett's rule (\cite{Arnett82}), derived
from analytic radiative transfer analysis assuming a grey opacity, states that
the maximum light luminosity is proportional to $^{56}Ni$ mass
 and has been used to constrain the latter (e.g. \cite{Stritzinger06b}): 
it varies by about one order of magnitude from very fast to very slow 
decliners, and by about a factor of 2 for the range of decline rates 
commonly included in cosmological analyses. 

Radiation transfer codes simulate, in an expanding and inhomogeneous medium, the
conversion of nuclear radiation into the eventually observed light;
their sophistication has followed the increase in available
computing power. The complexity arises from two causes: first,
opacities depend on the ionization states of the atoms, which
themselves depend on the radiation flux and temperature, which in turn depend on
opacities. One thus has to find a consistent solution, and the
convergence speed is a major concern (see \cite{Lucy99b} and
references therein). Second, the opacities are driven by interactions
with atomic lines (scattering and fluorescence), which are extremely
numerous. In particular, the iron group elements exhibit a wealth of
lines in the near UV that mostly shift light to the red, but also
produce most of the UV flux around 300~nm
(\cite{Mazzali00,Sauer08}). Both the details of ``atomic data'' (i.e.
the choice of lines involved in the calculation) and the treatment of
ionization are shown to have a large influence on the predicted
light curves and spectra (e.g. figure 7 of \cite{Kromer09}).

The current state of the art mostly consists in 3D solutions.
Monte Carlo techniques \cite{Lucy05} are efficiently parallelized
and the microscopic physics can be incrementally refined. The various 
implementations (e.g. \cite{Kasen06,Kromer09} and references therein) 
mostly differ in the details of the micro-physics approximations. 
Hydrodynamic solutions now also accomodate 3D (e.g. \cite{Jack11} 
and references therein). Going to three
dimensions allows one to account for possible anisotropies, which have been
proposed as a potential source of diversity through the random viewing
angle \cite{Kasen09,Kromer09}.

\subsection{Models at late times}
At about 100 days after maximum light, emission lines start to
dominate the spectrum, indicating that the luminous material is
scintillating because of the energy deposition of positrons and
$\gamma$ rays. This feature has been used to constrain both the amount
of ionizing radiation and the nature of material around it, using
models much simpler than in the early phases.  At t$> \sim$200~days,
the quasi-exponential slope of light curves is different from band to band
\cite{Milne01,Lair06}, possibly due to the increasing amount of
(stable) $^{56}Fe$ which scintillates significantly in the
$B$-band \cite{Milne01}, and also because more and more $\gamma$-rays and
positrons escape \cite{Colgate80,Milne99}. The positron escape fraction
constitutes a handle on the magnetic field \cite{RuizLapuente98}.
Simulations of late SNe~Ia \cite{Milne99} and $\gamma$-ray
observations \cite{Knodlseder05} indicate that ``late'' positrons from
SNe~Ia might constitute a dominant contribution to the 511 keV
emission in the Galaxy. From the difference of the bolometric late
light curve slope with the $^{56}Co$ decay rate, one can model the
escape rate of $\gamma$ rays and obtain constraints on the ejected
mass (e.g. \cite{Colgate80,Stritzinger06}) and on the amount of synthesized
$^{56}Ni$ (\cite{Stritzinger06b} and references therein).

\subsection{Explosion models and cosmological use of supernovae}

The value of $H_0$ has been inferred from a Hubble diagram of nearby
supernovae and numerical simulations to derive the intrinsic
luminosity of events \cite{Branch-H0-92,Hoeflich96,Stritzinger-H0-05},
and values fully compatible with conventional determinations (mostly
relying on the cosmic distance ladder \cite{FreedmanH010}). This
approach is probably the sole quantitative contribution to date of SN
explosion models for determining cosmological parameters. Because of
the complexity of the explosion and the consequent light production,
models nowadays only broadly reproduce light curves and spectra and
cannot be used as templates for fitting data. If the brighter-slower
relation is qualitatively understood and physically attributed to
synthetized $^{56}Ni$ mass variations (e.g. \cite{Mazzali01} and
references therein), there is no consensus on what causes the
diversity of the latter (e.g. \cite{Ropke06}). Models hence do
not yet provide strong guidance to implement the brighter-slower
relation. The brighter-bluer relation is a worse case because there is
no consensus on whether variability of supernovae (independently of the decline rate) 
contributes significantly to it (see \S \ref{sec:brighter-bluer}).

Cosmologists, however, hope to get some insight from further improving
the physical understanding of the initial state, its environment
and the explosion mechanism(s). Ideally, we would improve 
light curve modelling, reduce the distance scatter, and gain 
insights into redshift-dependent systematic biases of distances.
Such goals might look excessively ambitious, but we should
remark that qualitative indications might be sufficient. For example,
if models point out some measurable indicator of the WD metallicity,
the relation of metallicity to intrinsic luminosity can be inferred from data.

\section{Observing and reduction techniques for SN surveys \label{sec:obs-reduc}}

Up to now, most SNe~Ia events that were used to measure
distances went through three different observing phases: finding the event,
identifying it from a spectrum (spectroscopic follow-up), and measuring its light curve
(photometric follow-up), possibly in several bands.  We will now discuss how the required
observations are conducted and handled, with some historical
perspective.

\subsection{Finding supernovae \label{sec:finding-supernovae}}
\begin{figure} [h]
\begin{center}
\includegraphics[width=1.0\linewidth]{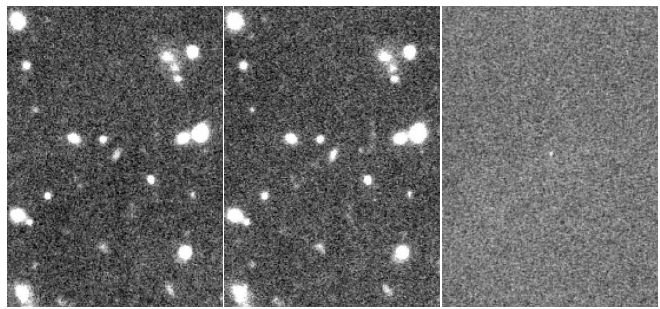}
\caption{Small portions of images of the same field. The leftmost was
  acquired on 8 September 1999 and gathers 3.6 h of integration
  on CFHT, using the CFH12k imager in I band. The central tile was
  taken 26 days later with the same instrument. The rightmost tile
  displays the image subtraction, where a positive excess is visible
  at the centre, at a brightness $I_{Vega}\simeq 24.7$ and about 6
  $\sigma$. This excess was spectroscopically identified as a type Ia
  event (SN1999fd) at z=0.87.
\label{fig:image-subtraction}}
\end{center}
\end{figure}

Essentially all supernovae are nowadays found using an apparently
simple technique: image subtraction. The concept consists in
subtracting a search image from a previously acquired image often
called reference. The subtraction is done pixel to pixel and point-like
positive excesses on the resulting image constitute supernova
candidates.  If the time gap between the two epochs is chosen to be around 2
weeks (in the restframe), found supernovae are likely to be still
rising and a prompt follow-up campaign allows one to capture the
maximum light (which allows one to safely estimate peak brightness,
and eases spectroscopic identification).  Prior to the subtraction
itself, both images have to be brought on the same pixel grid, on the
same flux scale and their point spread function (PSF, in practice the
shape of stars, which varies with atmospheric conditions) should be
homogenized. The process aims at making non-variable objects disappear
in the subtraction process. Since the concept of image subtraction was
initially proposed \cite{Hansen87} and demonstrated on the sky
\cite{Norgaard89}, the major improvement brought to the technique
consists in fitting the convolution kernel for PSF homogenization
\cite{Alard98}, by explicitly minimizing the subtraction
image. Subtractions are vulnerable to small defects of geometrical
alignment or PSF matching, or more generally, significantly different
observing conditions, and reaching the shot noise limit requires some
care. Image subtractions are then essentially always performed using
image pairs from the same instrument. One has to promptly find
candidates in order to rapidly obtain spectroscopic confirmations and
start the photometric follow-up as soon as possible.  The computing
load involved in finding supernovae has not decreased significantly in the two 
two decades since the area of silicon sensors (nowadays often
assembled into mosaics) used to image the sky, and the computing power
of silicon processors used to find supernovae have followed similar
slopes. An example of two images tiles and their subtraction
  is displayed in figure\ref{fig:image-subtraction}.

Searching supernovae requires an imager with as large a field of view
as possible.
The first sizable SNe~Ia sample (at z$<$0.1) with CCD-measured
light curves \cite{Hamuy96b} was found using photographic plates in the
early 1990s. Discovering supernovae in a somehow systematical way using
CCDs started around 1995: for example, the IAU telegram 6270 announced 11
supernovae at $0.35<z<0.65$, found using a 16'x16' single-CCD imager
on the CTIO-4m, and spectroscopically confirmed at the Keck (10 m).  We
note here that a sizable fraction of the distant events reported in the 
accelerated expansion papers \cite{Riess98b, Perlmutter99} (\S \ref{sec:scp-hi-z}) was
discovered using this CTIO-4m/Keck combination.

The SN finding techniques and instruments then improved rapidly: in
1999, the IAU telegram 7312 reports 20 spectroscopically confirmed
high-redshift supernovae (among which 3 events at $z>1$), discovered
using a 30'x40' imager at the Canada-France-Hawaii Telescope (CFHT, 3.6 m),
and confirmed at Keck.  Unfortunately, the photometric follow-up of
this wonderful harvest could not be fully conducted, illustrating
that actually measuring light curves of distant supernovae was 
becoming the heaviest part of the observing programme.

\subsection{Rolling searches}
In this 1999 discovery telegram (IAU 7312), one can easily check from
the event coordinates that several supernovae were found on average in
the same telescope pointing. This suggests the ``rolling
search'' approach, which consists in repeatedly imaging the same sky
patches, and using the image sequence both to find supernovae and
to measure their light curve. The spectroscopic identification phase
remains unchanged with respect to the ``traditional'' two-epoch search.

The technique becomes increasingly efficient as more supernovae are measured in
the same telescope exposure on average. In order to benefit from this
multiplex advantage, three ground-based surveys implemented this
technique (see \S \ref{sec:ESSENCE-SNLS-SDSS}): ESSENCE on
the CTIO-4m with the 36'x36' Mosaic camera (2002-2008), SNLS on CFHT
(3.6 m) using the 1 deg$^2$ Megacam imager (2003-2008) and the SDSS
SN search (2005-2007) on the SDSS 2.5 m with its 1.5 deg$^2$ fast
scanning imager. These three projects were successful in finding and
measuring events, essentially at the expected rate. Rolling
  searches using the Hubble Space Telescope (HST) imagers have also been implemented (\S
  \ref{sec:HST-searches}), however, with a lower multiplex advantage
  than the ground-based wide-field imagers.

The rolling search technique not only efficiently addresses the
observing time shortage that rules astronomic observation, but also
improves the data quality over the classical two-epoch search scheme
followed by photometry. First, SN light curves are sampled
independently of their phase, since observations are not conducted for
a specific event; in particular, SN light curves contain very early
points, taken even before the event was actually detected. Second, the
search is indeed easier than in the early times because photometry for
distance estimation requires deeper images than for mere search.  One
also usually has more than a single image and epoch to turn a light
excess into a candidate. Finally, these repeated images can be stacked
to reach considerable depths, especially if the survey lasts
several years. The wide-field stacks usually deliver the photometric
characterization of the supernova galaxy hosts (e.g. \cite{Sullivan06,
Lampeitl10}). They can also be used for science topics unrelated to 
supernovae, which can even be the main drivers for the survey, as
happened, in particular, for some heavy SN programmes on the HST 
(\S \ref{sec:HST-searches}).

\subsection{Spectroscopy of candidates}
Candidates from supernovae searches have to be confirmed, most of the
time using spectroscopy. From reported discoveries, one can draw the
crude rule that the telescope used for spectroscopy is twice as large
in diameter as the one used for the search via imaging. SNe~Ia
supernovae at $z \lesssim 1$ are now found using 4 m class telescopes,
and the spectroscopic follow-up is conducted on 8 m class facilities, where
a target at $0.8<z<1$ requires integrating for 1-2 h (see
e.g. table 1 in \cite{Howell05}).  For nearby supernovae at $z \lesssim 0.1$,
searches are typically conducted with 1 m telescopes, and spectroscopy
is done in 10-20 mn exposures on 2 m telescopes. In order
to use the observing time as efficiently as possible, spectroscopic 
observations aiming at classification are usually conducted around
maximum light, if possible. For distances to
SNe~Ia, besides positive identification of the event against other SN
types, spectroscopy is expected to provide the redshift value, in order
to place the event on the Hubble diagram.

The needed spectral resolution is modest: supernovae have no narrow lines,
and a resolution ($\lambda/\delta\lambda$) of about 100 is enough to sample
properly the peaks and troughs of SNe~Ia spectra. One, however, usually uses
spectral resolutions around 1000, in order to reduce the impact of 
telluric absorptions and sky line emissions, and in order to properly 
measure narrow lines of the host galaxy spectrum, which constitute the 
best redshift indicator.

At variance with imaging, there is currently no practical way to
efficiently multiplex spectroscopic observations of supernovae because
the field of view of most spectroscopic instruments is too small. The
events are then observed one at a time. When using a multi-object
spectrograph, one can still observe other targets (typically galaxies)
in parallel with the SN candidate.

\subsubsection{Host galaxy subtraction}
Supernovae are point sources while their host galaxy is not. The size
of the supernova does not depend on redshift while galaxies get
smaller as redshift increases. Hence, the fraction of host galaxy light mixed
with the supernova increases with redshift.  For photometry, this is
addressed by subtraction (or some equivalent technique, see
\S\ref{sec:sn-phot-techniques}), but for slit spectroscopy, subtracting
a supernova-free observed galaxy spectrum is not practical. Moreover,
most surveys do not allocate observing time to collect such spectra.

Increasingly sophisticated approaches to address the host galactic
contamination of supernova spectra have been developed, mostly for
high-redshift surveys. One can simply ignore the problem, thus
selecting against SNe on bright hosts. One can attempt to synthesize
the observed spectrum from template libraries of both supernova and
galaxies \cite{Howell05}. In \cite{Blondin07SNID}, both continuum and
narrow lines are filtered out (in wavelength space) from the acquired
spectrum, thus efficiently removing most of the contaminating host
galaxy features, however at the cost of altering the SN spectrum. In
long-slit spectroscopy, one can attempt to separate the point-like
component (the SN) from a more extended background (the host galaxy),
either with some spatial filtering \cite{Blondin05}, or by relying on
the distribution of galactic light along the slit estimated from
multi-band imaging \cite{Baumont08PHASE}. In \cite{Ellis08}, galaxy
models are fitted to supernova-free multi-band imaging measurements
of the host galaxy for subsequent subtraction from the observed spectrum.
\subsubsection{Type and redshift}
\begin{figure} [h]
\begin{center}
\begin{minipage}[b]{0.50\textwidth}
\includegraphics[width=\textwidth]{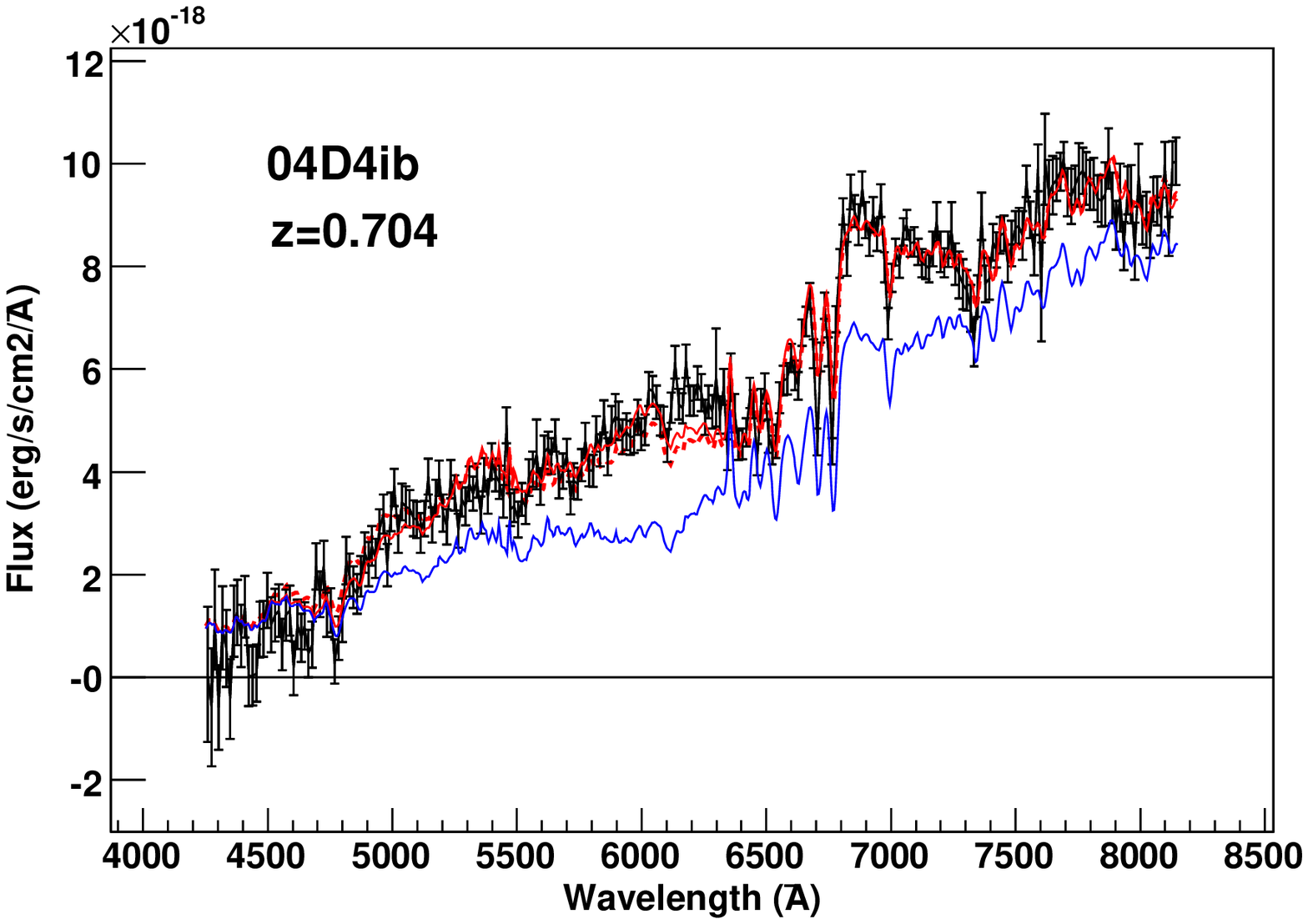}
\end{minipage} \begin{minipage}[b]{0.50\textwidth}
\includegraphics[width=\textwidth]{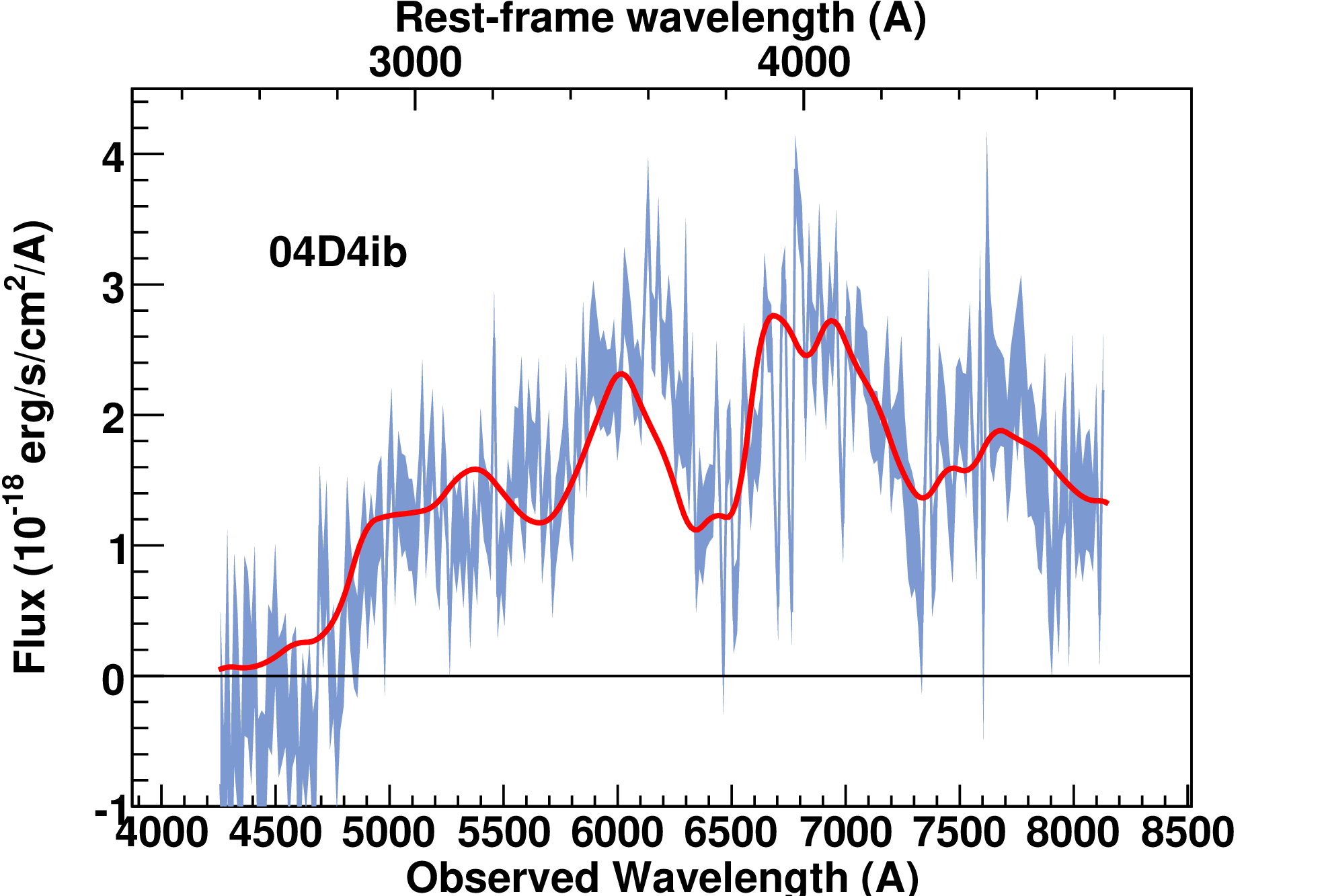}
\end{minipage}
\caption{On the left, measured spectrum (in black) of the SNLS-04D4ib
  supernova candidate on a relatively bright host, acquired with FORS1
  on the VLT, at maximum light.  The redshift $z=0.704$ is
  unambiguously obtained from the 6800~\AA (observer frame) jump due
  to the galaxy spectrum, confirmed by two H narrow absorption
  lines at 7000~\AA (H-$\delta$) and 7400~\AA (H-$\gamma$). A
  simultaneous fit to the measured spectrum and the light curves of the
  event is carried out using the SALT2 model (\S \ref{sec:SALT}),
  where the spectrum is modelled as the sum (red line) of a galaxy
  (blue line) and SN Ia templates. On the right, the SN spectrum
  obtained by subtracting the best fit galaxy template (S0, blue curve
  on the left) from the measured spectrum, compared to the fitted SN Ia
  model spectrum (red line). This spectrum displays a broad absorption
  (characteristic of SNe) around 3800\AA\ resframe, attributed to CaII,
  and was identified as a SN~Ia because the general shape of the
  spectrum leaves no ambiguity at this phase (known from
  photometry). Figure reproduced with permission 
from \cite{Baumont08PHASE}.
\label{fig:spectro-example}}
\end{center}
\end{figure}

At high S/N, the identification of a supernova from its spectrum does
not usually require any sophisticated method; problematic cases occur
for events departing significantly from the archetypes.  At higher
redshift (and smaller S/N), the SN identification is carried out by
comparing its spectrum to high-S/N archetypal templates, for various
phases and SN types, where the quality of the match is often
quantified via least squares
(e.g. \cite{Howell05,Blondin07SNID}), possibly matching both light curve and spectrum \cite{Balland09}.  When present, narrow
emission or absorption lines of the host galaxy provide the redshift or at least
redshift hypotheses, but one may also add the redshift to the searched
parameter space. Redshifts are typically measured to a precision of
0.001 from galactic features and to 0.01 from SNe features
(e.g. \cite{Howell05}, figure 1 in \cite{Matheson05}, figure 7 of
\cite{Foley09}). The quality of the measured spectrum obviously plays
a key role in the identification confidence level. Regarding SNe Ia, a
positive identification consists in the observation of Si and/or Ca
spectral features (see e.g. \cite{Howell05} for a practical
definition of identification confidence levels). 
figure~\ref{fig:spectro-example} illustrates
the separation of galaxy and SN spectra and the obtention of
redshift and type of a high-redshift SN candidate.

\subsection{Measuring light curves and photometry techniques}
\label{sec:sn-phot-techniques}

The collection of photometry points along the SN light curve can be
carried out with a small field instrument provided there are enough
stars in the images to serve as photometric and geometric anchors.
As indicated above, surveys targeting high redshifts now use the
repeated imaging approach and hence do not need dedicated and pointed
imaging observations. 
 
Supernova photometry consists in measuring the flux of the supernova
in the survey images. Since one cannot spatially separate the light of
the supernova from that of its host galaxy, images of the field before
the supernova explosion or at least one year after are usually
required for this task. Explicit or implicit subtraction of these
supernova-free images from supernova-on images is generally
necessary. Rolling searches obtain these supernova-free images within
their core programme if they monitor the same field for at least 
two years.

Astronomical photometry involves two measurements: one has to measure
the source of interest and some standard stars with the same
instrument. So, astronomical photometry
indeed delivers flux ratios, and we will discuss the calibration of
the standards themselves in \S \ref{sec:cal-standards}. For
ground-based observations, the transparency of the atmosphere has to
remain sufficiently constant between both sets of observations for the
calibration process to deliver anything useful, and astronomers label
a night as ``photometric'' if the ensemble of standard stars observed
throughout the night indicates comparable atmospheric extinctions. In
order to salvage non-photometric (or uncalibrated) observations of
supernovae, one can rely on the assumption that stars in the same
images as the supernova are on average non-variable. The task of SN
photometry hence consists in delivering the flux ratio of the
supernova to neighbouring stars, and the calibration process will have
to deliver, once and for all, the flux ratios of these neighbouring stars
to some standards. Space-based photometry can rely on the usually
excellent long-term stability of the instruments and may follow a
simpler calibration path (see e.g. \S 2.1 in \cite{Knop03}).

For faint sources, photon shot noise constitutes, in practice, the main
source of measurement uncertainty, and one will seek a statistically
optimal flux estimator, in order to maximize the number of measured
events within a fixed observing time allocation. In fact, observing
times are usually tailored to deliver some predefined signal-to-noise
ratio, assuming an optimal flux estimator. Since both the supernova
and surrounding stars are point sources, they have the same spatial
shape, namely the Point Spread Function (PSF), which can be modelled
on each image using bright stars. One can model the SN observations as
a time-variable point source at a fixed position on the sky, possibly
on top of a non-varying galaxy.  Fitting such a model to the image
series delivers statistically optimal estimators of the flux at all
epochs, together with their covariance matrix. Practically validated
implementations of a simultaneous fit to an image series are described
in at least two instances \cite{Astier06, Holtzman08}, where the
latter avoids image re-sampling. Other approaches to SN photometry
are bound to be statistically less or as efficient, as stated by the
Cram\`er-Rao inequality.

Statistical efficiency is usually not a serious concern for nearby
supernovae and the subtraction of supernova-free images from follow-up
images with subsequent photometry proves to be sufficient for
statistical uncertainties to be subdominant (see
e.g. \cite{Contreras10}).

Measuring point sources using their shape is referred to as PSF
photometry, and is obviously adversely affected by imperfections of
the PSF model. However, flux ratios are not affected by imperfections
of the PSF model, as long as the latter enters in the same way into
both terms of the ratio. This is not the case for standard PSF
photometry, optimized to fight shot noise (e.g. \cite{Irwin85}), which
makes the ratio of faint to bright stars depend on the fidelity of the
PSF modelling. In contrast \cite{Astier06,Guy10}, one might use
for all brightnesses the flux estimator optimal for faint stars. Faint
sources are still optimally measured, flux ratios become independent
of the PSF fidelity, and fluxes are linear functions of the data. One might
indeed regard the latter as a requirement when measuring fluxes for
luminosity distances.

One important concern for PSF flux measurement of faint sources is
that fitting simultaneously flux and position involves a flux bias at
low signal to noise (see e.g. appendix B in \cite{Guy10}), which might
have a non-negligible impact when supernovae become faint. The obvious
remedy consists in fitting a single common position to all images and
even to all bands. In order to limit systematic biases of flux ratios,
one should then enforce the same constraint for measuring fluxes of
the surrounding stars. For surveys spanning several years, one may
have to account for proper motions of stars \cite{Holtzman08}.

One might have noted that SN measurements involve explicit or implicit
subtractions, while star measurements do not. This difference might
systematically bias flux ratios.  For this reason and for mere method
validation, supernova photometry is commonly verified by inserting
artificial supernovae (either synthetic, or copies of
real stars) into real images, in order to bound the bias of recovered
flux ratios below dominant systematic
uncertainties \cite{Schmidt98,Astier06,Holtzman08}.  
Rolling searches allow one
to measure pre-explosion SN fluxes, and checking that those average to
zero indicates that galaxy fluxes are properly subtracted
\cite{Holtzman08,Guy10}.

Although astronomers usually report magnitudes, it is mandatory that
light curves points are reported in flux space, because some low-S/N 
measurements unavoidably happen to be negative, and retaining
those is mandatory for flux estimates to remain
unbiased. For the same reason, light curve fitters should operate in flux
space \cite{Guy07,ConleySifto08,Kessler09}, but might report their 
results in magnitudes (as commonly done).

\section{Light curve fitters, distance estimation \label{sec:fitters}}

\subsection{K-corrections}

In order to compare supernovae, their fluxes have to be expressed in
the same way, and in particular using the same restframe bands.
Different supernovae are generally not measured using the same photometric
setup: first, since the quality of cosmological constraints improves
with redshift lever arm, nearby and distant supernovae are measured
with different telescopes and hence different spectral bands. Second,
even if all supernovae were measured using the same telescope, the
filters used for the measurement map differently in the supernova
restframe when the redshift varies.  We hence need to somehow
transform the photometric measurements carried out with an arbitrary
setup into what some well-defined instrument, not necessarily real,
would have measured.

In the astronomical literature, transforming photometric data obtained
with some filter into what would have been measured with some other
filter is called K-correction. It is usually presented as a magnitude
correction and was originally designed to place galaxies on a Hubble
diagram \cite{Humason56}; these galaxies were measured in similar
filters but at different (and moderate) redshifts. The principle is
simple (e.g. equation (B7) in \cite{Humason56}): correct the flux
measurement $f(T_2)$ by a multiplicative factor computed using an
assumed spectrum and the involved band passes:
\begin{equation}
f(T_1) = f(T_2) \frac{\int S(\lambda)T_1(\lambda) d\lambda}
{\int S(\lambda*(1+z))T_2(\lambda) d\lambda}
\label{eq:k-corr-flux}
\end{equation}
where $T_2$ is the transmission function of the filter that was used
for the measurement, $T_1$ is the transmission function of the
filter that would have been used ideally, and $S$ is the assumed
restframe spectrum of the object (at redshift $z$). Originally, the
concept was developed in a framework where $T_1$ and $T_2$ denote the same band.
Obviously, the transformation depends more and more on
the knowledge of $S$ as the overlap between $T_1(\lambda)$ and
$T_2(\lambda/(1+z))$ decreases. Note that the actual spectrum of the
object is generally not known, because if it were, the required flux
would be extracted from this spectrum, or spectra would even be
compared directly. So, we are dealing with some approximation of the
spectrum, and we will now discuss how the K-correction depends on
inaccuracies of this spectrum.

Let us consider a supernova at $z=0.05$: transforming the
measurements from the observer filter to the same restframe involves a
filter shift of 1.05, while standard filters have a relative width (width of
the bandpass over central value) of about 0.25 (see Fig
\ref{fig:bands}). It means that the numerator and denominator of
equation \ref{eq:k-corr-flux} overlap for about 80\% of their values, and
hence the correction does not depend too much on the precise knowledge
of S.  This is the situation that prevails for low-redshift samples
(e.g. \cite{Hamuy96a}).  At $z\simeq 0.25$, the overlap between
$T_1(\lambda)$ and $T_1(\lambda/(1+z))$ becomes essentially null, but
using the next redder filter as $T_2$ restores essentially a full
overlap. For example, using the standard UBVRI filters 
(figure~\ref{fig:bands}),
V($\bar{\lambda}=5450$) is a very good approximation of restframe
B($\bar{\lambda}=4350$) at $z=0.25$.  So, as common sense indicates,
the measurement filter should follow as much as possible the red-shifted 
target filter. This gives rise to the cross-filter
K-correction concept proposed in \cite{Kim96}, where $T_1$ and $T_2$
are assumed to be different (in the same frame).  When K-corrections
are expressed in magnitudes, transforming for different filters
introduces an additional term accounting for the different definition
of magnitudes in the involved bands.

The worst case for K-corrections happens roughly when $T_1(\lambda)$
and $T_2(\lambda/(1+z))$ overlap by about half of their width: then
about half of the flux in the target filter has to be derived from the
assumed spectrum. In \cite{Nugent02}, a detailed study of
K-corrections for supernovae concludes that rather than trying to
obtain detailed knowledge of the SN spectrum, one should first pay
attention to the spectrum broadband slope (i.e. colour), which varies
from supernova to supernova. The scheme proposed in \cite{Nugent02}
consists in tilting a template spectrum in order to reproduce a
measured colour, and then using this tilted spectrum to compute
K-corrections. One can note that to first order, the details of how
the spectrum is tilted do not matter. In order to apply this tilting
scheme, one has to measure an SN colour, which happens to be essentially
mandatory for distance estimation.

The Supernova Factory survey  (\S~\ref{sec:nearby-samples})
collects spectro-photometric SN ``light curves'', or photometric 
spectral sequence, thus eliminating the
need for K-corrections of any kind when comparing supernovae.

\subsection{Brighter-bluer relation and extinction\label{sec:brighter-bluer}}
Astronomical observations in the visible are essentially {\it always}
subject to some sort of extinction, generally by dust. For example,
the Milky Way contains dust, and extragalactic observations (including
supernova measurements) have to be corrected for Milky Way dust
extinction, nowadays using the maps derived from dust emission in the
far infrared \cite{Schlegel98}.

The extinction by dust has been characterized mainly in the Milky Way
\cite{Cardelli89}. By comparing colours of stars of the same spectral
types, but affected by different line-of-sight extinction, the authors
could derive an empirical ``extinction law'' (called CCM law after Cardelli, Clayton and Mathis) which relates
extinction at various wavelengths.  The total extinction is found by
extrapolating the found wavelength dependence to a null photon
energy. The first obvious parameter that characterizes a line of sight
will then be its optical depth, i.e. the amount of absorbing or
scattering medium. This is conventionally parametrized (and in particular in
\cite{Cardelli89}) by the reddening of the $B-V$ colour (defined in \S \ref{sec:obs-tech}). Naming $E(B-V)$ the colour excess attributed to
dust, the extinction (in magnitude) at a given wavelength reads $A(\lambda)=
E(B-V)*R(\lambda)$, where $R(\lambda)$ is the ``extinction law'',
directly related to the cross section of photons on the dust. Once
$R(\lambda)$ is known, one may deduce the line-of-sight extinction
from the reddening of colour. Identical objects extinguished
by different amounts of intervening dust exhibit colours which lie
along lines in colour-colour spaces.

The dust extinction law should depend on the nature of the intervening dust 
and \cite{Cardelli89} proposes a one-parameter family of extinction curves,
parametrized by $R_V\equiv R(\lambda_V)$. The Milky Way dust is characterized
on average by $R_V=3.1$.

SNe~Ia exhibit a brighter-bluer relation (see e.g. \cite{Tripp98} and
references therein), as one would expect from dust, and also exhibit a
strong correlation of colour relations
\cite{LeibundgutPhd,ConleySifto08}, which constitutes a tantalizing
analogy. However, this justifies parametrizing the colour variations
of supernovae in the same way as dust extinction, but does not {\it
  impose} the assumption of CCM extinction laws.  As we will discuss later,
supernovae allow one to carry out a similar analysis as the one done
to derive CCM extinction laws, in order to derive an average colour
law for supernovae.  Using then the standard candle property of
SNe~Ia, one can extract the equivalent of $R_V$ from the supernova
data.  One should note that in order to measure an SN colour, one has
to carry out measurements in at least two bands.

\subsection{Supernova empirical light curve models \label{sec:lightcurve-models}}

We are now in a position to express some minimal requirements of
supernova models used to ``compress'' the photometric data
characterizing an event: in order to derive a distance we need a
brightness (anything that scales linearly with the observed flux),
some measure of the decline rate and some measure of colour.

If the measurements are dense enough along the light curve, there is
no crucial need for an explicit model, which was the approach
originally followed for the $\Delta m_{15}$ paradigm
\cite{Phillips93}. From well-sampled light curves, one can build smooth
discrete templates with different $\Delta m_{15}$ values, where this
quantity indexes the decline rate measured as the magnitude difference between peak
and 15 (restframe) days later.  An
interpolation scheme (figure 1 of \cite{Hamuy95}) allows one to fit possibly
sparse observations.  Templates have been continuously refined
\cite{Hamuy96d, Phillips99, Germany04,Prieto06}.  The $\Delta m_{15}$
model was in particular used in the Calan-Tololo survey analysis
(\cite{Hamuy96a,Hamuy96c} , \S \ref{sec:nearby-samples}), and one of
the acceleration discovery papers (\cite{Riess98b}, \S
\ref{sec:scp-hi-z}). Recently, the Snoopy model \cite{BurnsSnoopy11}
revisited the $\Delta m_{15}$ paradigm and extended it to NIR.

Light curve templates for standard visible and NIR bands were
determined in \cite{LeibundgutPhd}, mainly to assess the empirical
reproducibility of SNe~Ia, and could then be used to fit SN
photometry. In order to account for the decline-rate variation, the
``stretch'' paradigm \cite{Perlmutter97} proposed to stretch the time
axis of these templates in $B$ and $V$ standard bands (\S
\ref{sec:obs-tech}). As data improved in size and quality, these
templates were refined and extended to the $U$-band
\cite{Goldhaber01,Knop03}. The stretch paradigm works poorly for bands
redder than $R$. The stretch approach was used in one of the
accelerated expansion discovery papers (\cite{Perlmutter99}, \S
\ref{sec:scp-hi-z}).  All the models sketched above rely on light curve
templates in standard bands. 

The Bayesian Adapted Template Match (BATM) method \cite{Tonry03}
relies on about 20 well-measured events (spanning a decline-rate
range) and about 100 spectra to predict the observed light curves in a
given instrument at a given redshift, for a given extinction. Using
an extinction prior, it eventually delivers a distance estimate,
sensitive to the prior choice. It was used to extract cosmological constraints
from the first large SN compilation \cite{Tonry03}.

Currently, there are two main streams of SNe Ia empirical models in the 
visible used to derive cosmological constraints: MLCS on the one hand,
SALT, SALT2 and SiFTO on the other, and we will now describe those.

\subsubsection{The MLCS model \label{sec:MLCS}}

The Multi Light Curve Shape (MLCS) model consists in a one-parameter
family of light curve shapes for standard visible bands $BVRI$ (i.e. from
400 to 900~nm, figure \ref{fig:bands}). This complete model of SN
light curves depends on three parameters (plus a reference date): a
distance modulus, a brightness offset (due to the brighter-slower
relation, which indeed indexes light curve shapes), and an extinction
value (in the $V$-band). At variance with essentially all other SN
light curve models (BATM is another), MLCS incorporates the
brighter-slower and brighter-bluer relations, which imposes to train
it on events at known distances \cite{Riess95,Riess96a}. The initial
modest training sample of 12 events (by 1995) was later enlarged to
about 100 events (around 2002) and the model extended towards the blue
by adding the $U$-band (320-400~nm) \cite{Jha07-MLCS2k2}. This second
version is named MLCS2k2. Prior to fitting, measurements have to be
translated into standards restframe bands using K-corrections.

The MLCS(2k2)  model assumes that extinctions follow some known distribution
\cite{Riess96a} (the ``extinction prior''), which is not mandatory but
improves the formal uncertainty of colour measurements (figure 4 
of \cite{Riess96a}) and enforces
positive extinctions. The derivation of this prior in
\cite{Jha07-MLCS2k2} is sometimes described as a {\it statistical}
separation of intrinsic (internal to the supernova) and extrinsic (due
to dust extinction) colour variations, which means that there is no
separation on an event per event basis. This statistical separation
only matters for the shape of the prior. The other assumption of
the model is that after accounting for colour variations correlated with
light curve shape, the average colour relations are due to extinction
by dust, similar to the average Milky Way dust.

Both of these assumptions have been extensively discussed:
  first, the extinction prior can bias distances (as argued in,
  e.g., \cite{Knop03}, appendix of \cite{Perlmutter99}).
  Reference \cite{Kessler09} shows from simulations that distances are unbiased
  if the prior faithfully reproduces the actual extinction
  distribution, and the accuracy of colour measurements is accurately
  known.  This distribution is, however, not known from first
  principles, and one should expect some difference between the
  assumed and actual distributions, but in a cosmological context, a
  difference only matters if it is redshift-dependent.  So, using a
  prior assumes at least that the SN population at all redshifts is
  drawn from the same parent population, possibly with a
  redshift-dependent selection function (\S 2.3 of \cite{WoodVasey07},
  \S 5.1 of \cite{Kessler09}). Note that as any other light curve
  model, MLCS can be used without any extinction prior.  We will
  discuss the Milky Way extinction dust hypothesis in
  \S \ref{sec:LC-comp}.

\subsubsection{SALT and SALT2 models\label{sec:SALT}}

SALT \cite{Guy05} models supernova light curves by integrating a spectral template
(i.e. a spectral energy distribution as a function of SN phase) into
properly shifted observer filters. It implements
the light curve width variability mostly using the ``stretch'' paradigm
(\S \ref{sec:lightcurve-models} and \cite{Perlmutter97,Goldhaber01}).
 The training consists in
reproducing photometry of a training sample by adjusting a broadband
phase-dependent correction to a spectral time series from
\cite{Nugent02}. The colour
variations are modelled as an extinction factor $exp(-c C_l(\lambda))$
multiplying the flux, where $c$ is the colour parameter of each
supernova, and $C_l(\lambda)$ is the ``colour law'', common to all
supernovae. It is fitted during the training, and found to be
poorly compatible with extinction laws. There is no assumption on the
source on colour variations and colours that might reflect negative
extinctions are accepted. Note that, even if colour variations induced
by extinction are one-sided, the measured colour value is still an
unbiased estimate of true colour (e.g. \S 5.2 in \cite{Freedman09}).  The
SALT spectral template spans [300,800]~nm, which covers the restframe 
$U$-band, as required to fit high-redshift data from the Supernova Legacy
Survey, for which SALT was initially developed. As a spectral
template, SALT incorporates K-corrections into the model. SALT does
not attempt to model distances, and hence any sufficiently well-measured 
supernova can enter the training sample. SALT parameters are
$m_B^*$, the peak brightness in restframe $B$-band, $s$, the stretch
factor, and $c$, the (restframe) $B-V$ colour at maximum, plus some time reference.

SALT2 is a further development of SN spectral templates
\cite{Guy07,Guy10}: the variability of spectra (and broadband
light curves) of supernovae is modelled using principal component
analysis (PCA), and the whole spectral model and its variability is
fitted from a set of training spectra and light curves. As in SALT, the
``colour law'' is fitted to the data (and found again incompatible
with known extinction laws).  The first variability component
found is extremely similar to stretch (or decline rate) and no
significant further component is found. Because SALT2 does not model
distances, it can incorporate high-redshift events in its training sample, and
doing so allows it to go as blue as 250~nm, i.e. bluer than accessible
from the ground using low-redshift events. SALT2 delivers the average
spectral model and its first variability component, the colour law,
the uncertainties, and some estimate of the intrinsic variability of
supernovae around these averages. SALT2 finds that broadband
light curve amplitudes scatter (coined ``colour smearing'' in
\cite{Kessler09}) by at least 2.5\% around the average model, with a
rapid increase when going bluer than $\sim$360~nm.  At variance with
all other SN models, SALT2 training actually determines spectral
sequences, but the large number of required parameters constitutes the
main limitation of the model. This flexibility of the training process
imposes the necessity to resort to some regularization scheme in order to actually
define the template in regions of the phase/wavelength/decline-rate space 
poorly covered by the training sample. SALT2 delivers a model uncertainty.

\subsubsection{The SiFTO model \label{sec:sifto}}
SiFTO \cite{ConleySifto08} models SNe light curves from a SED template,
constructed through a smooth correction of the empirical template
from \cite{Hsiao07}. The light curve shape variability is accounted
for using the stretch paradigm, using a measured relation
between stretch factors in different bands. When fitting a supernova event
SiFTO adjusts a phase-independent multiplicative function of wavelength
in order to match exactly the measured light curve amplitudes. The obtained 
mangled SED is then used to evaluate the light curve amplitudes in 
the standard $UBVR$ bands plus a bluer band, covering in total [250,710]~nm.
This set of restframe magnitudes readily provides 
an overall amplitude $m_B^*$, and a single $B-V$ colour derived from an optimal
combination of all measured restframe colours. This combination
relies on restframe colour-colour diagrams obtained during the training,
which also delivers a model uncertainty, used when fitting. 
As SALT and SALT2, SiFTO does not predict
distances and can then use any well-measured event for training.
Its small number of parameters makes its training easy and precise.
As SALT and SALT2, SiFTO finds colour relations (e.g. $U-B$ versus $B-V$)
incompatible with extinction by dust, even allowing for a 
coloured variation of supernovae related to stretch.

\subsubsection{Tripp distance estimators\label{sec:Tripp}}

SALT, SALT2 and SiFTO do not directly deliver distances.  All propose
(but do not impose) to estimate distances using the most general
distance estimator linear in decline rate $s$ (e.g. stretch) and colour $c$:
\begin{equation}
\mu = m_B^*+\alpha(s-1) - \beta c + {\mathrm  Const}\label{eq:mu-alpha-beta}
\end{equation}
where $\mu$ is a distance modulus estimate (i.e. it varies as
$5\ log_{10} (d_L)$ up to an additive constant), $\alpha$ and $\beta$
are global parameters that characterize, respectively, the
brighter-slower and brighter-bluer relationships. These parameters
have to be empirically determined, typically along with the
cosmological parameters, as proposed by Tripp in \cite{Tripp98}.
$m_B^*$ refers to the peak magnitude that would be observed
in the filter matching $B$ restframe, where the choice of $B$ is
arbitrary. For most cosmological SN analyses, the measurements are
actually carried out around the B restframe band and the distance
uncertainty is dominated by the $\beta c$ term. Assuming
that colour variability at a fixed decline rate is due to Milky Way like
dust sets $\beta = R_B = 4.1$.

  The most common criticism to SALT, SALT2 and SiFTO is that the
  reported colour parameter mixes both intrinsic colour variations
  (slower supernovae tend to be bluer) and extrinsic ones (e.g. dust
  extinction), and the distance estimator defined in equation
  \ref{eq:mu-alpha-beta} conflates both effects via a single
  brighter-bluer coefficient $\beta$ (e.g. \S 2.1 of
  \cite{WoodVasey07}, \S 2 of \cite{Hicken09}). This is suboptimal if
  there are two or more sources of colour variations with different
  $\beta$ values, but more importantly biases cosmology if the
  admixture building up the colour distribution evolves with redshift.
  Let us first assume the extra source of colour variation is entirely
  related to decline rate: assuming we know the relation, we can then
  separate the contributions to colour for each event.  To first
  order, this separation does not change the form of
  equation \ref{eq:mu-alpha-beta}, but only alters the value of $\alpha$
  ($\beta$ is unaffected, see \S~4.1.1 in \cite{Guy05}, \S 4.2.3 in
  \cite{Guy10}). Hence, accounting for decline rate related colour
  variations does not change distances evaluated using
  equation~\ref{eq:mu-alpha-beta}. So, this criticism assumes that there
  are at least two contributions to colour variations, independent of the decline rate. This is a serious concern
  that can only be handled via more parameters per event and hence
  applies to all current photometric distance estimators. 
\subsubsection{Comparing light curve fitters}
\label{sec:LC-comp}

SALT2 and MLCS2k2 were compared thoroughly in \cite{Kessler09}, which
finds that they deliver systematically different distances as soon as
they make use of the restframe $U$-band ([320,400]~nm). This difference in
distances has a dramatic effect on cosmological fits, and is traced to
two causes: different predicted $U$-band peak values at given $B$ and $V$
and the effect of the MLCS2k2 extinction prior on high-redshift
ground-based events from the SNLS sample (see \S 11 of
\cite{Kessler09}). When MLCS2k2 uses SALT2 light curve templates, and
does not apply any extinction prior, the cosmological results become
essentially as when using SALT2 right away. So, \cite{Kessler09}
argues that some unidentified issue with the low-redshift
observer-frame $U$-band seriously affects MLCS2k2, and accounts for the
difference with SALT2 as a systematic uncertainty. On a similar data
set, \cite{Conley11} identifies that observer-frame $U$-band SN data 
are incompatible with higher redshift measurements of the same restframe 
spectral region, and ignores this chunk of data.

Regarding restframe U-band, the difference between MLCS2k2 and SALT2
models lies in their training strategy: while MLCS2k2 has to train
restframe $U$ from observer-frame $U$ SN data, SALT2 (and SiFTO) use
higher redshift events (typically at $z \sim 0.2-0.5$) to model
restframe $U$-band.  Observer-frame $U$ is notoriously difficult to
calibrate because the actual shape of the bandpass is determined by
the variable atmospheric absorption\footnote{Correction of the
  absorption by the Milky Way dust might also be large in observer $U$-band,
 and correspondingly uncertain \cite{Schlegel98}.}. This
difficulty is illustrated, in the supernova framework, by the observed
dispersion of the restframe $U$-band light curve amplitude, which is
much larger for nearby events (where restframe $U$ is measured from
observer $U$) than for higher redshift events (where redshift pushes
restframe $U$ into regions unaffected by the atmosphere (see figure 11 in
\cite{Astier06}, \S 2.6 in \cite{Conley11}).  For MLCS2k2, dropping
the observer-frame $U$-band at training implies that the resulting model
misses restframe $U$-band, which is unacceptable: all modern SN models
insist on modelling near UV SED of supernovae, because distances to
many high-redshift events rely on this modelling.

In \cite{Kessler09}, the second source of MLCS2k2/SALT2 difference is
traced to the use of an extinction prior for high-redshift SNLS
events. The fact that applying or not the extinction prior to a
measurable quantity (the event colour) questions the prior
(as discussed in \S \ref{sec:MLCS}), and
dropping the prior is technically possible. We also note that
for samples of several hundred events,
the uncertainties associated with the choice of the prior
(table 7 of \cite{WoodVasey07}, figure 32 of \cite{Kessler09})
are larger than the statistical uncertainties. 
Finally, as events are better and better measured,
the statistical benefit from constraining 
their parameters a priori is vanishing.

Colour variations independent from decline rate are typically assumed
to be due to extinction by dust or not. In this second case, both
colour relations (the colour law) and the brighter-bluer coefficient
($\beta$ of equation~\ref{eq:mu-alpha-beta}) should be extracted from the
data.  In \cite{Kessler09} (\S 11), it is shown that inserting the 
fitted SALT2 colour law into MLCS2k2 has no sizable effect on MLCS2k2
distances. Fitted brighter-bluer coefficients are generally smaller
than those for Milky Way dust (see e.g. \cite{Tripp98} and references
therein, \cite{Guy05}). The finding from MLCS2k2 distances that we
might be at the centre of a ``Hubble bubble'' \cite{Jha07-MLCS2k2}
(i.e. a spherically symmetric under-density) was shown to be a
consequence of assuming rather than fitting the brighter-bluer
relation \cite{Conley07,Hicken09}.

In \cite{Hicken09}, one can find a comparison of SALT, SALT2 and
MLCS, but the SALT and SALT2 packages were probably not used properly:
using the same photometric data, supernova fits differ from the ones
published by the authors of the packages. In \cite{WoodVasey07} SALT
and MLCS2k2 were compared and found to agree very well. This is not
very surprising as their training samples vastly overlap. 
In \cite{Guy10}, SALT2 and SiFTO are compared using exactly the same
event sample and the found differences are acceptable for a 
few hundred events. 

Comparing the scatter of the resulting Hubble diagram yields 
useful comparisons of distance estimation strategies. For large and 
similar samples, table 11 in \cite{Kessler09}
exhibits rms Hubble residuals of 0.21 mag (about 10\% in distance)
using MLCS2k2,
while SALT2/SiFTO combined distances deliver about 0.17 mag
(table 4 of \cite{Conley11}).

SALT2 and SiFTO are compared in \cite{Guy10} on exactly the same distant 
events, and SiFTO outperforms SALT2 by a small but significant 
amount, 0.150 versus 0.173 mag.

\subsection{Other distance estimators\label{sec:other-distance-estimators}}

Most distance estimators proposed for SNe~Ia follow the
form of equation~\ref{eq:mu-alpha-beta}. We review here two notable
exceptions.

The so-called ``Bailey ratio'' \cite{Bailey09} was optimized to obtain
the best distance estimator from a sample of 58 photometric spectra
from the SuperNova Factory (\S \ref{sec:nearby-samples}).  It uses the
flux ratio of two narrow spectral regions F(642~nm)/F(443~nm) as a
single intrinsic luminosity estimator, and achieves a smaller Hubble diagram
scatter than all other known distance estimators on the same sample. This
good performance remains unexplained, and was confirmed using long
slit spectra of nearby events \cite{Blondin09}. Note that at $z>0.6$,
most of the current SN spectra do not cover the 642~nm region, although
the Si line characteristic of SNe Ia is around 615~nm.   

The CMagic distance estimator \cite{Wang-CMagic-03} makes use
of the post-maximum trajectory of SNe~Ia in the colour-magnitude (B-V, B) plane,
in order to evaluate the brightness when a given colour is reached. 
The obtained brightness is then corrected for brighter-slower relation
to yield distances, which exhibit a reduced scatter with respect
to other distances, on the nearby training sample. The technique was 
applied to a cosmological analysis \cite{Conley-CMagic-06} and 
proved to be applicable to high-redshift events, provided that the latter
are measured in bands close to restframe $B$ and $V$, a demanding
constraint. The 
performance improvement on this sample is not large enough
to compensate for the loss in event statistics due to the requirements
in phase and wavelength coverage.

\section{Photometric calibration \label{sec:photom-calib}}

\subsection{The traditional way\label{sec:cal-standards}}
Stellar fluxes are essentially always measured relative to some other
star. The ratio of the studied object flux to a standard star is 
commonly reported as a magnitude. In order to measure luminosity distances,
one has to convert magnitudes into fluxes, relying
on a star that has both magnitudes (in the same system as the standard)
and a spectrum in physical units; one also needs to know the pass bands
of all involved photometric measurements.

 In order to report fluxes from relative measurements, one obviously needs
primary calibrators, in practice non-variable stars with known spectrum.
Establishing the spectrum of Vega (or $\alpha$-Lyrae) is described in
\cite{Hayes85}, where the astronomical instrument is calibrated
against artificial black bodies (or sources themselves calibrated to
black bodies). As Vega is far too bright to be measured in most modern
instruments\footnote{A SN~Ia at z$\sim$0.1 typically peaks at $m_B\simeq
  20$, i.e. $10^{-8}$ times Vega.}, anchoring photometry of
$m\simeq 20$ objects to Vega ($m=0$) goes in practice (see
\S 8.1 in \cite{Regnault09}) through two steps: the Landolt star network
\cite{Landolt83,Landolt92} ($<m> \simeq 12$), referenced to Johnson
\cite{Johnson53} and Cousins \cite{Cousins76} standards ($<m> \simeq
5$), themselves referenced to Vega \cite{Johnson51}. Although this lengthy
sequence was not conducted to ensure a robust conversion of magnitude
to fluxes, all supernova cosmology works until 2006 calibrate
supernova magnitudes against Landolt stars and convert those to fluxes
using the Vega spectrum (usually from \cite{Hayes85}), because they had 
no alternative.  One should note that the Landolt system is not a
natural system (\S 8.1 in \cite{Regnault09}), meaning that the reported magnitudes
are not directly measured, but rather derived from actual measurements
and colours, and hence that the filters defining the reported magnitudes
are unclear. The networks of standard stars were designed to allow astronomers
to precisely compare magnitudes of stars, but allowing one to convert
these magnitudes to high precision fluxes was not a primary goal.

\newcommand{\BDtruc}{BD~+17~4708}

A new and shorter path to convert magnitudes into fluxes was designed
in the framework of the HST flux calibration: the
Calspec project\footnote{\href{http://www.stsci.edu/hst/observatory/cdbs/calspec.html}{http://www.stsci.edu/hst/observatory/cdbs/calspec.html}}
reports spectra measured from the HST and carefully cross-calibrated.
The instruments used to gather these spectra are flux-calibrated from
stellar models of pure hydrogen white dwarfs (references in
\cite{Bohlin00}). The models make use of temperatures and gravity
obtained from Balmer line shapes from ground-based high-resolution
spectroscopy. The consistency with the calibration route sketched in
the above section was checked by bravely measuring Vega in the
same programme \cite{Bohlin04Vega}. As a few Calspec standards now have
broadband magnitudes in the widely used Landolt system
\cite{Landolt07b}, the numerous SNe~Ia distances calibrated on 
this system now have 
a direct link to this spectral library. The overall flux scale of the Calspec library is
set to match ground-based photometry, and dominates the uncertainty
budget at about 2\% \cite{Bohlin01}, but is irrelevant for cosmology with SNe~Ia.
Because the ratio of fluxes of supernovae at different redshifts depend on
the ratio of fluxes of the standards at different wavelengths, the ``colour
uncertainty'' of the spectral library is, in contrast, extremely relevant to our subject.
Two terms dominate this uncertainty: the model slope (dominated
in the visible by temperature uncertainty), and the reproducibility of the 
Calspec measurements. Both terms are estimated to typically 0.5\% or less (table 12
in \cite{Regnault09}) for the measurement of the star \BDtruc~ from \cite{Bohlin04BD17}.

\subsection{Instrumental calibration from calibrated laboratory sources}

Using artificial sources to calibrate instruments is not a modern idea
(see e.g. \cite{Hayes85} and references therein). Suggestions to
calibrate imagers from artificial light sources have been recently
formulated \cite{Stubbs06}, relying on photo-diodes calibrated by the US NIST. 
Results from a setup delivering a monochromatic illumination of a
screen in the telescope dome are provided in
\cite{Stubbs10}.  However, reflections on the numerous optical
surfaces in front of wide-field imagers (typically at least three
lenses, one filter and the cold chamber window) contribute to the
image obtained with a flat-field screen, but not to star photometry.
For the specific case of the MegaPrime instrument on CFHT, direct
evidence for these internal reflections is provided in
\cite{Juramy08}; \cite{Regnault09} evaluates to about 10\% or more
their contribution to twilight images. This justifies proposing a
different artificial light source setup in \cite{Juramy08,Barrelet08}.
The most recent results presenting calibration from an artificial source
\cite{Tonry12} are encouraging but do not yet outperform the stellar
approach.

There are great expectations from these illumination devices since 
most ground-based wide-field survey instruments either own one 
or plan to install one. We can sketch plausible outcomes of current efforts
roughly ordered in increasing difficulty and
reward: daily monitoring of the instrument response, 
measuring the bandpass profiles in situ, establishing a flat-field
correction, measuring the relative transmission of photometric bands, or
even absolute transmission of photometric bands.

\subsection{Ground-based observations and atmospheric extinction.}
The canonical procedure to evaluate atmospheric extinction consists 
in following sources at various elevations and extrapolating 
above the atmosphere, or more commonly around zenith. One then relies on repetition to beat down 
atmospheric variability.
A more precise technique, relying on spectroscopy and
atmospheric extinction models is implemented in \cite{Burke10}.

Atmospheric extinction is less critical when calibrating from stellar
standards, and again, repeating the calibration of
stars around supernovae over tens of nights  allows one to mitigate the atmospheric
variability (e.g. \cite{Regnault09}).  Ground-based rolling supernovae
searches incorporate these calibration observations in their programme.
 
\subsection{Linearity issues}

Since standard stars are much brighter than tertiary stars, themselves
often brighter than supernovae, one should wonder whether the reported
signals scale as fluxes. One first varies the exposure time from
typically 1 s to a few hundred seconds, in order to compare different
brightnesses in the same electronic range. Precise shutter timing is
required for CCDs and can be obtained (see e.g. \S 12.1 in
\cite{Regnault09}). The overall linearity of CCDs and their readout
electronics can be checked on the sky (App. C in \cite{Regnault09}),
alleviating concerns that quasi-uniform illumination (``flat fields'')
might not be representative of the instrument response to stars. CCDs
in space suffer from charge transfer inefficiency that builds up with
the integrated radiation dose, alters linearity, and should be
corrected for (e.g. \cite{Knop03,Massey10}). Observations with the
NICMOS NIR instrument on HST suffer from a poorly understood
non-linearity issue, which adversely affects fluxes, and hence distances to 
high-redshift
supernovae obtained with this instrument (see \S~5.1.3 in \cite{Conley11}).

\subsection{Prospects for improvements}
As calibration constitutes the leading systematic uncertainty of
current supernova cosmology results \cite{Conley11,Sullivan11},
improving over the present situation is mandatory. The weaknesses of
the Landolt system \cite{Regnault09,Conley11} have a decreasing impact
as new supernovae calibrated via alternative paths are
published: for example, the SDSS \cite{Holtzman08} directly ties its
magnitudes to solar analogues with CALSPEC spectra. Some pass bands 
of the CSP nearby sample \cite{Contreras10,Stritzinger11} are calibrated
to the Smith magnitude catalogue \cite{Smith02}, which is now anchored
as well to CALSPEC standards. The calibration of instruments against
laboratory standards can provide a significant breakthrough, provided
that parasitic reflections in the optics are controlled well enough
and atmospheric extinction accurately measured. These two complications
vanish if the artificial light source is mounted on a
satellite, as proposed in \cite{Albert09}.

\section{Astrophysical systematic uncertainties affecting SN distances\label{sec:systematics}}

\subsection{Evolution of supernovae}
Evolution of supernovae refers to the concept that supernovae could be
different at high redshift (i.e. early in the stellar history) than at
low redshift.  We already know that a demographic evolution exists (\S
\ref{sec:variability}), and our concern here is to constrain if
supernovae exhibiting the same observables could have a different
average brightness at low and high redshifts: the derivation of
distances used for the cosmological analysis makes the hypothesis that
it is not the case. One set of tests compares properties
of nearby and distant samples, and the other set compares properties of events
at similar redshifts in different environments, i.e. different stellar
ages.

\subsubsection{Light curve rise time}
Reference \cite{Arnett79} proposes the approximate rule 
that the maximum (bolometric) luminosity equals the instantaneous 
energy released by $\beta$ and $\gamma$ decays. This relates the rise time, peak luminosity
and produced $^{56}Ni$ mass, so that a change in rise time at a given nickel mass
implies a change in peak brightness. Rise times at a given decline rate hence 
constitute a plausible handle on intrinsic peak brightness. Even 
without theoretical guidance, the rise time of the light curve is 
sensitive to the explosion energy and the evolution of the opacity of 
the ejecta, and any evolution of rise time with redshift 
unrelated to usual decline-rate indicators is a direct threat to 
relative distances, as they are evaluated from a single light curve
shape modelling. 

In order to measure rise times accurately, authors devise methods that
scale the actual measurement of the rise time to that of the fall
time, so that variations of the decline rate do not smear out the
measurement. Three studies
\cite{Goldhaber01,ConleyRiseTime06,Hayden10} do not detect any sizable
difference between distant and nearby samples. One might note, however,
that different studies find different average rise time values because
of different definitions, and/or different fiducial supernova.
Reference \cite{Aldering00} discusses some difficulties of the measurement and
closes a dispute caused by the troublesome comparison of rise time
determinations obtained through different estimators.  Rise times are
also compared as a function of the
``stellar age'' of the host galaxy and they turn out to be in
excellent agreement (table 2 in \cite{ConleyRiseTime06}).

\subsubsection{Spectroscopic evolution tests}

Comparisons of supernova spectra across redshifts appear in the very
first SN cosmology papers (e.g. figure 11 in \cite{Riess98b}).
Quantitative comparisons involving line velocities \cite{Hook05}, line
velocities and profiles \cite{blondin06}, line velocities and
strengths \cite{Garavini07} did not reveal compelling differences.

More detailed comparisons with increasing statistics are presented in
\cite{Foley08}, where global agreement is found, with perhaps some
small differences in line strengths, however, without subtraction of
the host galaxy light. Studies of samples of similar size
\cite{Ellis08,Balland09,Walker11} do not
confirm these small differences, but they carry out 
galaxy light subtraction. 

   High-redshift samples allow one to study the UV part of the spectrum
(below $\sim$350~nm) absorbed by the Earth's atmosphere for nearby events,
and deemed to encode metallicity of the progenitor, although theoretical
predictions disagree on the sense of the effect \cite{Hoeflich98,Lentz00}.
These high-redshift studies agree
on a event-to-event scatter increasing towards blue wavelengths
(figure 7 in \cite{Guy07} shows the same trend using photometry only),
and stress that comparison at wavelengths below $\sim$350~nm lack
a nearby counterpart, which is nowadays being collected from space. A first 
sample of 12 events \cite{Cooke11} shows reasonable agreement in the UV part
with events at $z\sim 0.5$ and mildly suggests that the 
observed differences might reflect sampling variance. The large observed 
variability in UV, probably due to metallic features \cite{Cooke11},
suggests that cosmological analyses aiming at higher and higher
precision limit their use of restframe UV (below $\sim$ 350~nm) to measure
distances to high-redshift supernovae. 

To summarize, when differences between nearby and distance spectra are observed,
they can usually be interpreted as the already known demographic evolution,
i.e. that explosions tend to be more energetic as redshift grows.

\subsubsection{Segregation according to host galaxy types\label{sec:segregation}}

In order to compare brightnesses (or even derived distances) of events
at different stellar ages, one can take advantage that varying stellar
ages cohabit at the same cosmological age.  For example, the local
Universe hosts both young stellar systems (and even galaxies producing
stars), as well as old evolved heavy elliptical galaxies. One can then
question whether we derive identical average distances to events at the
same redshift but in different environments, thus directly addressing
the heart of the evolution question without strong assumptions on the
distance-redshift relationship.  Comparisons between spiral and
elliptical galaxies reported null differences (e.g. \S 5.3 in \cite{Schmidt98},
\cite{Sullivan03}).  Many works test the effects of host galaxy
metallicity (see \S 1. in \cite{Howell09}), and report null results,
with the exception of \cite{Gallagher08} which only tests a small sample
of early-type hosts.  A first indication of systematic difference of
distances as a function of the host galaxy stellar mass was proposed
in \cite{Kelly10}, and confirmed on independent samples by
\cite{Sullivan10,Lampeitl10}, and could perhaps be attributed to
systematic differences in metallicity.  Note that the galaxy stellar
mass is inferred from the wide-band photometry of the host galaxy, a
by-product of light curve measurements, and hence the effect is easily
corrected for. Ignoring this brightness difference of $\sim$8\% between low and
high stellar-mass hosts would bias the dark energy equation of state $w$ by $\sim$0.04 to 0.08 (table 9 in \cite{Sullivan10}).

\subsection{Gravitational lensing of supernovae \label{sec:lensing}}

Gravitation bends light rays along the transverse component of the
gravitational potential gradient (e.g. \cite{Schneider92,Narayan96}).
Mass concentrations act somehow as lenses, which coined this class of
phenomena.  Cosmic bending of light rays re-maps the image plane,
which alters the brightness, shape, size and position of sources. In
the framework of supernova observations, the gravitational lensing of
supernova light by foreground mass contrasts induces two main effects:
an increased scatter of the observed brightness, and when going
through strong lenses, the possibility of multiple images of the same
event, usually with different time delays. Strong lenses are
characterized by a singular mapping of the source plane onto the image plane, leading
to an infinite magnification on critical lines, and the possibility of
multiple images (see e.g. \cite{Schneider92,Fort94}). The lines of
sight subject to critical lines are very scarce and the vast majority
of supernovae incur asymptotically weak lensing. Because supernovae
are point sources, their geometrical shape is unaffected by lensing,
but their brightness follows the optical magnification. 

Lensing of supernovae causes an extra scatter of their apparent
brightness: the rare events observed through a mass concentration
will appear brighter, while the bulk of the events (going through
under-dense regions) will be slightly fainter than the
average. Lensing conserves the average intensity \cite{Peacock86}, and
does not constitute a first-order threat on luminosity distance
measurements \cite{Weinberg76}: the mass density that is probed through
these measurements is the actual average density, rather than that
of a smooth component\footnote{The most common parametrizations of
  gravitational magnification are obtained perturbatively, thus miss the
  large-angle de-magnification, and hence do not conserve the average
  brightness.}. Strong lensing might violate flux conservation in
the case of several images where one or more escapes detection. This is
commonly investigated through numerical simulations
(e.g. \cite{Holz98b,Bergstrom00,Goobar02SNOC}), and as expected, the
result strongly depends on the assumptions on the ``clumpiness'' of
matter distribution.  Monte-Carlo techniques also face practical
difficulties in re-summing multiple images that would be merged by an
observer \cite{Bergstrom00, Goobar02SNOC}. Fortunately, one may use
systematical searches for multiple images of distant radio sources
\cite{ClassSurvey}, in order to evaluate an upper bound to the average
amount of flux lost in faint secondary images of supernovae: in
\cite{Astier06} (\S 7.6), it is found that at $z=1$, less than 0.3\%
of the flux is lost to secondary images. This is far below other
systematic uncertainties, both for present and foreseen SN surveys.
Note that for very large statistics SN surveys, the frequency of
multiple images becomes measurable and this very small effect could
even be estimated and corrected for.

The expected magnification distribution is significantly skewed
\cite{Peacock86, Holz98a, Holz98b, Bergstrom00, Goobar02SNOC,
  Amanullah03}, because the magnification decays very rapidly with the
impact parameter to a lens. So, even if the average flux is not altered 
by lensing, the average magnitude is, and its distribution is no longer
Gaussian: least-squares estimates from high-statistics Hubble diagram fits 
made in magnitude space can be biased.
In \cite{Amanullah03}, it is shown that the maximum likelihood 
(rather than least squares) delivers unbiased cosmological parameters, even
for several thousands of events up to $z\sim 1.7$. 

The extra brightness scatter due to lensing also broadens distance residuals
and degrades the cosmological inferences. Again, different simulations
using different assumptions on the clumpiness of matter lead to 
different conclusions: while \cite{HolzLinder04} find an extra 
magnitude scatter of $\sim 0.09\times z$, \cite{Bergstrom00} find 
$0.04$ at $z=1$ for smooth halo profiles. 

Evidence of weak lensing of supernovae by foreground structures was
obtained recently, by correlating the Hubble line residuals of individual events
with the magnification expected from the foreground galaxies
\cite{Kronborg10}, and also shows that the amplifications agree with
the expectations from measured mass contrasts.  Stronger detections
are only expected from increased statistics.  From the same sample of
events, \cite{Johnsson10} extracts weak constraints on dark matter
halos. We also have evidence for a supernova at $z\simeq0.6$
experiencing strong magnification by $\sim$3.6 (1.4 mag.) through a foreground
galaxy cluster at $z=0.18$ \cite{Goobar09}.

Lensing of supernovae will not probably limit the cosmological
accuracy of future high-statistics and high-precision SN surveys.  For
large-scale SN surveys (typically targeting $10^4$ to $10^5$ events)
lensing will open two new avenues: constraining the shape of dark
matter halo from correlation of Hubble residuals with properties of
the line of sight (see \cite{Kronborg10,Johnsson10} and references
therein); large surveys will also detect multiple images of strongly
lensed events, where an accurate measurement of the distance, flux
ratio and time delay of the images provides valuable cosmological
information, assuming the lenses can be accurately modelled (e.g
\cite{Goobar02} and references therein). One should note here that any
kind of high-redshift time-variable source is eligible for this type
of study, typically any type of bright enough supernova, and to a
lesser extent quasars. However, the ``standard candle'' property of SNe~Ia
allows one to determine the lensing magnification factors (rather
than their sole ratio), which breaks some degeneracy between the lens
potential and the Hubble constant \cite{Oguri03}.  Simulations
conducted in \cite{Oguri10} indicate that the prospect for finding
multiply imaged supernovae are modest for a possible space mission
(about 10 events), where the ground-based LSST project would find
$O(100)$ events in its 10-year mission; the same space mission would
measure time delays of about 20 multiply imaged quasars, while LSST
would measure several thousands.

\section{SN surveys and their cosmological findings \label{sec:surveys}}

We review here the main SNe~Ia surveys. Table \ref{tab:SN-samples}
lists the main event samples we will present in the
following sections. We will start with nearby SN samples (typically $z<0.1$)
used in essentially all analyses. Most analyses presenting
new events also incorporate events from pre-existing samples. 
For exhaustive lists of SNe~Ia samples, we refer to compilations, 
rapidly presented in \S \ref{sec:compilations}.
\begin{table*}
\begin{center}
\begin{tabular}{|l|lllcll|}
\hline
Name   &  $z$ range & $N_{events}$ &  Dates &  $N_{bands}$ & \S       &  References \\
\hline
Calan-Tololo  &    $<$ 0.1 &      29 &   90-93  &  4 &  \ref{sec:nearby-samples} & \cite{Hamuy96c,Hamuy96b,Hamuy96a} \\
HZT        &      [0.2,0.6]  & 10  &95-97   &  2 &  \ref{sec:scp-hi-z} & \cite{Riess98b,Schmidt98,Garnavich98} \\
SCP        &  [0.2,0.8] &  42 &      92-97  & 1-2  & \ref{sec:scp-hi-z} &  \cite{Perlmutter99} \\
SCP (HST)  &  [0.4,0.8] &  11 &     97-00   & 2 &  \ref{sec:HST-first}   &     \cite{Knop03} \\
SN-Goods (1\&2)  &  [0.3,1.5] &  37  &    02-05  &  2-3 & \ref{sec:HST-searches} &  \cite{Riess04,Riess07} \\
SNLS (1\&3)      &     [0.2,1.0]      &  242 ($\sim$400) &  03-08 &  3-4 &\ref{sec:ESSENCE-SNLS-SDSS}  & \cite{Astier06,Guy10,Conley11,Sullivan11} \\
ESSENCE    &     [0.3,0.7]     &  60  ($\sim$200)  &   02-06 &  2 & \ref{sec:ESSENCE-SNLS-SDSS} & \cite{WoodVasey07} \\
SN-SDSSII    &    [0.1,0.4] &  103 ($\sim$500) &   05-07  &  3-5 & \ref{sec:ESSENCE-SNLS-SDSS} & \cite{Frieman08,Kessler09} \\
CFA (1,2 \& 3) &  $<$0.1   &   185     & 01-08 & 4-5 & \ref{sec:nearby-samples} & \cite{Riess99a,Jha06,Hicken09a} \\
CSP nearby   &   $<$0.1     &   85 (129)     &  04-09 & 5-8 & \ref{sec:nearby-samples} & \cite{Contreras10,Stritzinger11} \\
\hline
\end{tabular}
\caption{SN surveys presented in the following sections.
$N_{events}$ is the number of reported events, possibly
followed by the anticipated size of the final sample, in parentheses.
$N_{bands}$ indicates the number of different bands 
these events were measured in. A fair fraction of CSP events 
was measured in NIR, which makes 8 bands in total.  
\label{tab:SN-samples}}
\end{center}
\end{table*}

\subsection{Nearby samples \label{sec:nearby-samples}}

``Nearby'' here broadly refers to samples at $z\lesssim 0.1$, but at
$z\gtrsim 0.01$, because smaller redshifts are too contaminated by Doppler
shifts due to proper motions.  Practically, the photometry and
spectroscopy of these events are mostly carried out using respectively,
1 and 2 m class telescopes.  There are no practical examples of
cosmological constraints derived without a nearby sample, because it 
increases the redshift lever arm, and decouples the nuisance
parameter ${\mathcal L} H_o^2$ (see \S \ref{sec:standard-candles})
from cosmological parameters.

While wide-field imagers make it easier and easier to find and measure
distant supernovae, nearby events are still difficult to find.  As
discussed above, the Calan-Tololo nearby survey delivered the first
well-measured sample \cite{Hamuy96b} found from photographic
plates. The Lick Observatory Supernova Search (LOSS) \cite{LOSS-Li00}
has been a key provider of nearby events for about 15 years, and mostly
targets galaxies not in the Hubble flow, but also delivers events in
the Hubble flow. 

The Harvard Center for Astrophysics (CfA) has produced three releases
\cite{Riess99a,Jha06,Hicken09a} of spectra and light curves of nearby
events discovered by both professionals and amateurs. With 185 events,
the last release (CfA3) constitutes the largest homogeneous sample to
date, with systematic calibration uncertainties evaluated to 3\% (a
probably pessimistic value according to \cite{Conley11}). The CfA
data set has recently been complemented by a massive release of spectra
\cite{Blondin12}, which will
significantly contribute to training of light curve fitters.

The Carnegie Supernova Project (CSP) also follows up detections from
professionals and amateurs. It released two batches
\cite{Contreras10,Stritzinger11} of high-quality light curves reaching
a total of 85 events among which 70 have NIR light curves.

Regarding distance measurements the merits of the various samples are
evaluated in detail in \cite{Conley11} (\S 2.3 and 5.1.2), and turn
out to be comparable after strict quality cuts. The issues related to
the U-band of nearby supernovae are discussed as well in the same work
(\S 2.6) and the conclusion is that the U-band measurements of the CfA
and the CSP are incompatible, and the latter agree with colours of higher
redshift events.
 
The SuperNova Factory aims at collecting several hundreds of
spectro-photometric spectral sequences of SNe Ia \cite{SNFAldering},
using its own blank-field search. The reduction of spectra is very
involved and sequences have not yet been released. A low-scatter
distance estimator relying on photometric spectra has been proposed
(\S \ref{sec:other-distance-estimators} and \cite{Bailey09}).

When events are used for cosmology, one should evaluate the distance
bias caused by the search and identification steps possibly missing
faint events (the so-called Malmquist bias). For a nearby sample, only the
average distance bias matters for cosmology, and when caused by a
sharp flux limit, its value does not depend on the value of this limit
and only mildly depends on the shape of the cut (\S 7.4 in
\cite{Astier06} and \S 2.7.1 in \cite{Conley11}). Different works find
slightly different bias values, from 4\% (\cite{Perlmutter99}) down to
2.7\% (\cite{Astier06})) mainly because of different assumed values of
the brighter-bluer relation. Whether this flux limit model applies to
a given sample can be tested through the phase of the first
measurement of the found events \cite{Perlmutter99, Conley11}.

\subsection{The SCP and HZT pioneers: the twin papers \label{sec:scp-hi-z}}

In the mid 1990s, two groups embarked on the measurement of distances
to high-redshift supernovae (typically $z\sim 0.5$), the Supernova
Cosmology Project (SCP) and the High-Z team (HZT). The first campaigns allowed the teams to refine the
search strategy, and the ``batch discoveries'' of 1995 (\S
\ref{sec:finding-supernovae}) define the start of systematic programmes.
From there on, time allocation committees awarded observing time to both
teams on a regular basis.  Both teams even gained access to the HST 
for photometric follow-up of their supernovae.

The SCP produced an early analysis of 7 high-redshift events
\cite{Perlmutter97}, which when compared with the Calan-Tololo nearby sample
mildly favoured a matter-dominated Universe assumed to be flat. The quality
of distances was very poor, and some events lacked spectroscopic
identification. The HZT presented an early sample of 4 distant
events among which 3 are measured by the HST \cite{GarnavichHST98}, and
found $\Omega_M\simeq 0.3 \pm 0.3$.

\begin{figure} [h]
\begin{center}
\includegraphics[width=\linewidth]{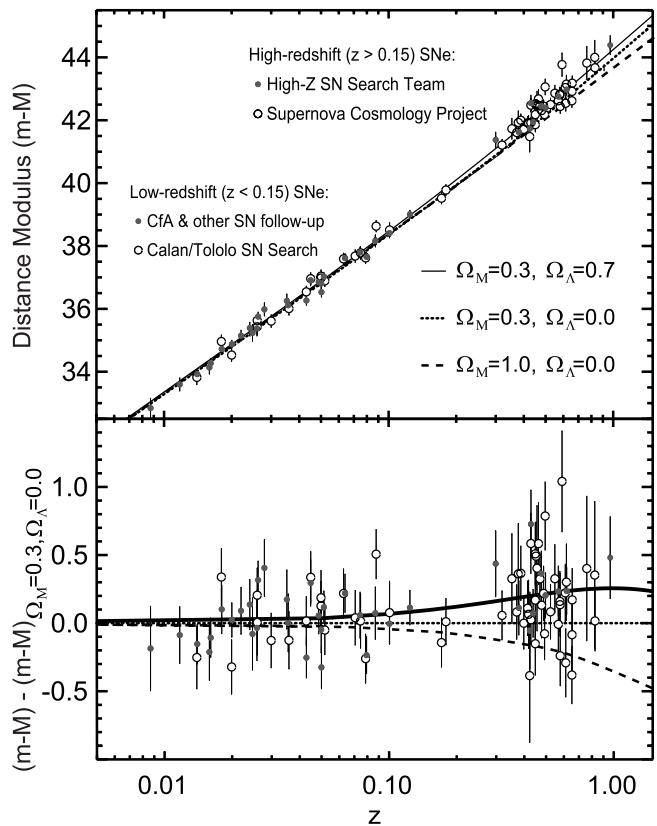}
\caption{Combined Hubble diagram (logarithm of a distance versus $z$) from the twin papers (top), 
and residuals to a $\Omega_M=0.3, \Omega_\Lambda=0$ cosmology (bottom).
The fit favours the thick line over the two other proposed hypotheses.
Note that the HZT events have more precise distances than those from the SCP.
Reproduced from \cite{PerlmutterSchmidt03}, with permission. 
\label{fig:scp-hzt-hd}}
\end{center}
\end{figure}

Both teams were in fact building up larger samples, and published
their twin analyses \cite{Riess98b,Perlmutter99} a few months apart.
The SCP analysis \cite{Perlmutter99} presents 42 distant
events, mostly with poor colour measurements, and uses 17 nearby
events (from the Calan-Tololo survey \cite{Hamuy96b}). The used
distance estimator ignores colour, however, after eliminating reddened
events and having checked that the distant and nearby samples have the
same average colour and understood spreads. The HZT analysis
\cite{Riess98b} makes use of 10 distant events and has colour
measurements for 9 of them, but colours are used through an extinction
prior, which limits their impact\footnote{The average distance modulus
  uncertainty without a prior would be $\sim$0.38 from the uncertainty of peak magnitudes, while the average reported the uncertainty of distance modulus
  is 0.18 (without intrinsic dispersion). This together with the
  fact that more than half of the events hit the prior hard limit
  $A_V$ = 0 indicates that the weight of the colour measurements in
  the distance estimation is modest.}. It also makes use of 27 nearby
events mostly from the Calan-Tololo survey, and finds similar results
using two different distance estimators (which, however, share the same
extinction prior). Both distant samples exhibit similar photometric
quality in restframe $B$-band with an average photometric uncertainty of
the light curve amplitude of 0.07 mag. The SCP sample
has poor or even missing restframe $V$-band measurements. Those from the
HZT compare well with their restframe $B$-band, but however do not
deliver a precise enough colour that might enter a distance estimator without
a prior. A Hubble diagram combining both sets is shown in figure \ref{fig:scp-hzt-hd}.

Both teams fitted their data to an $\Omega_M,\Omega_\Lambda$ Universe
and found extremely similar best fits and contours:
$Prob(\Omega_\Lambda>0)>99\%$, a flat matter-dominated Universe was
excluded at more than 7 $\sigma$, $\Omega_M=0.28 \pm 0.10$ for a flat
Universe.  Fitting a matter density with $\Omega_\Lambda = 0$ drives
$\Omega_M$ negative for both analyses, a certainly uncomfortable
situation, which strongly suggests acceleration. Both teams
reported a long study of systematic uncertainties, and did not find
room for large offsets: the SCP reported a systematic uncertainty
roughly half of the statistical one.

The significance of a non-zero cosmological constant
was in fact modest, typically 3 $\sigma$ or less, but there were two of
them, however statistically correlated because they share
a great part of their nearby samples. The significance of acceleration  
is even smaller. But taking the confidence levels at face value, 
matter-dominated universes with $\Omega_M>0.2$ were excluded at 
more than 4 $\sigma$ by each analysis.

Both teams also explored variants of the cosmological constant by
replacing it by a fluid X with unknown equation of state $w$ (i.e. $w$
is the ratio of pressure to density, see \S \ref{sec:Friedman}), so that
 $\rho_X(z) = \rho_X(0)(1+z)^{3(w+1)}$. In a flat Universe,
both analyses \cite{Garnavich98,Perlmutter99} found similar and almost
degenerate contours, which can be roughly approximated by
$\Omega_Xw=(1-\Omega_M)w \simeq -0.72 \pm 0.1 $.

These twin papers are usually regarded as the original key evidence for an
accelerated expansion, and Saul Perlmutter (SCP), Adam Riess and Brian
Schmidt (HZT) were awarded the 2011 Nobel Price in Physics for this discovery.

\subsection{First Hubble Space Telescope samples\label{sec:HST-first}}

Both pioneering analyses illustrate that the colour measurement quality
is the Achilles' heel of distance measurements, because both teams could
not use the bare colour measurements in their distance estimation. 
Measuring accurate light curves and accurate colours with the HST seems 
an efficient response to this limitation and HST photometric follow-up 
of ground-based searches started in 1997. One event from the SCP
sample has HST measurements and three events from the HZT sample 
\cite{GarnavichHST98}. 

The SCP team delivered \cite{Knop03} the first cosmological
analysis in 2003 where distances to high-redshift events use 
bare measured colours, from a mixture of 11 new HST well-measured events
and ground-based ones. Interestingly, three ways of
inserting colours in the distance estimator are compared (figure 9 of
\cite{Knop03}):
\begin{enumerate}
\item ignoring colours on a low reddening sample (as in the SCP original paper);
\item applying a one-sided extinction prior (as in the HZT original paper);
\item using bare measured colours.
\end{enumerate}

One realizes that with the third approach, the events with HST
colour measurements totally dominate the obtained cosmological
constraints, which are slightly worse than the ones drawn in each of the twin
papers. The ``ground-based'' events alone deliver very poor
constraints if their bare colour measurements are used.

Among these different approaches, one is not necessarily right and the
others wrong; the different obtained cosmological constraints rather
reflect different hypotheses. In the first method, one should check
(as done in SCP original analysis \cite{Perlmutter99}) that the average colour of distant and nearby events
agrees, but if the uncertainty in the difference between the average
colour of both samples is propagated to the end of the analysis, one
gets the same uncertainties as with the third method. So, not
propagating this uncertainty relies on the assumption that both
samples have the same average colour, and that the
measured difference reflects sampling variance, provided however that 
it is compatible with zero. The second method
assumes that the colour
distribution itself does not vary with redshift, i.e. that events
at all redshifts are drawn from the same parent population
(possibly with selection biases, see \S \ref{sec:MLCS}).
So, the first method assumes identical colour averages, the second
identical (or known) colour distributions and the third makes no assumption.
The third method is robust to unknown redshift-dependent
selection biases or demographic evolution, which should generally be expected to some level.

This 11-event sample of \cite{Knop03} also illustrates 
wavelength coverage issues: these events were measured in  
$R$ and $I$ filters (namely F675W and F814W of WFPC2), which
map almost exactly on restframe $B$ and $V$ at $z \simeq 0.5$ but on 
$U$ and $B$ at $z\sim 0.75$. The colours of three events at $z>0.7$ 
could only be measured by a comparison with $U-B$ of nearby events,
but there were very few nearby events to compare, and the 
restframe $U-B$ versus $B-V$ colour relation was logically the dominant
uncertainty of the study (table 9. of \cite{Knop03}).
Later, the $U-B$ versus $B-V$ restframe relation was
first established in \cite{Guy05} from nearby events
from \cite{Jha06}, and then using distant events \cite{Astier06,Guy07,
ConleySifto08}.

\subsection{HST searches \label{sec:HST-searches}}

In 2002, the Advanced Camera for Surveys (ACS) was installed on HST,
and opened the possibility of finding faint high-redshift events with
HST.  The first harvest of supernovae detected with this instrument
reported 16 events \cite{Riess04} (complemented by 2 from \cite{Blakeslee03})
among which 8 at z$>$1. The
identification came from a mixture of ground-based spectroscopy, ACS
low-resolution slit-less spectroscopy, and typing from light curves.  The
highest redshift light curves are measured in the NIR
using NICMOS aboard HST, using the F110W ($[0.8,1.4] \mu m$) and the
F160W ($[1.4,1.8] \mu m$) filters. This work gathers all SN
Ia events available by late 2003, selects a ``gold'' sample and estimates
distances in a consistent way, using MLCS2k2 (\S \ref{sec:MLCS}) 
with an extinction prior. 
Taking advantage of the increase in the
redshift lever arm offered by the $z>1$ events,
the key
conclusion of the paper is to provide evidence for past deceleration,
a distinctive feature of dark energies similar to $\Lambda$.  The
analysis thus eliminated some dust models as alternatives to the
cosmological constant. A standard cosmological fit
yield an the equation of state parameter $w$ very close to -1
(i.e. $\Lambda$) with a precision of $\sim 0.16$.

A second round with 22 more events was published 3 years later
\cite{Riess07}, using a better characterization of NICMOS (the key
instrument for events at $z>1$), establishing that the photometric
response of the instrument is non-linear. Uncertainties on corrections
are not propagated in the analysis, but however cause sizable distance
uncertainties common to all high-redshift events (see e.g. \S 5.1.3 of
\cite{Conley11} for details). The cosmological analysis
concentrates on variable equation of state fits, concludes that the
data do not require $w \neq -1$, and also indicates that $w(z>1)<0$
at $\sim$98\% CL. However, the determination of $w(z>1)$ still has a
variance of order 1 or more. 

More HST events have been collected or published
in the last few years within different
programmes. The six events (two at $z>1$) presented in \cite{Amanullah10} have 
HST data. The HST Cluster Supernova survey targets high-redshift
galaxy clusters in rolling search mode, in order to increase the SN
yield with respect to blank fields \cite{Dawson09}. Recently, the
programme delivered accurate distances to 14 SNe \cite{Suzuki12} among
which 10 are at $z>1$. The CLASH+CANDELS SN programme runs a rolling
search with the new WFC3-IR imager, targeting high-redshift events
mostly in galaxy clusters. It recently delivered an accurate distance
to an event at $z=1.55$ \cite{Rodney12}.

Although the redshift coverage of HST samples is unique,
they now have a modest weight in cosmological fits 
of the latest compilations, which typically gather
several hundred events (see e.g. \cite{Sullivan11}, and 
\S \ref{sec:compilations}).

\subsection{Ground-based rolling searches: ESSENCE, SNLS,  and SDSS-II}
\label{sec:ESSENCE-SNLS-SDSS}

The ESSENCE, SNLS and SDSS-II SN surveys are second-generation
``high-redshift'' surveys. They all aim at a significant increase in
both the amount and the quality of well-measured high-redshift SNe Ia,
in order to constrain the dark energy equation of state. They all
worked in the rolling search mode (\S \ref{sec:finding-supernovae}),
i.e. image repeatedly the same pointings and extract detections and
photometry of the events from the same image sequence.  All three
surveys benefited from large guaranteed multi-year allocations for
their imaging data. They spectroscopically identified their candidates
and measured redshifts using 4 and 8 m class telescopes.

The ESSENCE survey \cite{Miknaitis07} ran at the CTIO 4-m (Northern
Chile) on the 0.36 ${\rm deg^2}$ Mosaic-II imager. It monitored in $R$
and $I$ bands 36 pointings (i.e. $\sim 10 {\rm deg^2}$) every fourth
night, for a 3-month observing season, over 5 years (2003-2008). 
The 2002 season was essentially lost to weather and technical
problems.

The SNLS survey \cite{Astier06} ran at the CFHT 3.6 m in Hawaii,
equipped with the Megacam 1 ${\rm deg^2}$ imager. It monitored 4
pointings in $g$,$r$,$i$ and $z$ bands every fourth to fifth night, as long as
they remained visible from 2003 to 2008.  The observations were part
of the CFHT Legacy Survey\footnote{\href{http://www.cfht.hawaii.edu/CFHTLS}{http://www.cfht.hawaii.edu/CFHTLS}}
(CFHTLS) and were aiming also at galaxy studies in the very deep 
stacked images.

The SDSS-II Supernova Survey \cite{Holtzman08} used the SDSS 2.5-m
telescope at Apache Point (new Mexico), with its original 1.52 ${\rm
  deg^2}$ imaging camera. It monitored a 300 ${\rm deg^2}$ equatorial
stripe every second night for 3 months per year in $u$,$g$,$r$,$i$ and $z$
bands.

The deepest survey is SNLS, which delivered usable light curves at
$z>1$, thanks to $\sim$40 mn exposures in the red bands, with good IQ. 
Then comes ESSENCE, which used 200- and 400-s exposures at a
similar cadence, and is thus limited to $z \lesssim 0.7$. SDSS took
advantage of its fast scanning camera and used 54 s exposures and thus
efficiently addresses the $0.1 \lesssim z \lesssim 0.4$ ``redshift desert'' where
regular point-and-stare imaging with large telescopes is inefficient. ESSENCE and SNLS use
similar instrumentation and have made opposite choices: ESSENCE went
for a wide survey in order to obtain the smallest
statistical uncertainty on $w$ \cite{Miknaitis07}, while SNLS was
limited to four deep pointings by the observing time allocation
mechanism. It turned out that both surveys found about twice as many 
events than they could possibly identify within their sizable spectroscopic
observing time allocations.

\begin{figure*}
\begin{center}
\includegraphics[width=0.9\textwidth]{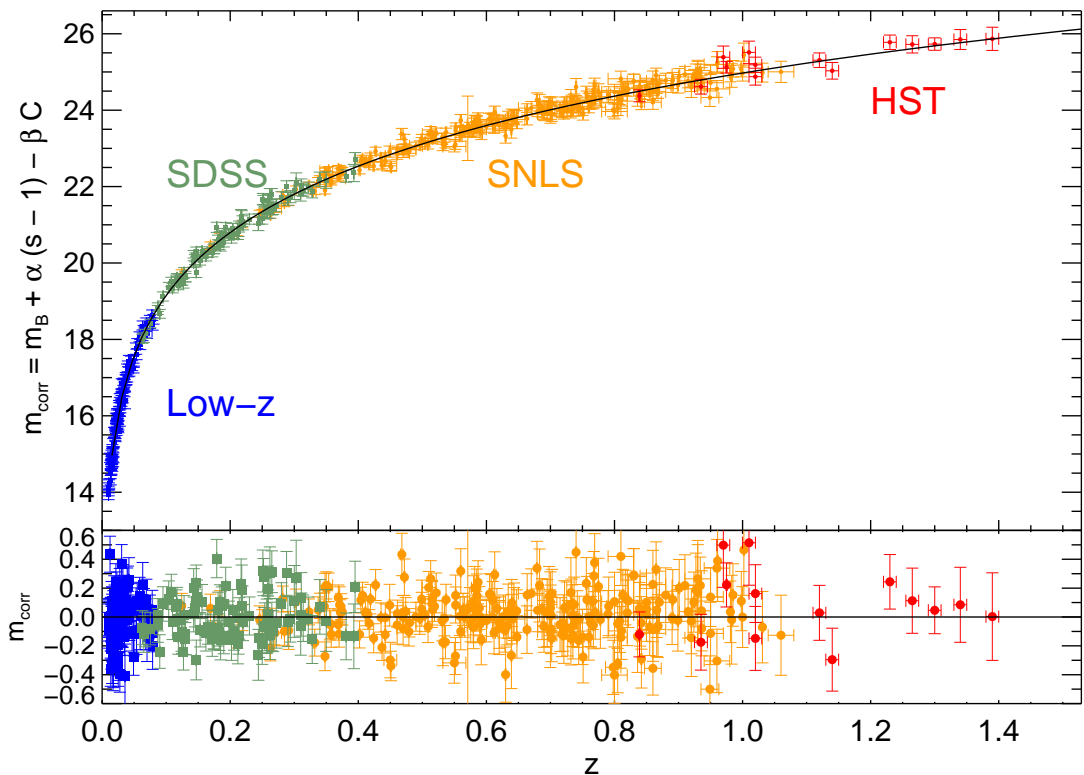}
\caption{One of the latest SNe~Ia Hubble diagrams (logarithm of
distance versus $z$), which gathers well-measured events by 2010. 
  The lower plot displays residuals to the best fit
  cosmology. A comparison with figure \ref{fig:scp-hzt-hd} shows that distance
uncertainties and scatter of high-redshift events were almost halved in about 10 years. Reproduced from \cite{Conley11}, with permission. \label{fig:snls3-hd}}
\end{center}
\end{figure*}

\begin{figure*}[t]
\begin{minipage}[t]{0.49\textwidth}
\begin{center}
\includegraphics[width=\textwidth]{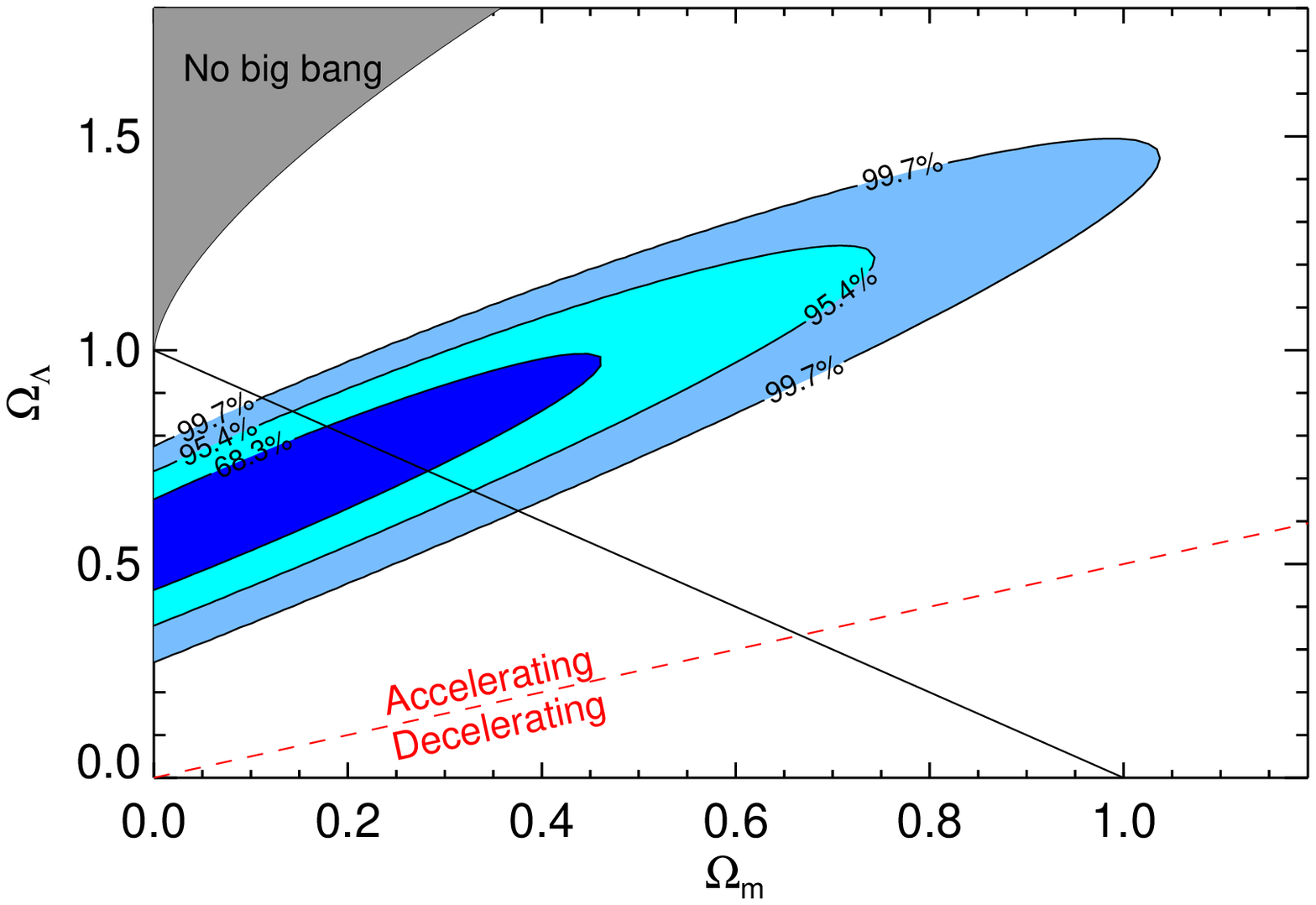}
\caption{Statistical constraints from supernovae alone (figure \ref{fig:snls3-hd})
on $(\Omega_{\rm M}, \Omega_\Lambda)$ cosmological models, including systematic 
uncertainties. Reproduced from \cite{Conley11}, with permission. \label{fig:snls3-omol}}
\end{center}
\end{minipage}
\hfill
\begin{minipage}[t]{0.49\textwidth}
\begin{center}
\includegraphics[width=\textwidth]{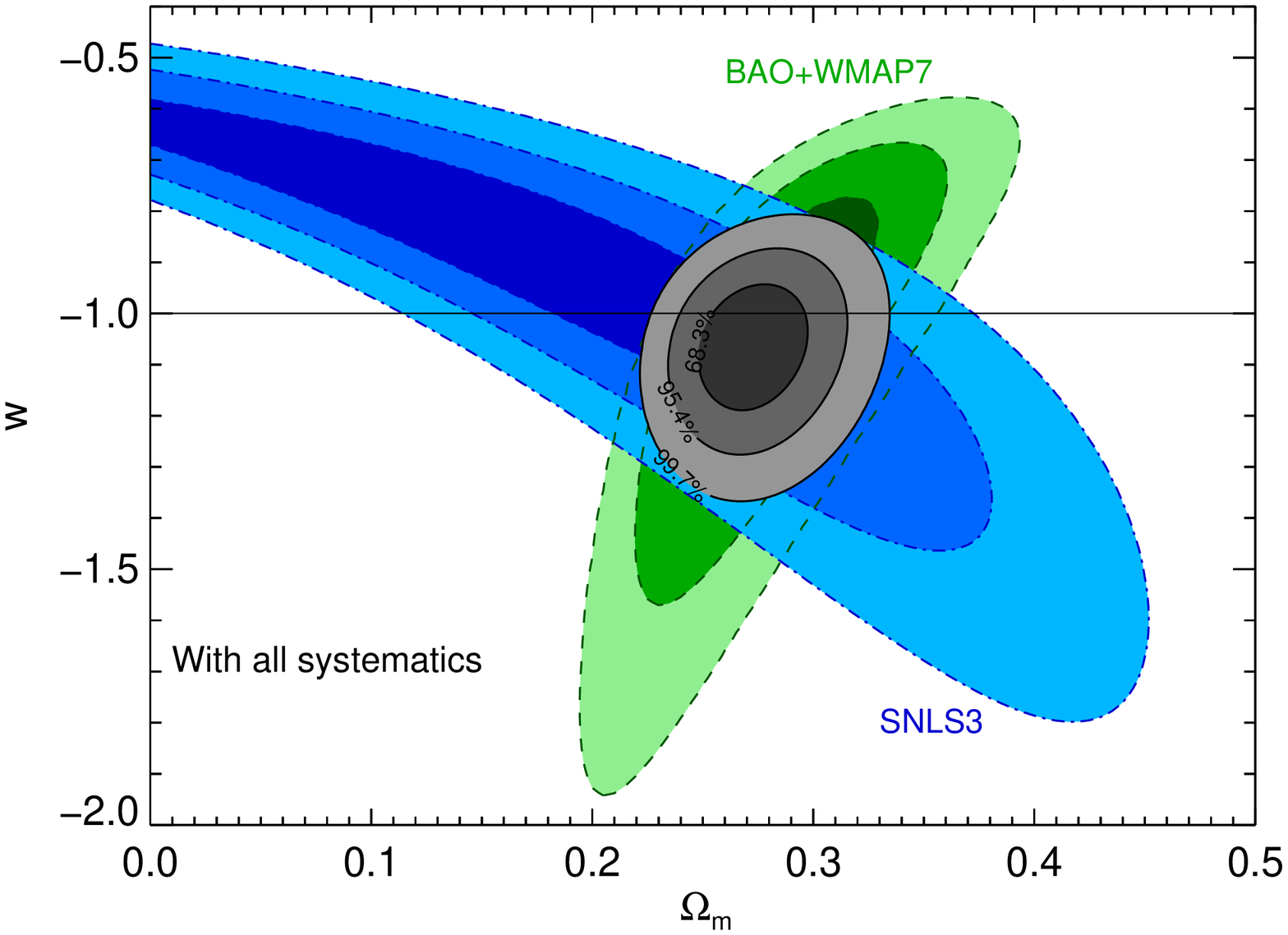}
\caption{Statistical joint constraints on $(\Omega_{\rm M},w)$ 
flat cosmological models from supernovae of figure \ref{fig:snls3-hd} 
(blue contours including systematics), 
CMB and BAOs (green). The joint contours (grey) favour models close to 
a cosmological constant, i.e. $w=-1$. Reproduced from \cite{Sullivan11},
with permission.
\label{fig:snls3-om-w}}
\end{center}
\end{minipage}
\end{figure*}

SNLS has produced its first \cite{Astier06} and third
\cite{Regnault09,Guy10,Conley11,Sullivan11} year analyses. The first
year analysis gathers 44 high quality nearby events and compares those
to 71 SNLS distant events, using distances estimated with SALT (\S \ref{sec:SALT}). Constraining supernova restframe $U-B$ versus $B-V$
colour relation is performed using the distant sample (measured in more
than 2 bands), where it is found to be much better defined than in
nearby samples. A constant equation of state fit to this data set
yields $w=-1.02 \pm 0.09 (stat) \pm 0.054 (syst)$, a result compatible
with $\Lambda$ and systematic uncertainties dominated by photometric
calibration issues. The third year SNLS analysis uses a much larger
sample of strictly selected events (table 1. in \cite{Conley11}): 123
nearby events, 242 SNLS events, 93 from the first year SDSS sample,
and 14 HST events (see figure \ref{fig:snls3-hd}).  The systematic
uncertainties correlate events and the introduced covariances are
propagated to the cosmological fits, and delivered with the data
\cite{Conley11}. The light curve fitters (SALT2 and SiFTO \S \ref{sec:SALT} \& \ref{sec:sifto}) training uses distant events in
order to model the UV behaviour of supernovae
\cite{Guy07,ConleySifto08}, which improves the distance scatter at
$z>0.8$ by about a factor of 2. For a flat Universe, a constant equation 
of state fit to the supernovae data, CMB, and BAO
yields \cite{Sullivan11} $w=-1.068 \pm 0.08(stat+sys)$, where
statistics contributes 0.054. Again, systematics are totally dominated
by several photometric calibration limitations (table 3 of \cite{Sullivan11}). 
Some could be overcome in a joint analysis with the SDSS full sample.
Fits allowing a variable
equation of state $w(z) = w_0+z w_a/(1+z)$ yield $w_a=-0.98 \pm 1.1 $ compatible with 0. These cosmological fits are fully compatible 
with the cosmological constant hypothesis (figures \ref{fig:snls3-omol} and \ref{fig:snls3-om-w}).

ESSENCE has produced its fourth year analysis
\cite{Miknaitis07,WoodVasey07} which presents 60 new events at
$z \lesssim 0.7$. These events are analysed using MLCS2k2 (\S \ref{sec:MLCS}), and considerable
care is devoted to choosing the prior\footnote {The same data set is
  analysed with a different extinction prior in \cite{Kessler09}},
which incorporates provision to compensate selection biases, because
the images are shallow. A constant equation of state fit yields
$w=-1.05 \pm 0.12(stat) \pm 0.13(syst)$.  The uncertain choice of the
extinction prior dominates systematics (0.08), followed by the
possible existence of a ``Hubble bubble'' (0.06) and the restframe $U-B$
versus $B-V$ relation (0.06). These three main contributions could be
reduced.

The SDSS-II SN survey has produced its first year analysis
\cite{Holtzman08,Kessler09} with 103 new events at $0.04<z<0.42$. The
presented analysis gathers nearby samples, the SNLS first year,
ESSENCE and HST samples. As already discussed (\S \ref{sec:LC-comp}),
the core of the cosmology paper \cite{Kessler09} is a comparison of
the outcome of MLCS2k2 and SALT2 frameworks, which finds
alarmingly different cosmologies. We just repeat here  (\S \ref{sec:LC-comp})
that
the main causes of the discrepancy are identified as the restframe
$U-B$ versus $B-V$ relation (likely induced by photometric calibration issues of
the training sample), and the effect of applying or not an extinction
prior to the SNLS sample. Both can be settled (and indeed are), and
hence do not threaten the use of supernovae as distance
indicators. This detailed and alarming comparison of approaches was
sometimes interpreted as a general failure of supernova cosmology. 
Alternatively, one might
argue that both the ESSENCE \cite{WoodVasey07} and SDSS
\cite{Kessler09} analyses have convincingly demonstrated that
uncertainties associated with the choice of an extinction prior 
are now too large for the present sample sizes.

Distances to 35 SDSS, ESSENCE and SNLS supernovae at $0.2<z<0.7$ were
measured in the restframe I-band (i.e. in NIR for observers) and with
different systematics, confirm the findings of the parent surveys
\cite{Freedman09}: $w = -1.05 \pm 0.13(stat) \pm 0.09(sys)$.

None of these second-generation surveys has finalised its full-sample
analysis yet, but efforts on cross-calibrating the surveys are
on-going, in order to tighten the relative flux scales. Because of their
multi-band coverage, SNLS and SDSS are expected to produce a precisely
cross-calibrated sample. In this scenario, nearby samples 
will not benefit from an improved SNLS-SDSS cross-calibration and their weight
in the determination of cosmological parameters will be diminished.

\subsection{Supernova compilations \label{sec:compilations}} 
There is no example of a single survey delivering interesting
supernova constraints on its own: covering a large redshift range
requires different instruments generally operated within different
projects. Compilations of supernovae primarily target the production
of cosmological constraints and mainly consist in producing a coherent
set of distances, possibly with their systematic uncertainties. In most of the cases,
compilations present some new events. By delivering state-of-the-art
and homogenized collections of events, these compilations receive a 
lot of attention from users primarily interested 
in cosmology constraints.

In \cite{Tonry03}, a new distance estimation method (BATM) is compared
to MLCS2k2 and two versions of $\Delta m_{15}$, and found to be
``generally consistent''. Using a combined distance, 172
events are then placed on a Hubble diagram, and the cosmological 
conclusions
strengthen earlier findings. In \cite{Riess04}, events found and measured
with the HST (see \S \ref{sec:HST-searches}) are merged into a
large compilation, ranking events into gold and silver samples,
analysed using MLCS2k2.
The {\it
  Union} sample \cite{Kowalski08} consists of 307 events selected from
more than 10 sources and analysed using SALT. The analysis accounts
for systematic uncertainties, in particular photometric calibration, and delivers
correlated uncertainties. This compilation was
augmented into the {\it Union~2} compilation \cite{Amanullah10}, mainly
adding the SDSS (\S \ref{sec:ESSENCE-SNLS-SDSS}) and CFA3 (\S
\ref{sec:nearby-samples}) samples, and reaches 557 events fitted using
SALT2, again evaluating systematic uncertainties in a correlated fashion. The {\it Union~2.1}
compilation \cite{Suzuki12} adds 23 events, in particular 14 HST events among
which 10 are at $z>1$.  The Union (1) compilation was also augmented
into the {\it Constitution} compilation \cite{Hicken09} by the
addition of the CFA 3 nearby SNe sample, analysed using MLCS2k2, SALT
and SALT2.  As noted in \S \ref{sec:LC-comp}, the comparison of light
curve fitters conducted there is very likely affected by an improper
use of SALT and SALT2.

\section{Prospects for supernova cosmology\label{sec:prospects}}
The fact that statistical and systematic uncertainties 
contribute similar amounts to current cosmological constraints 
from supernovae forces new supernova survey projects to
rely on new setups. Three fronts are usually simultaneously 
pursued: increasing the number and quality of events, 
improving the quality of the photometric calibration and
collecting higher redshift events. 

One should note that essentially all supernovae events involved in
cosmological analyses to date have been spectroscopically
confirmed. Given that samples already on disk amount today to about
1000 events usable for cosmology, future SN imaging surveys should
target at least a few thousand well-measured events. Spectroscopic
identification of such a large sample lies beyond likely allocations
on large-aperture telescopes.  Recent practical studies of photometric
identification (i.e. from light curves) propose techniques to reach a
purity of about 95\%, with an efficiency still greater than that of
spectroscopic identification \cite{KesslerChallenge10,Sako11,Bazin11}.  The required
quality of redshifts is still debated, and ``after the fact''
multi-object spectroscopy of host galaxies constitutes the most
obvious way to secure high-quality redshifts.  So, broadband
photometry of supernovae should aim in the future at providing
both distances and identification, and possibly even redshifts
\cite{Palanque10,Kessler-photoz-10}. Regarding dark energy
constraints, all current wide-field imaging large-scale projects aim
at combining probes, typically cluster counts, weak shear
correlations, BAOs, and distances to
supernovae.  Two reports \cite{DETF06,ESO-ESA} have reviewed in detail
the scientific and technical merits of these probes and note that
distances to supernovae is the most mature.  Both reports also stress
that the potential reach of each probe at constraining dark energy
primarily depends on systematic uncertainties, which especially for
forecasts are highly uncertain, and strongly suggest a multi-probe
approach. Coming back to results, one might note that five years after
these reports, dark energy constraints are still dominated by
supernova distances.

\subsection{Space missions concepts}

Although several mission concepts to measure supernovae from space
were developed in the last decade, none has yet passed the selection
processes. Working from space is mandatory at $z \gtrsim 1$, because
reliable distances should then be measured in the NIR,
in order to allow a direct comparison with nearby events measured in
blue bands. Ground-based measurements in the NIR suffer from both a
very bright atmospheric emission and a sizable and time-variable
atmospheric extinction. Space also offers a very good image quality
and a stable environment, which both significantly improve photometric
measurements with respect to ground.

Euclid at ESA is an approved dark energy project, to be launched in
2019, but has no compelling supernova programme yet: regarding dark
energy, the project concentrates on measuring galaxy shapes for weak
lensing and conducting a (spectroscopic) galaxy redshift survey in NIR
for BAOs \cite{EuclidRB}, on about one third of the sky. It will also
deliver NIR imaging in three bands of the surveyed fields.  The WFIRST
project at NASA is not yet approved and contains a high-redshift
supernova programme, targeting a few thousand events.

\subsection{Ground-based SN projects}

Planned ground-based SN surveys will enlarge the redshift span of
Hubble diagrams (in particular compared with SNLS) by using thick
high-resistivity CCD sensors (e.g. \cite{Holland02}), which typically
deliver quantum efficiencies above 50\% at 1 ${\rm \mu}$m (while the
silicon cutoff lies at 1.1 ${\rm \mu}$m). Ground-based projects are
including the four canonical dark energy probes in their science
agenda. High-statistics Hubble diagrams with a high efficiency up to
high redshifts ($z\sim 1$) is reserved to 4 and 8 m class
telescopes. Table \ref{tab:wide-field-projects} summarizes the key
figures of these projects.
\begin{table}
\begin{center}
\begin{tabular}{|l|rrrr|}
\hline
Project    & Mirror          &  Area          & First & Large survey\\
           & $\varnothing$ (m) & (${\rm deg^2}$) &  light & (nights)\\
\hline
CFHT/Megacam & 3.6  & 1.    & 2002 &  500 \\
Pan-STARRS & 1.8  & 7.  & 2009 &  $>$1000 \\
Blanco/DEC & 4.0  & 3.    & 2012 & 500 \\
Subaru/HSC & 8.2 & 1.8 & 2013 &  $\sim$500 \\
LSST      & 6.5  & 10.  & 2019 & 3500 \\
\hline
\end{tabular}
\caption{Key figures for the major past and future wide-field imaging
  facilities. The SNLS survey was part
  of the CFHTLS large survey. Pan-STARRS is currently constructing a
  second telescope and aims at eventually operating four of them. The
  500-night survey to be run at the CTIO-Blanco is called Dark Energy
  Survey (DES). LSST real diameter is 8.2~m but suffers from a 5~m central 
occultation. This facility will almost entirely observe in survey mode
  during its anticipated 10-year lifetime. \label{tab:wide-field-projects}
}
\end{center}
\end{table}

For DES, the promises of the SN survey were published
recently \cite{Bernstein11}: the available observing time will allow
one 
to measure distances to about 4000 events at $z<1.2$,
and tentatively relying on photometric redshift of both
host galaxies and supernovae to build a Hubble diagram.

\section{Summary and conclusions \label{sec:conclusion}}
During the 1990s, distances to SNe~Ia allowed observers to map the
expansion history of the Universe, and led to the discovery that this
expansion is currently accelerating. Dark energy has been inserted in
the standard cosmological model as the fluid causing the acceleration,
which has been confirmed independently of supernovae. Second-generation 
supernova surveys have delivered precise constraints on the
``nature'' of dark energy: the cosmological constant paradigm,
corresponding to a dark energy of constant density, describes the data very
well and leaves little room for dark energy density
evolution. The ${\rm \Lambda}$CDM cosmological model successfully
describes large-scale cosmological observables, including those
involving structure formation.

The nature of SNe~Ia explosion is still questioned, and both
progenitor scenarios have supporting evidence and face difficulties.
The paradigm of a WD exploding when reaching the
Chandrasekhar mass fulfils observational constraints, but is accepted
by default of an alternative rather than observational evidence, the
lack thereof being perfectly understood. Numerical explosion models
broadly match observations, even if they still cannot replace
empirical models used to fit observed light curves.  Cosmology is
expecting to gain insight into possible distance biases from the steady
improvement of numerical SN models.

Several second-generation surveys have collected samples of a few
hundred spectroscopically confirmed and well-measured SNe~Ia events up
to $z\sim 1$ over the last decade. This world sample amounts in total
to about 1000 events and the dark energy constraints already
extracted from about one half of it (\cite{Sullivan11}) show that
calibration and statistical uncertainties contribute similar amounts.
Improving the photometric calibration, improving the survey design to
make it less sensitive to calibration errors and increasing the
statistics (in particular at the highest possible redshift) are the
three main handles to improve over present constraints.  Since
spectroscopic identification of several thousand events involves too
much observing time, broadband photometry should now also deliver
identification of the events, which pushes for increased quality
measurements. Constraints from other dark energy probes will, hopefully,
soon become available and crossing results will tighten and secure the
obtained dark energy constraints.

Future SN surveys are planned but only secured on ground-based
facilities, although observing from space opens the possibility of
measuring precise distances to supernovae at $z>1$, a key handle on
dark energy constraints. Ground-based imaging surveys and space-based
imaging and spectroscopy surveys target high-quality dark energy
constraints.  For the next decade, the LSST (ground) and Euclid
(space) projects are aiming at surveying most of the extragalactic
sky, and expect to see their first light around 2020. Currently, only
LSST plans to constrain dark energy from supernovae distances.

\vspace{5mm}  
\begin{acknowledgments}
Writing this review would have been impossible without
numerous discussions with many colleagues, in particular C. Balland,
S. Beaumont, M. Betoule, S. Bongard, J. Guy, D. Hardin, R. Pain and
N. Regnault. It also benefited a lot from the remarks from the two
anonymous referees. Several illustrations were borrowed from
published papers and we are grateful to the original authors 
and publishers for granting their authorization to reproduce these figures.
\end{acknowledgments} 

\def\lt{$<$}
\def\gt{$>$}
\def\apj{ApJ}
\def\apjs{ApJS}
\def\apjl{ApJ Lett.}
\def\aj{AJ}
\def\pasp{PASP} 
\def\apss{Ap\&SS}
\def\araa{ARA\&A}
\def\aaps{A\&AS}
\def\na{New A}
\def\aa{A\&A}
\def\aap{A\&A}
\def\nat{Nature}
\def\mnras{MNRAS}
\def\memras{MmRAS}
\def\aapr{The Astron. and Astrop. Rev.}
\def\prd{Phys. Rev. D}
\def\sovast{Soviet Astronomy}
\def\jcap{J. Cosm. Astropart. P.}

\def\newblock{\ }
\bibliographystyle{hunsrt3}

\bibliography{biblio}

\begin{thebibliography}{100}

\bibitem{Lemaitre27}
{Lema{\^i}tre}, G.
\newblock {Un Univers homog{\`e}ne de masse constante et de rayon croissant
  rendant compte de la vitesse radiale des n{\'e}buleuses extra-galactiques}.
\newblock {\em {Annales de la Société Scientifique de Bruxelles}}, 47:49--59,
  1927\href{http://adsabs.harvard.edu/abs/1927ASSB...47...49L.
\newblock }{ (ADS)}.

\bibitem{Lemaitre27-translation}
{Lema{\^i}tre}, G.
\newblock {Expansion of the universe, A homogeneous universe of constant mass
  and increasing radius accounting for the radial velocity of extra-galactic
  nebulae}.
\newblock {\em \mnras}, 91:483--490, March
  1931\href{http://cdsads.u-strasbg.fr/abs/1931MNRAS..91..483L.
\newblock }{ (ADS)}.

\bibitem{Hubble29}
{Hubble}, E.
\newblock {A Relation between Distance and Radial Velocity among Extra-Galactic
  Nebulae}.
\newblock {\em Proceedings of the National Academy of Science},
  \href{http://dx.doi.org/10.1073/pnas.15.3.168}{15:168--173}, March
  1929\href{http://cdsads.u-strasbg.fr/abs/1929PNAS...15..168H.
\newblock }{ (ADS)}.

\bibitem{OstrikerTremaine75}
{Ostriker}, J.~P. \& {Tremaine}, S.~D.
\newblock {Another evolutionary correction to the luminosity of giant
  galaxies}.
\newblock {\em \apjl}, \href{http://dx.doi.org/10.1086/181992}{202:L113--L117},
  December 1975\href{http://cdsads.u-strasbg.fr/abs/1975ApJ...202L.113O.
\newblock }{ (ADS)}.

\bibitem{Kowal68}
{Kowal}, C.~T.
\newblock {Absolute magnitudes of supernovae.}
\newblock {\em \aj}, \href{http://dx.doi.org/10.1086/110763}{73:1021--1024},
  December 1968\href{http://cdsads.u-strasbg.fr/abs/1968AJ.....73.1021K.
\newblock }{ (ADS)}.

\bibitem{KK1974}
{Kirshner}, R.~P. \& {Kwan}, J.
\newblock {Distances to extragalactic supernovae}.
\newblock {\em \apj}, \href{http://dx.doi.org/10.1086/153123}{193:27--36},
  October 1974\href{http://cdsads.u-strasbg.fr/abs/1974ApJ...193...27K.
\newblock }{ (ADS)}.

\bibitem{Wagoner77}
{Wagoner}, R.~V.
\newblock {Determining q0 from Supernovae}.
\newblock {\em \apjl}, \href{http://dx.doi.org/10.1086/182429}{214:L5+}, May
  1977\href{http://cdsads.u-strasbg.fr/abs/1977ApJ...214L...5W.
\newblock }{ (ADS)}.

\bibitem{Hamuy96c}
{Hamuy}, M., {Phillips}, M.~M., {Suntzeff}, N.~B., {et~al.}
\newblock {The Hubble Diagram of the Calan/Tololo Type IA Supernovae and the
  Value of HO}.
\newblock {\em \aj}, \href{http://dx.doi.org/10.1086/118191}{112:2398--+},
  December 1996,
  arXiv:astro-ph/9609062\href{http://cdsads.u-strasbg.fr/abs/1996AJ....112.2398H.
\newblock }{ (ADS)}.

\bibitem{Hamuy96b}
{Hamuy}, M., {Phillips}, M.~M., {Suntzeff}, N.~B., {et~al.}
\newblock {BVRI Light Curves for 29 Type IA Supernovae}.
\newblock {\em AJ}, 112:2408--+, December
  1996\href{http://adsabs.harvard.edu/cgi-bin/nph-bib_query?bibcode=1996AJ....112.2408H&amp;db_key=AST.
\newblock }{ (ADS)}.

\bibitem{Riess98b}
{Riess}, A.~G., {Filippenko}, A.~V., {Challis}, P., {et~al.}
\newblock {Observational Evidence from Supernovae for an Accelerating Universe
  and a Cosmological Constant}.
\newblock {\em \aj}, 116:1009--1038, September
  1998\href{http://adsabs.harvard.edu/cgi-bin/nph-bib_query?bibcode=1998AJ....116.1009R&db_key=AST.
\newblock }{ (ADS)}.

\bibitem{Perlmutter99}
{Perlmutter}, S., {Aldering}, G., {Goldhaber}, G., {et~al.}
\newblock {Measurements of Omega and Lambda from 42 High-Redshift Supernovae}.
\newblock {\em \apj}, 517:565--586, June
  1999\href{http://adsabs.harvard.edu/cgi-bin/nph-bib_query?bibcode=1999ApJ...517..565P&db_key=AST.
\newblock }{ (ADS)}.

\bibitem{Efstathiou90}
{Efstathiou}, G., {Sutherland}, W.~J., \& {Maddox}, S.~J.
\newblock {The cosmological constant and cold dark matter}.
\newblock {\em \nat}, \href{http://dx.doi.org/10.1038/348705a0}{348:705--707},
  December 1990\href{http://cdsads.u-strasbg.fr/abs/1990Natur.348..705E.
\newblock }{ (ADS)}.

\bibitem{White93}
{White}, S.~D.~M., {Navarro}, J.~F., {Evrard}, A.~E., \& {Frenk}, C.~S.
\newblock {The baryon content of galaxy clusters: a challenge to cosmological
  orthodoxy}.
\newblock {\em \nat}, \href{http://dx.doi.org/10.1038/366429a0}{366:429--433},
  December 1993\href{http://cdsads.u-strasbg.fr/abs/1993Natur.366..429W.
\newblock }{ (ADS)}.

\bibitem{Spergel07}
{Spergel}, D.~N., {Bean}, R., {Dor{\'e}}, O., {et~al.}
\newblock {Three-Year Wilkinson Microwave Anisotropy Probe (WMAP) Observations:
  Implications for Cosmology}.
\newblock {\em \apjs}, \href{http://dx.doi.org/10.1086/513700}{170:377--408},
  June 2007,
  arXiv:astro-ph/0603449\href{http://cdsads.u-strasbg.fr/abs/2007ApJS..170..377S.
\newblock }{ (ADS)}.

\bibitem{Komatsu11}
{Komatsu}, E., {Smith}, K.~M., {Dunkley}, J., {et~al.}
\newblock {Seven-year Wilkinson Microwave Anisotropy Probe (WMAP) Observations:
  Cosmological Interpretation}.
\newblock {\em \apjs},
  \href{http://dx.doi.org/10.1088/0067-0049/192/2/18}{192:18--+}, February
  2011, 1001.4538\href{http://cdsads.u-strasbg.fr/abs/2011ApJS..192...18K.
\newblock }{ (ADS)}.

\bibitem{GoobarLeibundgut11}
{Goobar}, A. \& {Leibundgut}, B.
\newblock {Supernova Cosmology: Legacy and Future}.
\newblock {\em Annual Review of Nuclear and Particle Science},
  \href{http://dx.doi.org/10.1146/annurev-nucl-102010-130434}{61:251--279},
  November 2011,
  1102.1431\href{http://cdsads.u-strasbg.fr/abs/2011ARNPS..61..251G.
\newblock }{ (ADS)}.

\bibitem{Hillebrandt00}
{Hillebrandt}, W. \& {Niemeyer}, J.~C.
\newblock {Type IA Supernova Explosion Models}.
\newblock {\em \araa},
  \href{http://dx.doi.org/10.1146/annurev.astro.38.1.191}{38:191--230}, 2000,
  arXiv:astro-ph/0006305\href{http://cdsads.u-strasbg.fr/abs/2000ARA%26A..38..191H.
\newblock }{ (ADS)}.

\bibitem{HowellReview11}
{Howell}, D.~A.
\newblock {Type Ia supernovae as stellar endpoints and cosmological tools}.
\newblock {\em Nature Communications},
  \href{http://dx.doi.org/10.1038/ncomms1344}{2}, June 2011,
  1011.0441\href{http://cdsads.u-strasbg.fr/abs/2011NatCo...2E.350H.
\newblock }{ (ADS)}.

\bibitem{FreedmanH010}
{Freedman}, W.~L. \& {Madore}, B.~F.
\newblock {The Hubble Constant}.
\newblock {\em \araa},
  \href{http://dx.doi.org/10.1146/annurev-astro-082708-101829}{48:673--710},
  September 2010,
  1004.1856\href{http://cdsads.u-strasbg.fr/abs/2010ARA%26A..48..673F.
\newblock }{ (ADS)}.

\bibitem{FriemanRA&A}
{Frieman}, J.~A., {Turner}, M.~S., \& {Huterer}, D.
\newblock {Dark Energy and the Accelerating Universe}.
\newblock {\em \araa},
  \href{http://dx.doi.org/10.1146/annurev.astro.46.060407.145243}{46:385--432},
  September 2008,
  0803.0982\href{http://cdsads.u-strasbg.fr/abs/2008ARA%26A..46..385F.
\newblock }{ (ADS)}.

\bibitem{Weinberg12}
{Weinberg}, D.~H., {Mortonson}, M.~J., {Eisenstein}, D.~J., {et~al.}
\newblock {Observational Probes of Cosmic Acceleration}.
\newblock {\em ArXiv e-prints}, January 2012,
  1201.2434\href{http://cdsads.u-strasbg.fr/abs/2012arXiv1201.2434W.
\newblock }{ (ADS)}.

\bibitem{Smoot91}
{Smoot}, G.~F., {Bennett}, C.~L., {Kogut}, A., {et~al.}
\newblock {Preliminary results from the COBE differential microwave radiometers
  - Large angular scale isotropy of the cosmic microwave background}.
\newblock {\em \apjl}, \href{http://dx.doi.org/10.1086/185988}{371:L1--L5},
  April 1991\href{http://adsabs.harvard.edu/abs/1991ApJ...371L...1S.
\newblock }{ (ADS)}.

\bibitem{Hogg05}
{Hogg}, D.~W., {Eisenstein}, D.~J., {Blanton}, M.~R., {et~al.}
\newblock {Cosmic Homogeneity Demonstrated with Luminous Red Galaxies}.
\newblock {\em \apj}, \href{http://dx.doi.org/10.1086/429084}{624:54--58}, May
  2005,
  arXiv:astro-ph/0411197\href{http://cdsads.u-strasbg.fr/abs/2005ApJ...624...54H.
\newblock }{ (ADS)}.

\bibitem{Friedmann24}
Friedmann, A.
\newblock Über die möglichkeit einer welt mit konstanter negativer krümmung
  des raumes.
\newblock {\em Zeitschrift für Physik A Hadrons and Nuclei}, 21:326--332,
  1924.
\newblock 10.1007/BF01328280.

\bibitem{CPT92}
{Carroll}, S.~M., {Press}, W.~H., \& {Turner}, E.~L.
\newblock {The cosmological constant}.
\newblock {\em \araa},
  \href{http://dx.doi.org/10.1146/annurev.aa.30.090192.002435}{30:499--542},
  1992\href{http://cdsads.u-strasbg.fr/abs/1992ARA%26A..30..499C.
\newblock }{ (ADS)}.

\bibitem{Etherington33}
{Etherington}, I.~M.~H.
\newblock {On the Definition of Distance in General Relativity.}
\newblock {\em Philosophical Magazine}, 15:761--773,
  1933\href{http://cdsads.u-strasbg.fr/abs/1933PMag...15..761E.
\newblock }{ (ADS)}.

\bibitem{Riess11}
{Riess}, A.~G., {Macri}, L., {Casertano}, S., {et~al.}
\newblock {A 3\% Solution: Determination of the Hubble Constant with the Hubble
  Space Telescope and Wide Field Camera 3}.
\newblock {\em \apj},
  \href{http://dx.doi.org/10.1088/0004-637X/730/2/119}{730:119}, April 2011,
  1103.2976\href{http://cdsads.u-strasbg.fr/abs/2011ApJ...730..119R.
\newblock }{ (ADS)}.

\bibitem{Goobar95}
{Goobar}, A. \& {Perlmutter}, S.
\newblock {Feasibility of Measuring the Cosmological Constant Lambda and Mass
  Density Omega Using Type IA Supernovae}.
\newblock {\em \apj}, \href{http://dx.doi.org/10.1086/176113}{450:14},
  September 1995,
  arXiv:astro-ph/9505022\href{http://cdsads.u-strasbg.fr/abs/1995ApJ...450...14G.
\newblock }{ (ADS)}.

\bibitem{Riess04}
{Riess}, A.~G., {Strolger}, L., {Tonry}, J., {et~al.}
\newblock {Type Ia Supernova Discoveries at z $>$ 1 from the Hubble Space
  Telescope: Evidence for Past Deceleration and Constraints on Dark Energy
  Evolution}.
\newblock {\em \apj}, 607:665--687, June
  2004\href{http://adsabs.harvard.edu/cgi-bin/nph-bib_query?bibcode=2004ApJ...607..665R&db_key=AST.
\newblock }{ (ADS)}.

\bibitem{Cattoen07}
{Catto{\"e}n}, C. \& {Visser}, M.
\newblock {The Hubble series: convergence properties and redshift variables}.
\newblock {\em Classical and Quantum Gravity},
  \href{http://dx.doi.org/10.1088/0264-9381/24/23/018}{24:5985--5997}, December
  2007, 0710.1887\href{http://cdsads.u-strasbg.fr/abs/2007CQGra..24.5985C.
\newblock }{ (ADS)}.

\bibitem{Durrer01}
{Durrer}, R.
\newblock {The theory of CMB anisotropies}.
\newblock {\em Journal of Physical Studies}, 5:177--215, 2001,
  arXiv:astro-ph/0109522\href{http://adsabs.harvard.edu/abs/2001JPhSt...5..177D.
\newblock }{ (ADS)}.

\bibitem{SeoEisenstein03}
{Seo}, H.-J. \& {Eisenstein}, D.~J.
\newblock {Probing Dark Energy with Baryonic Acoustic Oscillations from Future
  Large Galaxy Redshift Surveys}.
\newblock {\em \apj}, \href{http://dx.doi.org/10.1086/379122}{598:720--740},
  December 2003,
  arXiv:astro-ph/0307460\href{http://cdsads.u-strasbg.fr/abs/2003ApJ...598..720S.
\newblock }{ (ADS)}.

\bibitem{Blake11BAO}
{Blake}, C., {Kazin}, E.~A., {Beutler}, F., {et~al.}
\newblock {The WiggleZ Dark Energy Survey: mapping the distance-redshift
  relation with baryon acoustic oscillations}.
\newblock {\em \mnras},
  \href{http://dx.doi.org/10.1111/j.1365-2966.2011.19592.x}{418:1707--1724},
  December 2011,
  1108.2635\href{http://cdsads.u-strasbg.fr/abs/2011MNRAS.418.1707B.
\newblock }{ (ADS)}.

\bibitem{Eisenstein05}
{Eisenstein}, D.~J., {Zehavi}, I., {Hogg}, D.~W., {et~al.}
\newblock {Detection of the Baryon Acoustic Peak in the Large-Scale Correlation
  Function of SDSS Luminous Red Galaxies}.
\newblock {\em \apj}, \href{http://dx.doi.org/10.1086/466512}{633:560--574},
  November 2005,
  arXiv:astro-ph/0501171\href{http://cdsads.u-strasbg.fr/abs/2005ApJ...633..560E.
\newblock }{ (ADS)}.

\bibitem{PeeblesRatra03}
{Peebles}, P.~J. \& {Ratra}, B.
\newblock {The cosmological constant and dark energy}.
\newblock {\em Reviews of Modern Physics},
  \href{http://dx.doi.org/10.1103/RevModPhys.75.559}{75:559--606}, April 2003,
  arXiv:astro-ph/0207347\href{http://cdsads.u-strasbg.fr/abs/2003RvMP...75..559P.
\newblock }{ (ADS)}.

\bibitem{Carrol01}
{Carroll}, S.~M.
\newblock {The Cosmological Constant}.
\newblock {\em Living Reviews in Relativity}, 4:1, February 2001,
  arXiv:astro-ph/0004075\href{http://cdsads.u-strasbg.fr/abs/2001LRR.....4....1C.
\newblock }{ (ADS)}.

\bibitem{Weinberg89}
{Weinberg}, S.
\newblock {The cosmological constant problem}.
\newblock {\em Reviews of Modern Physics},
  \href{http://dx.doi.org/10.1103/RevModPhys.61.1}{61:1--23}, January
  1989\href{http://adsabs.harvard.edu/abs/1989RvMP...61....1W.
\newblock }{ (ADS)}.

\bibitem{Odom06}
{Odom}, B., {Hanneke}, D., {D'Urso}, B., \& {Gabrielse}, G.
\newblock {New Measurement of the Electron Magnetic Moment Using a One-Electron
  Quantum Cyclotron}.
\newblock {\em Physical Review Letters},
  \href{http://dx.doi.org/10.1103/PhysRevLett.97.030801}{97(3):030801--+}, July
  2006\href{http://cdsads.u-strasbg.fr/abs/2006PhRvL..97c0801O.
\newblock }{ (ADS)}.

\bibitem{Caldwell98}
{Caldwell}, R.~R., {Dave}, R., \& {Steinhardt}, P.~J.
\newblock {Cosmological Imprint of an Energy Component with General Equation of
  State}.
\newblock {\em Physical Review Letters},
  \href{http://dx.doi.org/10.1103/PhysRevLett.80.1582}{80:1582--1585}, February
  1998,
  arXiv:astro-ph/9708069\href{http://cdsads.u-strasbg.fr/abs/1998PhRvL..80.1582C.
\newblock }{ (ADS)}.

\bibitem{Wetterich88}
{Wetterich}, C.
\newblock {Cosmology and the fate of dilatation symmetry}.
\newblock {\em Nuclear Physics B},
  \href{http://dx.doi.org/10.1016/0550-3213(88)90193-9}{302:668--696}, June
  1988\href{http://cdsads.u-strasbg.fr/abs/1988NuPhB.302..668W.
\newblock }{ (ADS)}.

\bibitem{RatraPeebles88}
{Ratra}, B. \& {Peebles}, P.~J.~E.
\newblock {Cosmological consequences of a rolling homogeneous scalar field}.
\newblock {\em \prd},
  \href{http://dx.doi.org/10.1103/PhysRevD.37.3406}{37:3406--3427}, June
  1988\href{http://cdsads.u-strasbg.fr/abs/1988PhRvD..37.3406R.
\newblock }{ (ADS)}.

\bibitem{FerreiraJoyce98}
{Ferreira}, P.~G. \& {Joyce}, M.
\newblock {Cosmology with a primordial scaling field}.
\newblock {\em \prd},
  \href{http://dx.doi.org/10.1103/PhysRevD.58.023503}{58(2):023503--+}, July
  1998,
  arXiv:astro-ph/9711102\href{http://cdsads.u-strasbg.fr/abs/1998PhRvD..58b3503F.
\newblock }{ (ADS)}.

\bibitem{Wetterich08}
{Wetterich}, C.
\newblock {Naturalness of exponential cosmon potentials and the cosmological
  constant problem}.
\newblock {\em \prd},
  \href{http://dx.doi.org/10.1103/PhysRevD.77.103505}{77(10):103505--+}, May
  2008, 0801.3208\href{http://cdsads.u-strasbg.fr/abs/2008PhRvD..77j3505W.
\newblock }{ (ADS)}.

\bibitem{BraxMartin99}
{Brax}, P.~H. \& {Martin}, J.
\newblock {Quintessence and supergravity}.
\newblock {\em Physics Letters B},
  \href{http://dx.doi.org/10.1016/S0370-2693(99)01209-5}{468:40--45}, November
  1999,
  arXiv:astro-ph/9905040\href{http://cdsads.u-strasbg.fr/abs/1999PhLB..468...40B.
\newblock }{ (ADS)}.

\bibitem{Linder06}
{Linder}, E.~V.
\newblock {Paths of quintessence}.
\newblock {\em \prd},
  \href{http://dx.doi.org/10.1103/PhysRevD.73.063010}{73(6):063010--+}, March
  2006,
  arXiv:astro-ph/0601052\href{http://cdsads.u-strasbg.fr/abs/2006PhRvD..73f3010L.
\newblock }{ (ADS)}.

\bibitem{Guzzo08}
{Guzzo}, L., {Pierleoni}, M., {Meneux}, B., {et~al.}
\newblock {A test of the nature of cosmic acceleration using galaxy redshift
  distortions}.
\newblock {\em \nat},
  \href{http://dx.doi.org/10.1038/nature06555}{451:541--544}, January 2008,
  0802.1944\href{http://cdsads.u-strasbg.fr/abs/2008Natur.451..541G.
\newblock }{ (ADS)}.

\bibitem{EuclidRB}
{Laureijs}, R., {Amiaux}, J., {Arduini}, S., {et~al.}
\newblock {Euclid Definition Study Report}.
\newblock {\em ArXiv e-prints}, October 2011,
  1110.3193\href{http://cdsads.u-strasbg.fr/abs/2011arXiv1110.3193L.
\newblock }{ (ADS)}.

\bibitem{Linder03}
{Linder}, E.~V. \& {Huterer}, D.
\newblock {Importance of supernovae at $z>1.5$ to probe dark energy}.
\newblock {\em \prd},
  \href{http://dx.doi.org/10.1103/PhysRevD.67.081303}{67(8):081303--+}, April
  2003,
  arXiv:astro-ph/0208138\href{http://cdsads.u-strasbg.fr/abs/2003PhRvD..67h1303L.
\newblock }{ (ADS)}.

\bibitem{Jaekel11}
{Jaekel}, M.-T. \& {Reynaud}, S.
\newblock {Tests of gravity at the solar system scale}.
\newblock {\em ArXiv e-prints}, July 2011,
  1107.4987\href{http://cdsads.u-strasbg.fr/abs/2011arXiv1107.4987J.
\newblock }{ (ADS)}.

\bibitem{Anderson02}
{Anderson}, J.~D., {Laing}, P.~A., {Lau}, E.~L., {et~al.}
\newblock {Study of the anomalous acceleration of Pioneer 10 and 11}.
\newblock {\em \prd},
  \href{http://dx.doi.org/10.1103/PhysRevD.65.082004}{65(8):082004--+}, April
  2002,
  arXiv:gr-qc/0104064\href{http://cdsads.u-strasbg.fr/abs/2002PhRvD..65h2004A.
\newblock }{ (ADS)}.

\bibitem{Levy09}
{Levy}, A., {Christophe}, B., {B{\'e}rio}, P., {et~al.}
\newblock {Pioneer 10 Doppler data analysis: Disentangling periodic and secular
  anomalies}.
\newblock {\em Advances in Space Research},
  \href{http://dx.doi.org/10.1016/j.asr.2009.01.003}{43:1538--1544}, May 2009,
  0809.2682\href{http://cdsads.u-strasbg.fr/abs/2009AdSpR..43.1538L.
\newblock }{ (ADS)}.

\bibitem{Francisco11}
{Francisco}, F., {Bertolami}, O., {Gil}, P.~J.~S., \& {P{\'a}ramos}, J.
\newblock {Modelling the reflective thermal contribution to the acceleration of
  the Pioneer spacecraft}.
\newblock {\em ArXiv e-prints}, March 2011,
  1103.5222\href{http://adsabs.harvard.edu/abs/2011arXiv1103.5222F.
\newblock }{ (ADS)}.

\bibitem{DGP00}
{Dvali}, G., {Gabadadze}, G., \& {Porrati}, M.
\newblock {4D gravity on a brane in 5D Minkowski space}.
\newblock {\em Physics Letters B},
  \href{http://dx.doi.org/10.1016/S0370-2693(00)00669-9}{485:208--214}, July
  2000,
  arXiv:hep-th/0005016\href{http://adsabs.harvard.edu/abs/2000PhLB..485..208D.
\newblock }{ (ADS)}.

\bibitem{Deffayet02}
{Deffayet}, C., {Landau}, S.~J., {Raux}, J., {Zaldarriaga}, M., \& {Astier}, P.
\newblock {Supernovae, CMB, and gravitational leakage into extra dimensions}.
\newblock {\em \prd}, 66(2):024019--+, July
  2002\href{http://adsabs.harvard.edu/cgi-bin/nph-bib_query?bibcode=2002PhRvD..66b4019D&db_key=AST.
\newblock }{ (ADS)}.

\bibitem{Fairbairn06}
{Fairbairn}, M. \& {Goobar}, A.
\newblock {Supernova limits on brane world cosmology}.
\newblock {\em Physics Letters B},
  \href{http://dx.doi.org/10.1016/j.physletb.2006.07.048}{642:432--435},
  November 2006,
  arXiv:astro-ph/0511029\href{http://cdsads.u-strasbg.fr/abs/2006PhLB..642..432F.
\newblock }{ (ADS)}.

\bibitem{Fang08}
{Fang}, W., {Wang}, S., {Hu}, W., {et~al.}
\newblock {Challenges to the DGP model from horizon-scale growth and geometry}.
\newblock {\em \prd},
  \href{http://dx.doi.org/10.1103/PhysRevD.78.103509}{78(10):103509--+},
  November 2008,
  0808.2208\href{http://adsabs.harvard.edu/abs/2008PhRvD..78j3509F.
\newblock }{ (ADS)}.

\bibitem{Amendola07a}
{Amendola}, L., {Polarski}, D., \& {Tsujikawa}, S.
\newblock {Are f(R) Dark Energy Models Cosmologically Viable?}
\newblock {\em Physical Review Letters},
  \href{http://dx.doi.org/10.1103/PhysRevLett.98.131302}{98(13):131302--+},
  March 2007,
  arXiv:astro-ph/0603703\href{http://cdsads.u-strasbg.fr/abs/2007PhRvL..98m1302A.
\newblock }{ (ADS)}.

\bibitem{Bean07}
{Bean}, R., {Bernat}, D., {Pogosian}, L., {Silvestri}, A., \& {Trodden}, M.
\newblock {Dynamics of linear perturbations in f(R) gravity}.
\newblock {\em \prd},
  \href{http://dx.doi.org/10.1103/PhysRevD.75.064020}{75(6):064020--+}, March
  2007,
  arXiv:astro-ph/0611321\href{http://cdsads.u-strasbg.fr/abs/2007PhRvD..75f4020B.
\newblock }{ (ADS)}.

\bibitem{Amendola07b}
{Amendola}, L., {Gannouji}, R., {Polarski}, D., \& {Tsujikawa}, S.
\newblock {Conditions for the cosmological viability of f(R) dark energy
  models}.
\newblock {\em \prd},
  \href{http://dx.doi.org/10.1103/PhysRevD.75.083504}{75(8):083504--+}, April
  2007,
  arXiv:gr-qc/0612180\href{http://cdsads.u-strasbg.fr/abs/2007PhRvD..75h3504A.
\newblock }{ (ADS)}.

\bibitem{Buchert00}
{Buchert}, T.
\newblock {On Average Properties of Inhomogeneous Fluids in General Relativity:
  Dust Cosmologies}.
\newblock {\em General Relativity and Gravitation}, 32:105--126, January 2000,
  arXiv:gr-qc/9906015\href{http://cdsads.u-strasbg.fr/abs/2000GReGr..32..105B.
\newblock }{ (ADS)}.

\bibitem{Marra08}
{Marra}, V., {Kolb}, E.~W., \& {Matarrese}, S.
\newblock {Light-cone averages in a Swiss-cheese universe}.
\newblock {\em \prd},
  \href{http://dx.doi.org/10.1103/PhysRevD.77.023003}{77(2):023003--+}, January
  2008, 0710.5505\href{http://cdsads.u-strasbg.fr/abs/2008PhRvD..77b3003M.
\newblock }{ (ADS)}.

\bibitem{Vanderveld08}
{Vanderveld}, R.~A., {Flanagan}, {\'E}.~{\'E}., \& {Wasserman}, I.
\newblock {Luminosity distance in ``Swiss cheese'' cosmology with randomized
  voids. I. Single void size}.
\newblock {\em \prd},
  \href{http://dx.doi.org/10.1103/PhysRevD.78.083511}{78(8):083511--+}, October
  2008, 0808.1080\href{http://cdsads.u-strasbg.fr/abs/2008PhRvD..78h3511V.
\newblock }{ (ADS)}.

\bibitem{Mukhanov97}
{Mukhanov}, V.~F., {Abramo}, L.~R.~W., \& {Brandenberger}, R.~H.
\newblock {Backreaction Problem for Cosmological Perturbations}.
\newblock {\em Physical Review Letters},
  \href{http://dx.doi.org/10.1103/PhysRevLett.78.1624}{78:1624--1627}, March
  1997,
  arXiv:gr-qc/9609026\href{http://cdsads.u-strasbg.fr/abs/1997PhRvL..78.1624M.
\newblock }{ (ADS)}.

\bibitem{Nambu02}
{Nambu}, Y.
\newblock {Back reaction and the effective Einstein equation for the universe
  with ideal fluid cosmological perturbations}.
\newblock {\em \prd},
  \href{http://dx.doi.org/10.1103/PhysRevD.65.104013}{65(10):104013--+}, May
  2002,
  arXiv:gr-qc/0203023\href{http://cdsads.u-strasbg.fr/abs/2002PhRvD..65j4013N.
\newblock }{ (ADS)}.

\bibitem{Ishibashi06}
{Ishibashi}, A. \& {Wald}, R.~M.
\newblock {Can the acceleration of our universe be explained by the effects of
  inhomogeneities?}
\newblock {\em Classical and Quantum Gravity},
  \href{http://dx.doi.org/10.1088/0264-9381/23/1/012}{23:235--250}, January
  2006,
  arXiv:gr-qc/0509108\href{http://cdsads.u-strasbg.fr/abs/2006CQGra..23..235I.
\newblock }{ (ADS)}.

\bibitem{Behrend08}
{Behrend}, J., {Brown}, I.~A., \& {Robbers}, G.
\newblock {Cosmological backreaction from perturbations}.
\newblock {\em \jcap},
  \href{http://dx.doi.org/10.1088/1475-7516/2008/01/013}{1:13--+}, January
  2008, 0710.4964\href{http://cdsads.u-strasbg.fr/abs/2008JCAP...01..013B.
\newblock }{ (ADS)}.

\bibitem{Buchert08}
{Buchert}, T.
\newblock {Dark Energy from structure: a status report}.
\newblock {\em General Relativity and Gravitation},
  \href{http://dx.doi.org/10.1007/s10714-007-0554-8}{40:467--527}, February
  2008, 0707.2153\href{http://cdsads.u-strasbg.fr/abs/2008GReGr..40..467B.
\newblock }{ (ADS)}.

\bibitem{Lena98}
{L{\'e}na}, P., {Lebrun}, F., \& {Mignard}, F.
\newblock {\em {Observational astrophysics.}}
\newblock Springer, Berlin (Germany), 1998.

\bibitem{Bessel90}
{Bessell}, M.~S.
\newblock {UBVRI passbands}.
\newblock {\em \pasp}, \href{http://dx.doi.org/10.1086/132749}{102:1181--1199},
  October 1990\href{http://cdsads.u-strasbg.fr/abs/1990PASP..102.1181B.
\newblock }{ (ADS)}.

\bibitem{Fukugita96}
{Fukugita}, M., {Ichikawa}, T., {Gunn}, J.~E., {et~al.}
\newblock {The Sloan Digital Sky Survey Photometric System}.
\newblock {\em AJ}, 111:1748--+, April
  1996\href{http://cdsads.u-strasbg.fr/cgi-bin/nph-bib_query?bibcode=1996AJ....111.1748F&amp;db_key=AST.
\newblock }{ (ADS)}.

\bibitem{Oke83}
{Oke}, J.~B. \& {Gunn}, J.~E.
\newblock {Secondary standard stars for absolute spectrophotometry}.
\newblock {\em \apj}, \href{http://dx.doi.org/10.1086/160817}{266:713--717},
  March
  1983\href{http://adsabs.harvard.edu/cgi-bin/nph-bib_query?bibcode=1983ApJ...266..713O&db_key=AST.
\newblock }{ (ADS)}.

\bibitem{Gunn-SDSS-telescope06}
{Gunn}, J.~E., {Siegmund}, W.~A., {Mannery}, E.~J., {et~al.}
\newblock {The 2.5 m Telescope of the Sloan Digital Sky Survey}.
\newblock {\em \aj}, \href{http://dx.doi.org/10.1086/500975}{131:2332--2359},
  April 2006,
  arXiv:astro-ph/0602326\href{http://cdsads.u-strasbg.fr/abs/2006AJ....131.2332G.
\newblock }{ (ADS)}.

\bibitem{Sauron00}
{de Zeeuw}, P.~T., {Allington-Smith}, J.~R., {Bacon}, R., {et~al.}
\newblock {The One Eye that Sees All: Integral Field Spectroscopy with SAURON
  on the WHT}.
\newblock {\em The Newsletter of the Isaac Newton Group of Telescopes},
  2:11--15, March 2000\href{http://cdsads.u-strasbg.fr/abs/2000INGN....2...11D.
\newblock }{ (ADS)}.

\bibitem{SnifsSPIE04}
{Lantz}, B., {Aldering}, G., {Antilogus}, P., {et~al.}
\newblock {SNIFS: a wideband integral field spectrograph with microlens
  arrays}.
\newblock In {L.~Mazuray, P.~J.~Rogers, \& R.~Wartmann}, editor, {\em Society
  of Photo-Optical Instrumentation Engineers (SPIE) Conference Series}, volume
  5249 of {\em Society of Photo-Optical Instrumentation Engineers (SPIE)
  Conference Series}, pages 146--155, February
  2004\href{http://adsabs.harvard.edu/abs/2004SPIE.5249..146L.
\newblock }{ (ADS)}.

\bibitem{Hodapp96}
{Hodapp}, K.-W., {Hora}, J.~L., {Hall}, D.~N.~B., {et~al.}
\newblock {The HAWAII Infrared Detector Arrays: testing and astronomical
  characterization of prototype and science-grade devices}.
\newblock {\em \na},
  \href{http://dx.doi.org/10.1016/S1384-1076(96)00013-9}{1:177--196}, October
  1996\href{http://cdsads.u-strasbg.fr/abs/1996NewA....1..177H.
\newblock }{ (ADS)}.

\bibitem{Leinert98}
{Leinert}, C., {Bowyer}, S., {Haikala}, L.~K., {et~al.}
\newblock {The 1997 reference of diffuse night sky brightness}.
\newblock {\em \aaps}, \href{http://dx.doi.org/10.1051/aas:1998105}{127:1--99},
  January 1998\href{http://cdsads.u-strasbg.fr/abs/1998A%26AS..127....1L.
\newblock }{ (ADS)}.

\bibitem{BaadeZwicky34a}
{Baade}, W. \& {Zwicky}, F.
\newblock {On Super-novae}.
\newblock {\em Proceedings of the National Academy of Science},
  \href{http://dx.doi.org/10.1073/pnas.20.5.254}{20:254--259}, May
  1934\href{http://adsabs.harvard.edu/abs/1934PNAS...20..254B.
\newblock }{ (ADS)}.

\bibitem{Minkowski41}
{Minkowski}, R.
\newblock {Spectra of Supernovae}.
\newblock {\em \pasp}, \href{http://dx.doi.org/10.1086/125315}{53:224--+},
  August 1941\href{http://cdsads.u-strasbg.fr/abs/1941PASP...53..224M.
\newblock }{ (ADS)}.

\bibitem{WheelerLevreault85}
{Wheeler}, J.~C. \& {Levreault}, R.
\newblock {The peculiar Type I supernova in NGC 991}.
\newblock {\em \apjl}, \href{http://dx.doi.org/10.1086/184500}{294:L17--L20},
  July 1985\href{http://adsabs.harvard.edu/abs/1985ApJ...294L..17W.
\newblock }{ (ADS)}.

\bibitem{DaSilva93}
{da Silva}, L.~A.~L.
\newblock {The classification of supernovae}.
\newblock {\em \apss},
  \href{http://dx.doi.org/10.1007/BF00626878}{202:215--236}, April
  1993\href{http://cdsads.u-strasbg.fr/abs/1993Ap%26SS.202..215D.
\newblock }{ (ADS)}.

\bibitem{Filippenko97}
{Filippenko}, A.~V.
\newblock {Optical Spectra of Supernovae}.
\newblock {\em \araa},
  \href{http://dx.doi.org/10.1146/annurev.astro.35.1.309}{35:309--355},
  1997\href{http://adsabs.harvard.edu/abs/1997ARA%26A..35..309F.
\newblock }{ (ADS)}.

\bibitem{Wheeler03}
{Wheeler}, J.~C.
\newblock {Resource Letter: OTS-1: Observations and theory of supernovae}.
\newblock {\em American Journal of Physics},
  \href{http://dx.doi.org/10.1119/1.1523076}{71:11--22}, January 2003,
  arXiv:astro-ph/0209514\href{http://cdsads.u-strasbg.fr/abs/2003AmJPh..71...11W.
\newblock }{ (ADS)}.

\bibitem{Woosley86}
{Woosley}, S.~E. \& {Weaver}, T.~A.
\newblock {The physics of supernova explosions}.
\newblock {\em \araa},
  \href{http://dx.doi.org/10.1146/annurev.aa.24.090186.001225}{24:205--253},
  1986\href{http://cdsads.u-strasbg.fr/abs/1986ARA%26A..24..205W.
\newblock }{ (ADS)}.

\bibitem{Smartt09}
{Smartt}, S.~J.
\newblock {Progenitors of Core-Collapse Supernovae}.
\newblock {\em \araa},
  \href{http://dx.doi.org/10.1146/annurev-astro-082708-101737}{47:63--106},
  September 2009,
  0908.0700\href{http://cdsads.u-strasbg.fr/abs/2009ARA%26A..47...63S.
\newblock }{ (ADS)}.

\bibitem{Woosley05}
{Woosley}, S. \& {Janka}, T.
\newblock {The physics of core-collapse supernovae}.
\newblock {\em Nature Physics},
  \href{http://dx.doi.org/10.1038/nphys172}{1:147--154}, December 2005,
  arXiv:astro-ph/0601261\href{http://adsabs.harvard.edu/abs/2005NatPh...1..147W.
\newblock }{ (ADS)}.

\bibitem{Richardson02}
{Richardson}, D., {Branch}, D., {Casebeer}, D., {et~al.}
\newblock {A Comparative Study of the Absolute Magnitude Distributions of
  Supernovae}.
\newblock {\em \aj}, \href{http://dx.doi.org/10.1086/338318}{123:745--752},
  February
  2002\href{http://adsabs.harvard.edu/cgi-bin/nph-bib_query?bibcode=2002AJ....123..745R&db_key=AST.
\newblock }{ (ADS)}.

\bibitem{Li10II}
{Li}, W., {Leaman}, J., {Chornock}, R., {et~al.}
\newblock {Nearby supernova rates from the Lick Observatory Supernova Search -
  II. The observed luminosity functions and fractions of supernovae in a
  complete sample}.
\newblock {\em \mnras},
  \href{http://dx.doi.org/10.1111/j.1365-2966.2011.18160.x}{412:1441--1472},
  April 2011,
  1006.4612\href{http://cdsads.u-strasbg.fr/abs/2011MNRAS.412.1441L.
\newblock }{ (ADS)}.

\bibitem{Minkowski64}
{Minkowski}, R.
\newblock {Supernovae and Supernova Remnants}.
\newblock {\em \araa},
  \href{http://dx.doi.org/10.1146/annurev.aa.02.090164.001335}{2:247--+},
  1964\href{http://cdsads.u-strasbg.fr/abs/1964ARA%26A...2..247M.
\newblock }{ (ADS)}.

\bibitem{ConleyRiseTime06}
{Conley}, A., {Howell}, D.~A., {Howes}, A., {et~al.}
\newblock {The Rise Time of Type Ia Supernovae from the Supernova Legacy
  Survey}.
\newblock {\em \aj}, \href{http://dx.doi.org/10.1086/507788}{132:1707--1713},
  October 2006,
  arXiv:astro-ph/0607363\href{http://cdsads.u-strasbg.fr/abs/2006AJ....132.1707C.
\newblock }{ (ADS)}.

\bibitem{Colgate69}
{Colgate}, S.~A. \& {McKee}, C.
\newblock {Early Supernova Luminosity}.
\newblock {\em \apj}, \href{http://dx.doi.org/10.1086/150102}{157:623}, August
  1969\href{http://cdsads.u-strasbg.fr/abs/1969ApJ...157..623C.
\newblock }{ (ADS)}.

\bibitem{Colgate80}
{Colgate}, S.~A., {Petschek}, A.~G., \& {Kriese}, J.~T.
\newblock {The luminosity of type I supernovae}.
\newblock {\em \apjl}, \href{http://dx.doi.org/10.1086/183239}{237:L81--L85},
  May 1980\href{http://cdsads.u-strasbg.fr/abs/1980ApJ...237L..81C.
\newblock }{ (ADS)}.

\bibitem{RuizLapuente98}
{Ruiz-Lapuente}, P. \& {Spruit}, H.~C.
\newblock {Bolometric Light Curves of Supernovae and Postexplosion Magnetic
  Fields}.
\newblock {\em \apj}, \href{http://dx.doi.org/10.1086/305697}{500:360--+}, June
  1998,
  arXiv:astro-ph/9711248\href{http://cdsads.u-strasbg.fr/abs/1998ApJ...500..360R.
\newblock }{ (ADS)}.

\bibitem{Stritzinger06}
{Stritzinger}, M., {Leibundgut}, B., {Walch}, S., \& {Contardo}, G.
\newblock {Constraints on the progenitor systems of type Ia supernovae}.
\newblock {\em \aap},
  \href{http://dx.doi.org/10.1051/0004-6361:20053652}{450:241--251}, April
  2006,
  arXiv:astro-ph/0506415\href{http://cdsads.u-strasbg.fr/abs/2006A%26A...450..241S.
\newblock }{ (ADS)}.

\bibitem{BranchRev92}
{Branch}, D. \& {Tammann}, G.~A.
\newblock {Type IA supernovae as standard candles}.
\newblock {\em \araa},
  \href{http://dx.doi.org/10.1146/annurev.aa.30.090192.002043}{30:359--389},
  1992\href{http://cdsads.u-strasbg.fr/abs/1992ARA%26A..30..359B.
\newblock }{ (ADS)}.

\bibitem{Riess96a}
{Riess}, A.~G., {Press}, W.~H., \& {Kirshner}, R.~P.
\newblock {A Precise Distance Indicator: Type IA Supernova Multicolor
  Light-Curve Shapes}.
\newblock {\em \apj}, 473:88--+, December
  1996\href{http://adsabs.harvard.edu/cgi-bin/nph-bib_query?bibcode=1996ApJ...473...88R&db_key=AST.
\newblock }{ (ADS)}.

\bibitem{Branch88}
{Branch}, D., {Drucker}, W., \& {Jeffery}, D.~J.
\newblock {Differences among expansion velocities of Type IA supernovae}.
\newblock {\em \apjl}, \href{http://dx.doi.org/10.1086/185217}{330:L117+}, July
  1988\href{http://cdsads.u-strasbg.fr/abs/1988ApJ...330L.117B.
\newblock }{ (ADS)}.

\bibitem{Benetti05}
{Benetti}, S., {Cappellaro}, E., {Mazzali}, P.~A., {et~al.}
\newblock {The Diversity of Type Ia Supernovae: Evidence for Systematics?}
\newblock {\em \apj}, \href{http://dx.doi.org/10.1086/428608}{623:1011--1016},
  April 2005,
  arXiv:astro-ph/0411059\href{http://cdsads.u-strasbg.fr/abs/2005ApJ...623.1011B.
\newblock }{ (ADS)}.

\bibitem{Contreras10}
{Contreras}, C., {Hamuy}, M., {Phillips}, M.~M., {et~al.}
\newblock {The Carnegie Supernova Project: First Photometry Data Release of
  Low-Redshift Type Ia Supernovae}.
\newblock {\em \aj},
  \href{http://dx.doi.org/10.1088/0004-6256/139/2/519}{139:519--539}, February
  2010, 0910.3330\href{http://adsabs.harvard.edu/abs/2010AJ....139..519C.
\newblock }{ (ADS)}.

\bibitem{Stanishev-SN2003du-07}
{Stanishev}, V., {Goobar}, A., {Benetti}, S., {et~al.}
\newblock {SN 2003du: 480 days in the life of a normal type Ia supernova}.
\newblock {\em \aap},
  \href{http://dx.doi.org/10.1051/0004-6361:20066020}{469:645--661}, July 2007,
  0704.1244\href{http://cdsads.u-strasbg.fr/abs/2007A%26A...469..645S.
\newblock }{ (ADS)}.

\bibitem{Pskovskii77}
{Pskovskii}, I.~P.
\newblock {Light curves, color curves, and expansion velocity of type I
  supernovae as functions of the rate of brightness decline}.
\newblock {\em \sovast}, 21:675--682, December
  1977\href{http://cdsads.u-strasbg.fr/abs/1977SvA....21..675P.
\newblock }{ (ADS)}.

\bibitem{Pskovskii84}
{Pskovskii}, Y.~P.
\newblock {Photometric classification and basic parameters of type I
  supernovae}.
\newblock {\em \sovast}, 28:658--+, December
  1984\href{http://cdsads.u-strasbg.fr/abs/1984SvA....28..658P.
\newblock }{ (ADS)}.

\bibitem{Phillips93}
{Phillips}, M.~M.
\newblock {The absolute magnitudes of Type Ia supernovae}.
\newblock {\em \apjl}, \href{http://dx.doi.org/10.1086/186970}{413:L105--L108},
  August 1993\href{http://cdsads.u-strasbg.fr/abs/1993ApJ...413L.105P.
\newblock }{ (ADS)}.

\bibitem{Perlmutter97}
{Perlmutter}, S., {Gabi}, S., {Goldhaber}, G., {et~al.}
\newblock {Measurements of the Cosmological Parameters Omega and Lambda from
  the First Seven Supernovae at $z > 0.35$}.
\newblock {\em \apj}, 483:565--+, July
  1997\href{http://cdsads.u-strasbg.fr/cgi-bin/nph-bib_query?bibcode=1997ApJ...483..565P&db_key=AST.
\newblock }{ (ADS)}.

\bibitem{Goldhaber01}
{Goldhaber}, G., {Groom}, D.~E., {Kim}, A., {et~al.}
\newblock {Timescale Stretch Parameterization of Type Ia Supernova B-Band Light
  Curves}.
\newblock {\em \apj}, \href{http://dx.doi.org/10.1086/322460}{558:359--368},
  September 2001,
  arXiv:astro-ph/0104382\href{http://cdsads.u-strasbg.fr/abs/2001ApJ...558..359G.
\newblock }{ (ADS)}.

\bibitem{Hayden10}
{Hayden}, B.~T., {Garnavich}, P.~M., {Kessler}, R., {et~al.}
\newblock {The Rise and Fall of Type Ia Supernova Light Curves in the SDSS-II
  Supernova Survey}.
\newblock {\em \apj},
  \href{http://dx.doi.org/10.1088/0004-637X/712/1/350}{712:350--366}, March
  2010, 1001.3428\href{http://cdsads.u-strasbg.fr/abs/2010ApJ...712..350H.
\newblock }{ (ADS)}.

\bibitem{Guy10}
{Guy}, J., {Sullivan}, M., {Conley}, A., {et~al.}
\newblock {The Supernova Legacy Survey 3-year sample: Type Ia supernovae
  photometric distances and cosmological constraints}.
\newblock {\em \aap},
  \href{http://dx.doi.org/10.1051/0004-6361/201014468}{523:A7+}, November 2010,
  1010.4743\href{http://cdsads.u-strasbg.fr/abs/2010A%26A...523A...7G.
\newblock }{ (ADS)}.

\bibitem{GonzalesGaitan11}
{Gonz{\'a}lez-Gait{\'a}n}, S., {Perrett}, K., {Sullivan}, M., {et~al.}
\newblock {Subluminous Type Ia Supernovae at High Redshift from the Supernova
  Legacy Survey}.
\newblock {\em \apj},
  \href{http://dx.doi.org/10.1088/0004-637X/727/2/107}{727:107--+}, February
  2011, 1011.4531\href{http://cdsads.u-strasbg.fr/abs/2011ApJ...727..107G.
\newblock }{ (ADS)}.

\bibitem{Guy05}
{Guy}, J., {Astier}, P., {Nobili}, S., {Regnault}, N., \& {Pain}, R.
\newblock {SALT: a spectral adaptive light curve template for type Ia
  supernovae}.
\newblock {\em \aap},
  \href{http://dx.doi.org/10.1051/0004-6361:20053025}{443:781--791}, December
  2005,
  arXiv:astro-ph/0506583\href{http://cdsads.u-strasbg.fr/abs/2005A%26A...443..781G.
\newblock }{ (ADS)}.

\bibitem{ConleySifto08}
{Conley}, A., {Sullivan}, M., {Hsiao}, E.~Y., {et~al.}
\newblock {SiFTO: An Empirical Method for Fitting SN Ia Light Curves}.
\newblock {\em \apj}, \href{http://dx.doi.org/10.1086/588518}{681:482--498},
  July 2008, 0803.3441\href{http://cdsads.u-strasbg.fr/abs/2008ApJ...681..482C.
\newblock }{ (ADS)}.

\bibitem{Phillips99}
{Phillips}, M.~M., {Lira}, P., {Suntzeff}, N.~B., {et~al.}
\newblock {The Reddening-Free Decline Rate Versus Luminosity Relationship for
  Type IA Supernovae}.
\newblock {\em \aj}, \href{http://dx.doi.org/10.1086/301032}{118:1766--1776},
  October 1999,
  arXiv:astro-ph/9907052\href{http://cdsads.u-strasbg.fr/abs/1999AJ....118.1766P.
\newblock }{ (ADS)}.

\bibitem{Riess96b}
{Riess}, A.~G., {Press}, W.~H., \& {Kirshner}, R.~P.
\newblock {Is the Dust Obscuring Supernovae in Distant Galaxies the Same as
  Dust in the Milky Way?}
\newblock {\em \apj}, 473:588--+, December
  1996\href{http://adsabs.harvard.edu/cgi-bin/nph-bib_query?bibcode=1996ApJ...473..588R&db_key=AST.
\newblock }{ (ADS)}.

\bibitem{Nobili08}
{Nobili}, S. \& {Goobar}, A.
\newblock {The colour-lightcurve shape relation of type Ia supernovae and the
  reddening law}.
\newblock {\em \aap},
  \href{http://dx.doi.org/10.1051/0004-6361:20079292}{487:19--31}, August 2008,
  0712.1155\href{http://cdsads.u-strasbg.fr/abs/2008A%26A...487...19N.
\newblock }{ (ADS)}.

\bibitem{Blondin09}
{Blondin}, S., {Prieto}, J.~L., {Patat}, F., {et~al.}
\newblock {A Second Case of Variable Na I D Lines in a Highly Reddened Type Ia
  Supernova}.
\newblock {\em \apj},
  \href{http://dx.doi.org/10.1088/0004-637X/693/1/207}{693:207--215}, March
  2009, 0811.0002\href{http://cdsads.u-strasbg.fr/abs/2009ApJ...693..207B.
\newblock }{ (ADS)}.

\bibitem{Guy07}
{Guy}, J., {Astier}, P., {Baumont}, S., {et~al.}
\newblock {SALT2: using distant supernovae to improve the use of type Ia
  supernovae as distance indicators}.
\newblock {\em \aap},
  \href{http://dx.doi.org/10.1051/0004-6361:20066930}{466:11--21}, April 2007,
  arXiv:astro-ph/0701828\href{http://cdsads.u-strasbg.fr/abs/2007A%26A...466...11G.
\newblock }{ (ADS)}.

\bibitem{Ellis08}
{Ellis}, R.~S., {Sullivan}, M., {Nugent}, P.~E., {et~al.}
\newblock {Verifying the Cosmological Utility of Type Ia Supernovae:
  Implications of a Dispersion in the Ultraviolet Spectra}.
\newblock {\em \apj}, \href{http://dx.doi.org/10.1086/524981}{674:51--69},
  February 2008,
  0710.3896\href{http://adsabs.harvard.edu/abs/2008ApJ...674...51E.
\newblock }{ (ADS)}.

\bibitem{Balland09}
{Balland}, C., {Baumont}, S., {Basa}, S., {et~al.}
\newblock {The ESO/VLT 3rd year Type Ia supernova data set from the supernova
  legacy survey}.
\newblock {\em \aap},
  \href{http://dx.doi.org/10.1051/0004-6361/200912246}{507:85--103}, November
  2009, 0909.3316\href{http://cdsads.u-strasbg.fr/abs/2009A%26A...507...85B.
\newblock }{ (ADS)}.

\bibitem{Milne10}
{Milne}, P.~A., {Brown}, P.~J., {Roming}, P.~W.~A., {et~al.}
\newblock {Near-ultraviolet Properties of a Large Sample of Type Ia Supernovae
  as Observed with the Swift UVOT}.
\newblock {\em \apj},
  \href{http://dx.doi.org/10.1088/0004-637X/721/2/1627}{721:1627--1655},
  October 2010,
  1007.5279\href{http://cdsads.u-strasbg.fr/abs/2010ApJ...721.1627M.
\newblock }{ (ADS)}.

\bibitem{Foley11}
{Foley}, R.~J. \& {Kasen}, D.
\newblock {Measuring Ejecta Velocity Improves Type Ia Supernova Distances}.
\newblock {\em \apj},
  \href{http://dx.doi.org/10.1088/0004-637X/729/1/55}{729:55}, March 2011,
  1011.4517\href{http://cdsads.u-strasbg.fr/abs/2011ApJ...729...55F.
\newblock }{ (ADS)}.

\bibitem{Tripp98}
{Tripp}, R.
\newblock {A two-parameter luminosity correction for Type IA supernovae}.
\newblock {\em \aap}, 331:815--820, March
  1998\href{http://adsabs.harvard.edu/cgi-bin/nph-bib_query?bibcode=1998A%26A...331..815T&db_key=AST.
\newblock }{ (ADS)}.

\bibitem{Chotard11}
{Chotard}, N., {Gangler}, E., {Aldering}, G., {et~al.}
\newblock {The reddening law of type Ia supernovae: separating intrinsic
  variability from dust using equivalent widths}.
\newblock {\em \aap},
  \href{http://dx.doi.org/10.1051/0004-6361/201116723}{529:L4}, May 2011,
  1103.5300\href{http://cdsads.u-strasbg.fr/abs/2011A%26A...529L...4C.
\newblock }{ (ADS)}.

\bibitem{Hamuy96a}
{Hamuy}, M., {Phillips}, M.~M., {Suntzeff}, N.~B., {et~al.}
\newblock {The Absolute Luminosities of the Calan/Tololo Type IA Supernovae}.
\newblock {\em \aj}, \href{http://dx.doi.org/10.1086/118190}{112:2391},
  December 1996,
  arXiv:astro-ph/9609059\href{http://cdsads.u-strasbg.fr/abs/1996AJ....112.2391H.
\newblock }{ (ADS)}.

\bibitem{Krisciunas04}
{Krisciunas}, K., {Phillips}, M.~M., \& {Suntzeff}, N.~B.
\newblock {Hubble Diagrams of Type Ia Supernovae in the Near-Infrared}.
\newblock {\em \apjl}, \href{http://dx.doi.org/10.1086/382731}{602:L81--L84},
  February 2004,
  arXiv:astro-ph/0312626\href{http://cdsads.u-strasbg.fr/abs/2004ApJ...602L..81K.
\newblock }{ (ADS)}.

\bibitem{WoodVasey08}
{Wood-Vasey}, W.~M., {Friedman}, A.~S., {Bloom}, J.~S., {et~al.}
\newblock {Type Ia Supernovae Are Good Standard Candles in the Near Infrared:
  Evidence from PAIRITEL}.
\newblock {\em \apj}, \href{http://dx.doi.org/10.1086/592374}{689:377--390},
  December 2008,
  0711.2068\href{http://cdsads.u-strasbg.fr/abs/2008ApJ...689..377W.
\newblock }{ (ADS)}.

\bibitem{Kattner12}
{Kattner}, S., {Leonard}, D.~C., {Burns}, C.~R., {et~al.}
\newblock {The Standardizability of Type Ia Supernovae in the Near-Infrared:
  Evidence for a Peak-Luminosity Versus Decline-Rate Relation in the
  Near-Infrared}.
\newblock {\em \pasp}, \href{http://dx.doi.org/10.1086/664734}{124:114--127},
  February 2012,
  1201.2913\href{http://cdsads.u-strasbg.fr/abs/2012PASP..124..114K.
\newblock }{ (ADS)}.

\bibitem{Kawabata10}
{Kawabata}, K.~S., {Maeda}, K., {Nomoto}, K., {et~al.}
\newblock {A massive star origin for an unusual helium-rich supernova in an
  elliptical galaxy}.
\newblock {\em \nat},
  \href{http://dx.doi.org/10.1038/nature09055}{465:326--328}, May 2010,
  0906.2811\href{http://cdsads.u-strasbg.fr/abs/2010Natur.465..326K.
\newblock }{ (ADS)}.

\bibitem{Mannucci05}
{Mannucci}, F., {Della Valle}, M., {Panagia}, N., {et~al.}
\newblock {The supernova rate per unit mass}.
\newblock {\em \aap},
  \href{http://dx.doi.org/10.1051/0004-6361:20041411}{433:807--814}, April
  2005,
  arXiv:astro-ph/0411450\href{http://cdsads.u-strasbg.fr/abs/2005A%26A...433..807M.
\newblock }{ (ADS)}.

\bibitem{Sullivan06}
{Sullivan}, M., {Le Borgne}, D., {Pritchet}, C.~J., {et~al.}
\newblock {Rates and Properties of Type Ia Supernovae as a Function of Mass and
  Star Formation in Their Host Galaxies}.
\newblock {\em \apj}, \href{http://dx.doi.org/10.1086/506137}{648:868--883},
  September 2006,
  arXiv:astro-ph/0605455\href{http://cdsads.u-strasbg.fr/abs/2006ApJ...648..868S.
\newblock }{ (ADS)}.

\bibitem{Lampeitl10}
{Lampeitl}, H., {Smith}, M., {Nichol}, R.~C., {et~al.}
\newblock {The Effect of Host Galaxies on Type Ia Supernovae in the SDSS-II
  Supernova Survey}.
\newblock {\em \apj},
  \href{http://dx.doi.org/10.1088/0004-637X/722/1/566}{722:566--576}, October
  2010, 1005.4687\href{http://cdsads.u-strasbg.fr/abs/2010ApJ...722..566L.
\newblock }{ (ADS)}.

\bibitem{Brandt10}
{Brandt}, T.~D., {Tojeiro}, R., {Aubourg}, {\'E}., {et~al.}
\newblock {The Ages of Type Ia Supernova Progenitors}.
\newblock {\em \aj},
  \href{http://dx.doi.org/10.1088/0004-6256/140/3/804}{140:804--816}, September
  2010, 1002.0848\href{http://cdsads.u-strasbg.fr/abs/2010AJ....140..804B.
\newblock }{ (ADS)}.

\bibitem{Pskovskii67}
{Pskovskii}, Y.~P.
\newblock {The Photometric Properties of Supernovae.}
\newblock {\em \sovast}, 11:63--+, August
  1967\href{http://cdsads.u-strasbg.fr/abs/1967SvA....11...63P.
\newblock }{ (ADS)}.

\bibitem{Howell07}
{Howell}, D.~A., {Sullivan}, M., {Conley}, A., \& {Carlberg}, R.
\newblock {Predicted and Observed Evolution in the Mean Properties of Type Ia
  Supernovae with Redshift}.
\newblock {\em \apjl}, \href{http://dx.doi.org/10.1086/522030}{667:L37--L40},
  September 2007,
  arXiv:astro-ph/0701912\href{http://cdsads.u-strasbg.fr/abs/2007ApJ...667L..37H.
\newblock }{ (ADS)}.

\bibitem{Sullivan10}
{Sullivan}, M., {Conley}, A., {Howell}, D.~A., {et~al.}
\newblock {The dependence of Type Ia Supernovae luminosities on their host
  galaxies}.
\newblock {\em \mnras},
  \href{http://dx.doi.org/10.1111/j.1365-2966.2010.16731.x}{406:782--802},
  August 2010,
  1003.5119\href{http://cdsads.u-strasbg.fr/abs/2010MNRAS.406..782S.
\newblock }{ (ADS)}.

\bibitem{Tran01}
{Tran}, H.~D., {Tsvetanov}, Z., {Ford}, H.~C., {et~al.}
\newblock {Dusty Nuclear Disks and Filaments in Early-Type Galaxies}.
\newblock {\em \aj}, \href{http://dx.doi.org/10.1086/321072}{121:2928--2942},
  June 2001,
  arXiv:astro-ph/0102292\href{http://cdsads.u-strasbg.fr/abs/2001AJ....121.2928T.
\newblock }{ (ADS)}.

\bibitem{Temi05}
{Temi}, P., {Brighenti}, F., \& {Mathews}, W.~G.
\newblock {Mid-Infrared Emission from Elliptical Galaxies: Sensitivity to
  Stellar Age}.
\newblock {\em \apjl}, \href{http://dx.doi.org/10.1086/499239}{635:L25--L28},
  December 2005,
  arXiv:astro-ph/0511328\href{http://cdsads.u-strasbg.fr/abs/2005ApJ...635L..25T.
\newblock }{ (ADS)}.

\bibitem{Smith12}
{Smith}, M., {Nichol}, R.~C., {Dilday}, B., {et~al.}
\newblock {The SDSS-II Supernova Survey: Parameterizing the Type Ia Supernova
  Rate as a Function of Host Galaxy Properties}.
\newblock {\em \apj},
  \href{http://dx.doi.org/10.1088/0004-637X/755/1/61}{755:61}, August 2012,
  1108.4923\href{http://cdsads.u-strasbg.fr/abs/2012ApJ...755...61S.
\newblock }{ (ADS)}.

\bibitem{Hicken09}
{Hicken}, M., {Wood-Vasey}, W.~M., {Blondin}, S., {et~al.}
\newblock {Improved Dark Energy Constraints from \~{}100 New CfA Supernova Type
  Ia Light Curves}.
\newblock {\em \apj},
  \href{http://dx.doi.org/10.1088/0004-637X/700/2/1097}{700:1097--1140}, August
  2009, 0901.4804\href{http://cdsads.u-strasbg.fr/abs/2009ApJ...700.1097H.
\newblock }{ (ADS)}.

\bibitem{Chandrasekhar31}
{Chandrasekhar}, S.
\newblock {The Maximum Mass of Ideal White Dwarfs}.
\newblock {\em \apj}, \href{http://dx.doi.org/10.1086/143324}{74:81--+}, July
  1931\href{http://adsabs.harvard.edu/abs/1931ApJ....74...81C.
\newblock }{ (ADS)}.

\bibitem{Hoyle60}
{Hoyle}, F. \& {Fowler}, W.~A.
\newblock {Nucleosynthesis in Supernovae.}
\newblock {\em \apj}, \href{http://dx.doi.org/10.1086/146963}{132:565--+},
  November 1960\href{http://cdsads.u-strasbg.fr/abs/1960ApJ...132..565H.
\newblock }{ (ADS)}.

\bibitem{Livio00}
{Livio}, M.
\newblock {The Progenitors of Type Ia Supernovae}.
\newblock In {J.~C.~Niemeyer \& J.~W.~Truran}, editor, {\em Type Ia Supernovae,
  Theory and Cosmology (Cambridge Univ. Press)}, pages 33--+. {Cambridge Univ.
  Press}, 2000,
  arXiv:astro-ph/9903264\href{http://cdsads.u-strasbg.fr/abs/2000tias.conf...33L.
\newblock }{ (ADS)}.

\bibitem{Kuchner94}
{Kuchner}, M.~J., {Kirshner}, R.~P., {Pinto}, P.~A., \& {Leibundgut}, B.
\newblock {Evidence for Ni-56 yields Co-56 yields Fe-56 decay in type IA
  supernovae}.
\newblock {\em \apjl}, \href{http://dx.doi.org/10.1086/187347}{426:L89+}, May
  1994\href{http://cdsads.u-strasbg.fr/abs/1994ApJ...426L..89K.
\newblock }{ (ADS)}.

\bibitem{Nomoto84III}
{Nomoto}, K., {Thielemann}, F.-K., \& {Yokoi}, K.
\newblock {Accreting white dwarf models of Type I supernovae. III - Carbon
  deflagration supernovae}.
\newblock {\em \apj}, \href{http://dx.doi.org/10.1086/162639}{286:644--658},
  November 1984\href{http://cdsads.u-strasbg.fr/abs/1984ApJ...286..644N.
\newblock }{ (ADS)}.

\bibitem{Holberg08}
{Holberg}, J.~B., {Sion}, E.~M., {Oswalt}, T., {et~al.}
\newblock {A New Look at the Local White Dwarf Population}.
\newblock {\em \aj},
  \href{http://dx.doi.org/10.1088/0004-6256/135/4/1225}{135:1225--1238}, April
  2008\href{http://cdsads.u-strasbg.fr/abs/2008AJ....135.1225H.
\newblock }{ (ADS)}.

\bibitem{Kilic11}
{Kilic}, M., {Brown}, W.~R., {Allende Prieto}, C., {et~al.}
\newblock {The ELM Survey. II. Twelve Binary White Dwarf Merger Systems}.
\newblock {\em \apj},
  \href{http://dx.doi.org/10.1088/0004-637X/727/1/3}{727:3--+}, January 2011,
  1011.4073\href{http://adsabs.harvard.edu/abs/2011ApJ...727....3K.
\newblock }{ (ADS)}.

\bibitem{Whelan73}
{Whelan}, J. \& {Iben}, Jr., I.
\newblock {Binaries and Supernovae of Type I}.
\newblock {\em \apj}, \href{http://dx.doi.org/10.1086/152565}{186:1007--1014},
  December 1973\href{http://cdsads.u-strasbg.fr/abs/1973ApJ...186.1007W.
\newblock }{ (ADS)}.

\bibitem{Gilfanov10}
{Gilfanov}, M. \& {Bogd{\'a}n}, {\'A}.
\newblock {An upper limit on the contribution of accreting white dwarfs to the
  typeIa supernova rate}.
\newblock {\em \nat},
  \href{http://dx.doi.org/10.1038/nature08685}{463:924--925}, February 2010,
  1002.3359\href{http://cdsads.u-strasbg.fr/abs/2010Natur.463..924G.
\newblock }{ (ADS)}.

\bibitem{Hachisu10}
{Hachisu}, I., {Kato}, M., \& {Nomoto}, K.
\newblock {Supersoft X-ray Phase of Single Degenerate Type Ia Supernova
  Progenitors in Early-type Galaxies}.
\newblock {\em \apjl},
  \href{http://dx.doi.org/10.1088/2041-8205/724/2/L212}{724:L212--L216},
  December 2010,
  1010.5860\href{http://cdsads.u-strasbg.fr/abs/2010ApJ...724L.212H.
\newblock }{ (ADS)}.

\bibitem{Cassisi98}
{Cassisi}, S., {Iben}, Jr., I., \& {Tornambe}, A.
\newblock {Hydrogen-accreting Carbon-Oxygen White Dwarfs}.
\newblock {\em \apj}, \href{http://dx.doi.org/10.1086/305381}{496:376--+},
  March 1998\href{http://cdsads.u-strasbg.fr/abs/1998ApJ...496..376C.
\newblock }{ (ADS)}.

\bibitem{Kato99}
{Kato}, M. \& {Hachisu}, I.
\newblock {A New Estimation of Mass Accumulation Efficiency in Helium Shell
  Flashes toward Type IA Supernova Explosions}.
\newblock {\em \apjl}, \href{http://dx.doi.org/10.1086/311893}{513:L41--L44},
  March 1999,
  arXiv:astro-ph/9901080\href{http://cdsads.u-strasbg.fr/abs/1999ApJ...513L..41K.
\newblock }{ (ADS)}.

\bibitem{Maoz08}
{Maoz}, D. \& {Mannucci}, F.
\newblock {A search for the progenitors of two TypeIa Supernovae in NGC 1316}.
\newblock {\em \mnras},
  \href{http://dx.doi.org/10.1111/j.1365-2966.2008.13403.x}{388:421--428}, July
  2008, 0801.2898\href{http://cdsads.u-strasbg.fr/abs/2008MNRAS.388..421M.
\newblock }{ (ADS)}.

\bibitem{Li11}
{Li}, W., {Bloom}, J.~S., {Podsiadlowski}, P., {et~al.}
\newblock {Exclusion of a luminous red giant as a companion star to the
  progenitor of supernova SN 2011fe}.
\newblock {\em \nat},
  \href{http://dx.doi.org/10.1038/nature10646}{480:348--350}, December 2011,
  1109.1593\href{http://adsabs.harvard.edu/abs/2011Natur.480..348L.
\newblock }{ (ADS)}.

\bibitem{Schaefer12}
{Schaefer}, B.~E. \& {Pagnotta}, A.
\newblock {An absence of ex-companion stars in the type Ia supernova remnant
  SNR 0509-67.5}.
\newblock {\em \nat},
  \href{http://dx.doi.org/10.1038/nature10692}{481:164--166}, January
  2012\href{http://cdsads.u-strasbg.fr/abs/2012Natur.481..164S.
\newblock }{ (ADS)}.

\bibitem{Brown11}
{Brown}, P.~J., {Dawson}, K.~S., {de Pasquale}, M., {et~al.}
\newblock {A Swift Look at SN 2011fe: The Earliest Ultraviolet Observations of
  a Type Ia Supernova}.
\newblock {\em ArXiv e-prints}, October 2011,
  1110.2538\href{http://cdsads.u-strasbg.fr/abs/2011arXiv1110.2538B.
\newblock }{ (ADS)}.

\bibitem{Brown12}
{Brown}, P.~J., {Dawson}, K.~S., {Harris}, D.~W., {et~al.}
\newblock {Constraints on Type Ia Supernova Progenitor Companions from Early
  Ultraviolet Observations with Swift}.
\newblock {\em \apj},
  \href{http://dx.doi.org/10.1088/0004-637X/749/1/18}{749:18}, April 2012,
  1203.5315\href{http://cdsads.u-strasbg.fr/abs/2012ApJ...749...18B.
\newblock }{ (ADS)}.

\bibitem{Iben84}
{Iben}, Jr., I. \& {Tutukov}, A.~V.
\newblock {Supernovae of type I as end products of the evolution of binaries
  with components of moderate initial mass (M not greater than about 9 solar
  masses)}.
\newblock {\em \apjs}, \href{http://dx.doi.org/10.1086/190932}{54:335--372},
  February 1984\href{http://cdsads.u-strasbg.fr/abs/1984ApJS...54..335I.
\newblock }{ (ADS)}.

\bibitem{Webbink84}
{Webbink}, R.~F.
\newblock {Double white dwarfs as progenitors of R Coronae Borealis stars and
  Type I supernovae}.
\newblock {\em \apj}, \href{http://dx.doi.org/10.1086/161701}{277:355--360},
  February 1984\href{http://cdsads.u-strasbg.fr/abs/1984ApJ...277..355W.
\newblock }{ (ADS)}.

\bibitem{Howell06}
{Howell}, D.~A., {Sullivan}, M., {Nugent}, P.~E., {et~al.}
\newblock {The type Ia supernova SNLS-03D3bb from a super-Chandrasekhar-mass
  white dwarf star}.
\newblock {\em \nat},
  \href{http://dx.doi.org/10.1038/nature05103}{443:308--311}, September 2006,
  arXiv:astro-ph/0609616\href{http://cdsads.u-strasbg.fr/abs/2006Natur.443..308H.
\newblock }{ (ADS)}.

\bibitem{Hachisu12}
{Hachisu}, I., {Kato}, M., {Saio}, H., \& {Nomoto}, K.
\newblock {A Single Degenerate Progenitor Model for Type Ia Supernovae Highly
  Exceeding the Chandrasekhar Mass Limit}.
\newblock {\em \apj},
  \href{http://dx.doi.org/10.1088/0004-637X/744/1/69}{744:69}, January 2012,
  1106.3510\href{http://adsabs.harvard.edu/abs/2012ApJ...744...69H.
\newblock }{ (ADS)}.

\bibitem{Saio85}
{Saio}, H. \& {Nomoto}, K.
\newblock {Evolution of a merging pair of C + O white dwarfs to form a single
  neutron star}.
\newblock {\em \aap}, 150:L21--L23, September
  1985\href{http://cdsads.u-strasbg.fr/abs/1985A%26A...150L..21S.
\newblock }{ (ADS)}.

\bibitem{Saio98}
{Saio}, H. \& {Nomoto}, K.
\newblock {Inward Propagation of Nuclear-burning Shells in Merging C-O and He
  White Dwarfs}.
\newblock {\em \apj}, \href{http://dx.doi.org/10.1086/305696}{500:388--+}, June
  1998,
  arXiv:astro-ph/9801084\href{http://cdsads.u-strasbg.fr/abs/1998ApJ...500..388S.
\newblock }{ (ADS)}.

\bibitem{Pakmor10}
{Pakmor}, R., {Kromer}, M., {R{\"o}pke}, F.~K., {et~al.}
\newblock {Sub-luminous type Ia supernovae from the mergers of equal-mass white
  dwarfs with mass $\sim 0.9 M_\odot$}.
\newblock {\em \nat}, \href{http://dx.doi.org/10.1038/nature08642}{463:61--64},
  January 2010,
  0911.0926\href{http://cdsads.u-strasbg.fr/abs/2010Natur.463...61P.
\newblock }{ (ADS)}.

\bibitem{Arnett79}
{Arnett}, W.~D.
\newblock {On the theory of Type I supernovae}.
\newblock {\em \apjl}, \href{http://dx.doi.org/10.1086/182957}{230:L37--L40},
  May 1979\href{http://cdsads.u-strasbg.fr/abs/1979ApJ...230L..37A.
\newblock }{ (ADS)}.

\bibitem{Nomoto84}
{Nomoto}, K., {Thielemann}, F.-K., \& {Yokoi}, K.
\newblock {Accreting white dwarf models of Type I supernovae. III - Carbon
  deflagration supernovae}.
\newblock {\em \apj}, \href{http://dx.doi.org/10.1086/162639}{286:644--658},
  November 1984\href{http://cdsads.u-strasbg.fr/abs/1984ApJ...286..644N.
\newblock }{ (ADS)}.

\bibitem{Khokhlov91}
{Khokhlov}, A.~M.
\newblock {Delayed detonation model for type IA supernovae}.
\newblock {\em \aap}, 245:114--128, May
  1991\href{http://adsabs.harvard.edu/abs/1991A%26A...245..114K.
\newblock }{ (ADS)}.

\bibitem{Nomoto76}
{Nomoto}, K., {Sugimoto}, D., \& {Neo}, S.
\newblock {Carbon deflagration supernova, an alternative to carbon detonation}.
\newblock {\em \apss},
  \href{http://dx.doi.org/10.1007/BF00648354}{39:L37--L42}, February
  1976\href{http://cdsads.u-strasbg.fr/abs/1976Ap%26SS..39L..37N.
\newblock }{ (ADS)}.

\bibitem{Smirnov95}
Smirnov, N.~N. \& Tyurnikov, M.~V.
\newblock Experimental investigation of deflagration to detonation transition
  in hydrocarbon-air gaseous mixtures.
\newblock {\em Combustion and Flame}, \href{http://dx.doi.org/DOI:
  10.1016/0010-2180(94)00151-H}{100(4):661 -- 668}, 1995.

\bibitem{Hoeflich96}
{Hoeflich}, P. \& {Khokhlov}, A.
\newblock {Explosion Models for Type IA Supernovae: A Comparison with Observed
  Light Curves, Distances, H 0, and Q 0}.
\newblock {\em \apj}, \href{http://dx.doi.org/10.1086/176748}{457:500--+},
  February 1996,
  arXiv:astro-ph/9602025\href{http://cdsads.u-strasbg.fr/abs/1996ApJ...457..500H.
\newblock }{ (ADS)}.

\bibitem{Hoeflich96b}
{Hoeflich}, P., {Khokhlov}, A., {Wheeler}, J.~C., {et~al.}
\newblock {Maximum Brightness and Postmaximum Decline of Light Curves of Type
  IA Supernovae: A Comparison of Theory and Observations}.
\newblock {\em \apjl}, \href{http://dx.doi.org/10.1086/310363}{472:L81+},
  December 1996,
  arXiv:astro-ph/9609070\href{http://cdsads.u-strasbg.fr/abs/1996ApJ...472L..81H.
\newblock }{ (ADS)}.

\bibitem{Arnett82}
{Arnett}, W.~D.
\newblock {Type I supernovae. I - Analytic solutions for the early part of the
  light curve}.
\newblock {\em \apj}, \href{http://dx.doi.org/10.1086/159681}{253:785--797},
  February 1982\href{http://cdsads.u-strasbg.fr/abs/1982ApJ...253..785A.
\newblock }{ (ADS)}.

\bibitem{Stritzinger06b}
{Stritzinger}, M., {Mazzali}, P.~A., {Sollerman}, J., \& {Benetti}, S.
\newblock {Consistent estimates of $^{56}$Ni yields for type Ia supernovae}.
\newblock {\em \aap},
  \href{http://dx.doi.org/10.1051/0004-6361:20065514}{460:793--798}, December
  2006,
  arXiv:astro-ph/0609232\href{http://cdsads.u-strasbg.fr/abs/2006A%26A...460..793S.
\newblock }{ (ADS)}.

\bibitem{Lucy99b}
{Lucy}, L.~B.
\newblock {Improved Monte Carlo techniques for the spectral synthesis of
  supernovae}.
\newblock {\em \aap}, 345:211--220, May
  1999\href{http://cdsads.u-strasbg.fr/abs/1999A%26A...345..211L.
\newblock }{ (ADS)}.

\bibitem{Mazzali00}
{Mazzali}, P.~A.
\newblock {Applications of an improved Monte Carlo code to the synthesis of
  early-time Supernova spectra}.
\newblock {\em \aap}, 363:705--716, November
  2000\href{http://cdsads.u-strasbg.fr/abs/2000A%26A...363..705M.
\newblock }{ (ADS)}.

\bibitem{Sauer08}
{Sauer}, D.~N., {Mazzali}, P.~A., {Blondin}, S., {et~al.}
\newblock {Properties of the ultraviolet flux of Type Ia supernovae: an
  analysis with synthetic spectra of SN 2001ep and SN 2001eh}.
\newblock {\em \mnras},
  \href{http://dx.doi.org/10.1111/j.1365-2966.2008.14018.x}{391:1605--1618},
  December 2008,
  0803.0871\href{http://cdsads.u-strasbg.fr/abs/2008MNRAS.391.1605S.
\newblock }{ (ADS)}.

\bibitem{Kromer09}
{Kromer}, M. \& {Sim}, S.~A.
\newblock {Time-dependent three-dimensional spectrum synthesis for Type Ia
  supernovae}.
\newblock {\em \mnras},
  \href{http://dx.doi.org/10.1111/j.1365-2966.2009.15256.x}{398:1809--1826},
  October 2009,
  0906.3152\href{http://cdsads.u-strasbg.fr/abs/2009MNRAS.398.1809K.
\newblock }{ (ADS)}.

\bibitem{Lucy05}
{Lucy}, L.~B.
\newblock {Monte Carlo techniques for time-dependent radiative transfer in 3-D
  supernovae}.
\newblock {\em \aap},
  \href{http://dx.doi.org/10.1051/0004-6361:20041656}{429:19--30}, January
  2005,
  arXiv:astro-ph/0409249\href{http://cdsads.u-strasbg.fr/abs/2005A%26A...429...19L.
\newblock }{ (ADS)}.

\bibitem{Kasen06}
{Kasen}, D.
\newblock {Secondary Maximum in the Near-Infrared Light Curves of Type Ia
  Supernovae}.
\newblock {\em \apj}, \href{http://dx.doi.org/10.1086/506588}{649:939--953},
  October 2006,
  arXiv:astro-ph/0606449\href{http://cdsads.u-strasbg.fr/abs/2006ApJ...649..939K.
\newblock }{ (ADS)}.

\bibitem{Jack11}
{Jack}, D., {Hauschildt}, P.~H., \& {Baron}, E.
\newblock {Theoretical light curves of type Ia supernovae}.
\newblock {\em \aap},
  \href{http://dx.doi.org/10.1051/0004-6361/201014778}{528:A141}, April 2011,
  1105.3330\href{http://cdsads.u-strasbg.fr/abs/2011A%26A...528A.141J.
\newblock }{ (ADS)}.

\bibitem{Kasen09}
{Kasen}, D., {R{\"o}pke}, F.~K., \& {Woosley}, S.~E.
\newblock {The diversity of type Ia supernovae from broken symmetries}.
\newblock {\em \nat},
  \href{http://dx.doi.org/10.1038/nature08256}{460:869--872}, August 2009,
  0907.0708\href{http://cdsads.u-strasbg.fr/abs/2009Natur.460..869K.
\newblock }{ (ADS)}.

\bibitem{Milne01}
{Milne}, P.~A., {The}, L.-S., \& {Leising}, M.~D.
\newblock {Late Light Curves of Type Ia Supernovae}.
\newblock {\em \apj}, \href{http://dx.doi.org/10.1086/322352}{559:1019--1031},
  October 2001,
  arXiv:astro-ph/0104185\href{http://cdsads.u-strasbg.fr/abs/2001ApJ...559.1019M.
\newblock }{ (ADS)}.

\bibitem{Lair06}
{Lair}, J.~C., {Leising}, M.~D., {Milne}, P.~A., \& {Williams}, G.~G.
\newblock {Late Light Curves of Normal Type Ia Supernovae}.
\newblock {\em \aj}, \href{http://dx.doi.org/10.1086/508322}{132:2024--2033},
  November 2006,
  arXiv:astro-ph/0610084\href{http://cdsads.u-strasbg.fr/abs/2006AJ....132.2024L.
\newblock }{ (ADS)}.

\bibitem{Milne99}
{Milne}, P.~A., {The}, L.-S., \& {Leising}, M.~D.
\newblock {Positron Escape from Type IA Supernovae}.
\newblock {\em \apjs}, \href{http://dx.doi.org/10.1086/313262}{124:503--526},
  October 1999,
  arXiv:astro-ph/9901206\href{http://cdsads.u-strasbg.fr/abs/1999ApJS..124..503M.
\newblock }{ (ADS)}.

\bibitem{Knodlseder05}
{Kn{\"o}dlseder}, J., {Jean}, P., {Lonjou}, V., {et~al.}
\newblock {The all-sky distribution of 511 keV electron-positron annihilation
  emission}.
\newblock {\em \aap},
  \href{http://dx.doi.org/10.1051/0004-6361:20042063}{441:513--532}, October
  2005,
  arXiv:astro-ph/0506026\href{http://cdsads.u-strasbg.fr/abs/2005A%26A...441..513K.
\newblock }{ (ADS)}.

\bibitem{Branch-H0-92}
{Branch}, D.
\newblock {The Hubble constant from nickel radioactivity in type IA
  supernovae}.
\newblock {\em \apj}, \href{http://dx.doi.org/10.1086/171401}{392:35--40}, June
  1992\href{http://cdsads.u-strasbg.fr/abs/1992ApJ...392...35B.
\newblock }{ (ADS)}.

\bibitem{Stritzinger-H0-05}
{Stritzinger}, M. \& {Leibundgut}, B.
\newblock {Lower limits on the Hubble constant from models of type Ia
  supernovae}.
\newblock {\em \aap},
  \href{http://dx.doi.org/10.1051/0004-6361:20041630}{431:423--431}, February
  2005,
  arXiv:astro-ph/0410686\href{http://cdsads.u-strasbg.fr/abs/2005A%26A...431..423S.
\newblock }{ (ADS)}.

\bibitem{Mazzali01}
{Mazzali}, P.~A., {Nomoto}, K., {Cappellaro}, E., {et~al.}
\newblock {Can Differences in the Nickel Abundance in Chandrasekhar-Mass Models
  Explain the Relation between the Brightness and Decline Rate of Normal Type
  IA Supernovae?}
\newblock {\em \apj}, \href{http://dx.doi.org/10.1086/318428}{547:988--994},
  February 2001,
  arXiv:astro-ph/0009490\href{http://adsabs.harvard.edu/abs/2001ApJ...547..988M.
\newblock }{ (ADS)}.

\bibitem{Ropke06}
{R{\"o}pke}, F.~K., {Gieseler}, M., {Reinecke}, M., {Travaglio}, C., \&
  {Hillebrandt}, W.
\newblock {Type Ia supernova diversity in three-dimensional models}.
\newblock {\em \aap},
  \href{http://dx.doi.org/10.1051/0004-6361:20053430}{453:203--217}, July 2006,
  arXiv:astro-ph/0506107\href{http://esoads.eso.org/abs/2006A%26A...453..203R.
\newblock }{ (ADS)}.

\bibitem{Hansen87}
{Hansen}, L., {Jorgensen}, H.~E., \& {Norgaard-Nielsen}, H.~U.
\newblock {Search for supernovae in distant clusters of galaxies}.
\newblock {\em The Messenger}, 47:46--49, March
  1987\href{http://adsabs.harvard.edu/abs/1987Msngr..47...46H.
\newblock }{ (ADS)}.

\bibitem{Norgaard89}
{Norgaard-Nielsen}, H.~U., {Hansen}, L., {Jorgensen}, H.~E., {Aragon
  Salamanca}, A., \& {Ellis}, R.~S.
\newblock {The discovery of a type IA supernova at a redshift of 0.31}.
\newblock {\em \nat}, \href{http://dx.doi.org/10.1038/339523a0}{339:523--525},
  June 1989\href{http://adsabs.harvard.edu/abs/1989Natur.339..523N.
\newblock }{ (ADS)}.

\bibitem{Alard98}
{Alard}, C. \& {Lupton}, R.~H.
\newblock {A Method for Optimal Image Subtraction}.
\newblock {\em \apj}, 503:325--+, August
  1998\href{http://adsabs.harvard.edu/cgi-bin/nph-bib_query?bibcode=1998ApJ...503..325A&db_key=AST.
\newblock }{ (ADS)}.

\bibitem{Howell05}
{Howell}, D.~A., {Sullivan}, M., {Perrett}, K., {et~al.}
\newblock {Gemini Spectroscopy of Supernovae from the Supernova Legacy Survey:
  Improving High-Redshift Supernova Selection and Classification}.
\newblock {\em \apj}, \href{http://dx.doi.org/10.1086/497119}{634:1190--1201},
  December 2005,
  arXiv:astro-ph/0509195\href{http://adsabs.harvard.edu/abs/2005ApJ...634.1190H.
\newblock }{ (ADS)}.

\bibitem{Blondin07SNID}
{Blondin}, S. \& {Tonry}, J.~L.
\newblock {Determining the Type, Redshift, and Age of a Supernova Spectrum}.
\newblock {\em \apj}, \href{http://dx.doi.org/10.1086/520494}{666:1024--1047},
  September 2007,
  0709.4488\href{http://adsabs.harvard.edu/abs/2007ApJ...666.1024B.
\newblock }{ (ADS)}.

\bibitem{Blondin05}
{Blondin}, S., {Walsh}, J.~R., {Leibundgut}, B., \& {Sainton}, G.
\newblock {Extracting clean supernova spectra. Towards a quantitative analysis
  of high-redshift Type Ia supernova spectra}.
\newblock {\em \aap},
  \href{http://dx.doi.org/10.1051/0004-6361:20042009}{431:757--771}, February
  2005,
  arXiv:astro-ph/0410406\href{http://adsabs.harvard.edu/abs/2005A%26A...431..757B.
\newblock }{ (ADS)}.

\bibitem{Baumont08PHASE}
{Baumont}, S., {Balland}, C., {Astier}, P., {et~al.}
\newblock {PHotometry Assisted Spectral Extraction (PHASE) and identification
  of SNLS supernovae}.
\newblock {\em \aap},
  \href{http://dx.doi.org/10.1051/0004-6361:200810210}{491:567--585}, November
  2008, 0809.4407\href{http://adsabs.harvard.edu/abs/2008A%26A...491..567B.
\newblock }{ (ADS)}.

\bibitem{Matheson05}
{Matheson}, T., {Blondin}, S., {Foley}, R.~J., {et~al.}
\newblock {Spectroscopy of High-Redshift Supernovae from the ESSENCE Project:
  The First 2 Years}.
\newblock {\em \aj}, \href{http://dx.doi.org/10.1086/429679}{129:2352--2375},
  May 2005,
  arXiv:astro-ph/0411357\href{http://cdsads.u-strasbg.fr/abs/2005AJ....129.2352M.
\newblock }{ (ADS)}.

\bibitem{Foley09}
{Foley}, R.~J., {Matheson}, T., {Blondin}, S., {et~al.}
\newblock {Spectroscopy of High-Redshift Supernovae from the Essence Project:
  The First Four Years}.
\newblock {\em \aj},
  \href{http://dx.doi.org/10.1088/0004-6256/137/4/3731}{137:3731--3742}, April
  2009, 0811.4424\href{http://cdsads.u-strasbg.fr/abs/2009AJ....137.3731F.
\newblock }{ (ADS)}.

\bibitem{Knop03}
{Knop}, R.~A., {Aldering}, G., {Amanullah}, R., {et~al.}
\newblock {New Constraints on {$\Omega$}$_{M}$, {$\Omega$}$_{\Lambda}$, and w
  from an Independent Set of 11 High-Redshift Supernovae Observed with the
  Hubble Space Telescope}.
\newblock {\em \apj}, \href{http://dx.doi.org/10.1086/378560}{598:102--137},
  November 2003,
  arXiv:astro-ph/0309368\href{http://adsabs.harvard.edu/abs/2003ApJ...598..102K.
\newblock }{ (ADS)}.

\bibitem{Astier06}
{Astier}, P., {Guy}, J., {Regnault}, N., {et~al.}
\newblock {The Supernova Legacy Survey: measurement of $\Omega_{M}$,
  $\Omega_\Lambda$ and $w$ from the first year data set}.
\newblock {\em \aap},
  \href{http://dx.doi.org/10.1051/0004-6361:20054185}{447:31--48}, February
  2006,
  astro-ph/0510447\href{http://adsabs.harvard.edu/cgi-bin/nph-bib_query?bibcode=2006A%26A...447...31A&db_key=AST.
\newblock }{ (ADS)}.

\bibitem{Holtzman08}
{Holtzman}, J.~A., {Marriner}, J., {Kessler}, R., {et~al.}
\newblock {The Sloan Digital Sky Survey-II Photometry and Supernova IA Light
  Curves from the 2005 Data}.
\newblock {\em \aj},
  \href{http://dx.doi.org/10.1088/0004-6256/136/6/2306}{136:2306--2320},
  December 2008\href{http://adsabs.harvard.edu/abs/2008AJ....136.2306H.
\newblock }{ (ADS)}.

\bibitem{Irwin85}
{Irwin}, M.~J.
\newblock {Automatic analysis of crowded fields}.
\newblock {\em \mnras}, 214:575--604, June
  1985\href{http://adsabs.harvard.edu/abs/1985MNRAS.214..575I.
\newblock }{ (ADS)}.

\bibitem{Schmidt98}
{Schmidt}, B.~P., {Suntzeff}, N.~B., {Phillips}, M.~M., {et~al.}
\newblock {The High-Z Supernova Search: Measuring Cosmic Deceleration and
  Global Curvature of the Universe Using Type IA Supernovae}.
\newblock {\em \apj}, 507:46--63, November
  1998\href{http://adsabs.harvard.edu/cgi-bin/nph-bib_query?bibcode=1998ApJ...507...46S&db_key=AST.
\newblock }{ (ADS)}.

\bibitem{Kessler09}
{Kessler}, R., {Becker}, A.~C., {Cinabro}, D., {et~al.}
\newblock {First-Year Sloan Digital Sky Survey-II Supernova Results: Hubble
  Diagram and Cosmological Parameters}.
\newblock {\em \apjs},
  \href{http://dx.doi.org/10.1088/0067-0049/185/1/32}{185:32--84}, November
  2009, 0908.4274\href{http://cdsads.u-strasbg.fr/abs/2009ApJS..185...32K.
\newblock }{ (ADS)}.

\bibitem{Humason56}
{Humason}, M.~L., {Mayall}, N.~U., \& {Sandage}, A.~R.
\newblock {Redshifts and magnitudes of extragalactic nebulae.}
\newblock {\em \aj}, \href{http://dx.doi.org/10.1086/107297}{61:97--162},
  1956\href{http://cdsads.u-strasbg.fr/abs/1956AJ.....61...97H.
\newblock }{ (ADS)}.

\bibitem{Kim96}
{Kim}, A., {Goobar}, A., \& {Perlmutter}, S.
\newblock {A Generalized K Correction for Type IA Supernovae: Comparing R-band
  Photometry beyond z=0.2 with B, V, and R-band Nearby Photometry}.
\newblock {\em \pasp}, \href{http://dx.doi.org/10.1086/133709}{108:190},
  February 1996,
  arXiv:astro-ph/9505024\href{http://cdsads.u-strasbg.fr/abs/1996PASP..108..190K.
\newblock }{ (ADS)}.

\bibitem{Nugent02}
{Nugent}, P., {Kim}, A., \& {Perlmutter}, S.
\newblock {K-Corrections and Extinction Corrections for Type Ia Supernovae}.
\newblock {\em \pasp}, \href{http://dx.doi.org/10.1086/341707}{114:803--819},
  August 2002,
  arXiv:astro-ph/0205351\href{http://cdsads.u-strasbg.fr/abs/2002PASP..114..803N.
\newblock }{ (ADS)}.

\bibitem{Schlegel98}
{Schlegel}, D.~J., {Finkbeiner}, D.~P., \& {Davis}, M.
\newblock {Maps of Dust Infrared Emission for Use in Estimation of Reddening
  and Cosmic Microwave Background Radiation Foregrounds}.
\newblock {\em \apj}, 500:525--+, June
  1998\href{http://cdsads.u-strasbg.fr/cgi-bin/nph-bib_query?bibcode=1998ApJ...500..525S&db_key=AST.
\newblock }{ (ADS)}.

\bibitem{Cardelli89}
{Cardelli}, J.~A., {Clayton}, G.~C., \& {Mathis}, J.~S.
\newblock {The relationship between infrared, optical, and ultraviolet
  extinction}.
\newblock {\em APJ}, 345:245--256, October
  1989\href{http://adsabs.harvard.edu/cgi-bin/nph-bib_query?bibcode=1989ApJ...345..245C&amp;db_key=AST.
\newblock }{ (ADS)}.

\bibitem{LeibundgutPhd}
{Leibundgut}, B.
\newblock {\em {Light curves of supernovae type, I.}}
\newblock PhD thesis, PhD thesis.~Univ.~Basel.137 pp.~, (1988),
  1988\href{http://cdsads.u-strasbg.fr/abs/1988PhDT.......171L.
\newblock }{ (ADS)}.

\bibitem{Hamuy95}
{Hamuy}, M., {Phillips}, M.~M., {Maza}, J., {et~al.}
\newblock {A Hubble diagram of distant type IA supernovae}.
\newblock {\em \aj}, \href{http://dx.doi.org/10.1086/117251}{109:1--13},
  January 1995\href{http://cdsads.u-strasbg.fr/abs/1995AJ....109....1H.
\newblock }{ (ADS)}.

\bibitem{Hamuy96d}
{Hamuy}, M., {Phillips}, M.~M., {Suntzeff}, N.~B., {et~al.}
\newblock {The Morphology of Type IA Supernovae Light Curves}.
\newblock {\em \aj}, \href{http://dx.doi.org/10.1086/118193}{112:2438},
  December 1996,
  arXiv:astro-ph/9609063\href{http://cdsads.u-strasbg.fr/abs/1996AJ....112.2438H.
\newblock }{ (ADS)}.

\bibitem{Germany04}
{Germany}, L.~M., {Reiss}, D.~J., {Schmidt}, B.~P., {Stubbs}, C.~W., \&
  {Suntzeff}, N.~B.
\newblock {Results of the Mount Stromlo Abell cluster supernova search}.
\newblock {\em \aap},
  \href{http://dx.doi.org/10.1051/0004-6361:20031616}{415:863--878}, March
  2004\href{http://cdsads.u-strasbg.fr/abs/2004A%26A...415..863G.
\newblock }{ (ADS)}.

\bibitem{Prieto06}
{Prieto}, J.~L., {Rest}, A., \& {Suntzeff}, N.~B.
\newblock {A New Method to Calibrate the Magnitudes of Type Ia Supernovae at
  Maximum Light}.
\newblock {\em \apj}, \href{http://dx.doi.org/10.1086/504307}{647:501--512},
  August 2006,
  arXiv:astro-ph/0603407\href{http://cdsads.u-strasbg.fr/abs/2006ApJ...647..501P.
\newblock }{ (ADS)}.

\bibitem{BurnsSnoopy11}
{Burns}, C.~R., {Stritzinger}, M., {Phillips}, M.~M., {et~al.}
\newblock {The Carnegie Supernova Project: Light-curve Fitting with SNooPy}.
\newblock {\em \aj},
  \href{http://dx.doi.org/10.1088/0004-6256/141/1/19}{141:19}, January 2011,
  1010.4040\href{http://adsabs.harvard.edu/abs/2011AJ....141...19B.
\newblock }{ (ADS)}.

\bibitem{Tonry03}
{Tonry}, J.~L., {Schmidt}, B.~P., {Barris}, B., {et~al.}
\newblock {Cosmological Results from High-z Supernovae}.
\newblock {\em \apj}, \href{http://dx.doi.org/10.1086/376865}{594:1--24},
  September 2003,
  arXiv:astro-ph/0305008\href{http://cdsads.u-strasbg.fr/abs/2003ApJ...594....1T.
\newblock }{ (ADS)}.

\bibitem{Riess95}
{Riess}, A.~G., {Press}, W.~H., \& {Kirshner}, R.~P.
\newblock {Using Type IA supernova light curve shapes to measure the Hubble
  constant}.
\newblock {\em \apjl}, 438:L17--L20, January
  1995\href{http://adsabs.harvard.edu/cgi-bin/nph-bib_query?bibcode=1995ApJ...438L..17R&db_key=AST.
\newblock }{ (ADS)}.

\bibitem{Jha07-MLCS2k2}
{Jha}, S., {Riess}, A.~G., \& {Kirshner}, R.~P.
\newblock {Improved Distances to Type Ia Supernovae with Multicolor Light-Curve
  Shapes: MLCS2k2}.
\newblock {\em \apj}, \href{http://dx.doi.org/10.1086/512054}{659:122--148},
  April 2007,
  arXiv:astro-ph/0612666\href{http://cdsads.u-strasbg.fr/abs/2007ApJ...659..122J.
\newblock }{ (ADS)}.

\bibitem{WoodVasey07}
{Wood-Vasey}, W.~M., {Miknaitis}, G., {Stubbs}, C.~W., {et~al.}
\newblock {Observational Constraints on the Nature of Dark Energy: First
  Cosmological Results from the ESSENCE Supernova Survey}.
\newblock {\em \apj}, \href{http://dx.doi.org/10.1086/518642}{666:694--715},
  September 2007,
  arXiv:astro-ph/0701041\href{http://cdsads.u-strasbg.fr/abs/2007ApJ...666..694W.
\newblock }{ (ADS)}.

\bibitem{Freedman09}
{Freedman}, W.~L., {Burns}, C.~R., {Phillips}, M.~M., {et~al.}
\newblock {The Carnegie Supernova Project: First Near-Infrared Hubble Diagram
  to z \~{} 0.7}.
\newblock {\em \apj},
  \href{http://dx.doi.org/10.1088/0004-637X/704/2/1036}{704:1036--1058},
  October 2009,
  0907.4524\href{http://adsabs.harvard.edu/abs/2009ApJ...704.1036F.
\newblock }{ (ADS)}.

\bibitem{Hsiao07}
{Hsiao}, E.~Y., {Conley}, A., {Howell}, D.~A., {et~al.}
\newblock {K-Corrections and Spectral Templates of Type Ia Supernovae}.
\newblock {\em \apj}, \href{http://dx.doi.org/10.1086/518232}{663:1187--1200},
  July 2007,
  arXiv:astro-ph/0703529\href{http://cdsads.u-strasbg.fr/abs/2007ApJ...663.1187H.
\newblock }{ (ADS)}.

\bibitem{Conley11}
{Conley}, A., {Guy}, J., {Sullivan}, M., {et~al.}
\newblock {Supernova Constraints and Systematic Uncertainties from the First
  Three Years of the Supernova Legacy Survey}.
\newblock {\em \apjs},
  \href{http://dx.doi.org/10.1088/0067-0049/192/1/1}{192:1--+}, January 2011,
  1104.1443\href{http://cdsads.u-strasbg.fr/abs/2011ApJS..192....1C.
\newblock }{ (ADS)}.

\bibitem{Conley07}
{Conley}, A., {Carlberg}, R.~G., {Guy}, J., {et~al.}
\newblock {Is There Evidence for a Hubble Bubble? The Nature of Type Ia
  Supernova Colors and Dust in External Galaxies}.
\newblock {\em \apjl}, \href{http://dx.doi.org/10.1086/520625}{664:L13--L16},
  July 2007, 0705.0367\href{http://cdsads.u-strasbg.fr/abs/2007ApJ...664L..13C.
\newblock }{ (ADS)}.

\bibitem{Bailey09}
{Bailey}, S., {Aldering}, G., {Antilogus}, P., {et~al.}
\newblock {Using spectral flux ratios to standardize SN Ia luminosities}.
\newblock {\em \aap},
  \href{http://dx.doi.org/10.1051/0004-6361/200911973}{500:L17--L20}, June
  2009, 0905.0340\href{http://cdsads.u-strasbg.fr/abs/2009A%26A...500L..17B.
\newblock }{ (ADS)}.

\bibitem{Wang-CMagic-03}
{Wang}, L., {Goldhaber}, G., {Aldering}, G., \& {Perlmutter}, S.
\newblock {Multicolor Light Curves of Type Ia Supernovae on the Color-Magnitude
  Diagram: A Novel Step toward More Precise Distance and Extinction Estimates}.
\newblock {\em \apj}, \href{http://dx.doi.org/10.1086/375020}{590:944--970},
  June 2003,
  arXiv:astro-ph/0302341\href{http://adsabs.harvard.edu/abs/2003ApJ...590..944W.
\newblock }{ (ADS)}.

\bibitem{Conley-CMagic-06}
{Conley}, A., {Goldhaber}, G., {Wang}, L., {et~al.}
\newblock {Measurement of {$\Omega$}$_{m}$, {$\Omega$}$_{\Lambda}$ from a Blind
  Analysis of Type Ia Supernovae with CMAGIC: Using Color Information to Verify
  the Acceleration of the Universe}.
\newblock {\em \apj}, \href{http://dx.doi.org/10.1086/503533}{644:1--20}, June
  2006,
  arXiv:astro-ph/0602411\href{http://cdsads.u-strasbg.fr/abs/2006ApJ...644....1C.
\newblock }{ (ADS)}.

\bibitem{Hayes85}
{Hayes}, D.~S.
\newblock {Stellar absolute fluxes and energy distributions from 0.32 to 4.0
  microns}.
\newblock In {Hayes}, D.~S., {Pasinetti}, L.~E., \& {Philip}, A.~G.~D.,
  editors, {\em Calibration of Fundamental Stellar Quantities}, volume 111 of
  {\em IAU Symposium}, pages 225--249,
  1985\href{http://cdsads.u-strasbg.fr/abs/1985IAUS..111..225H.
\newblock }{ (ADS)}.

\bibitem{Regnault09}
{Regnault}, N., {Conley}, A., {Guy}, J., {et~al.}
\newblock {Photometric calibration of the Supernova Legacy Survey fields}.
\newblock {\em \aap},
  \href{http://dx.doi.org/10.1051/0004-6361/200912446}{506:999--1042}, November
  2009, 0908.3808\href{http://cdsads.u-strasbg.fr/abs/2009A%26A...506..999R.
\newblock }{ (ADS)}.

\bibitem{Landolt83}
{Landolt}, A.~U.
\newblock {UBVRI photometric standard stars around the celestial equator}.
\newblock {\em AJ}, 88:439--460, March
  1983\href{http://cdsads.u-strasbg.fr/cgi-bin/nph-bib_query?bibcode=1983AJ.....88..439L&amp;db_key=AST.
\newblock }{ (ADS)}.

\bibitem{Landolt92}
{Landolt}, A.~U.
\newblock {UBVRI photometric standard stars in the magnitude range 11.5-16.0
  around the celestial equator}.
\newblock {\em AJ}, 104:340--371, July
  1992\href{http://cdsads.u-strasbg.fr/cgi-bin/nph-bib_query?bibcode=1992AJ....104..340L&amp;db_key=AST.
\newblock }{ (ADS)}.

\bibitem{Johnson53}
{Johnson}, H.~L. \& {Morgan}, W.~W.
\newblock {Fundamental stellar photometry for standards of spectral type on the
  revised system of the Yerkes spectral atlas}.
\newblock {\em \apj}, 117:313--+, May
  1953\href{http://adsabs.harvard.edu/cgi-bin/nph-bib_query?bibcode=1953ApJ...117..313J&db_key=AST.
\newblock }{ (ADS)}.

\bibitem{Cousins76}
{Cousins}, A.~W.~J.
\newblock {VRI standards in the E regions}.
\newblock {\em \memras}, 81:25--+,
  1976\href{http://adsabs.harvard.edu/abs/1976MmRAS..81...25C.
\newblock }{ (ADS)}.

\bibitem{Johnson51}
{Johnson}, H.~L. \& {Morgan}, W.~W.
\newblock {On the Color-Magnitude Diagram of the Pleiades.}
\newblock {\em \apj}, 114:522--+, November
  1951\href{http://adsabs.harvard.edu/abs/1951ApJ...114..522J.
\newblock }{ (ADS)}.

\bibitem{Bohlin00}
{Bohlin}, R.~C.
\newblock {Comparison of White Dwarf Models with STIS Spectrophotometry}.
\newblock {\em \aj}, \href{http://dx.doi.org/10.1086/301431}{120:437--446},
  July
  2000\href{http://adsabs.harvard.edu/cgi-bin/nph-bib_query?bibcode=2000AJ....120..437B&db_key=AST.
\newblock }{ (ADS)}.

\bibitem{Bohlin04Vega}
{Bohlin}, R.~C. \& {Gilliland}, R.~L.
\newblock {Hubble Space Telescope Absolute Spectrophotometry of Vega from the
  Far-Ultraviolet to the Infrared}.
\newblock {\em \aj}, 127:3508--3515, June
  2004\href{http://cdsads.u-strasbg.fr/cgi-bin/nph-bib_query?bibcode=2004AJ....127.3508B&db_key=AST.
\newblock }{ (ADS)}.

\bibitem{Landolt07b}
{Landolt}, A.~U. \& {Uomoto}, A.~K.
\newblock {Optical Multicolor Photometry of Spectrophotometric Standard Stars}.
\newblock {\em \aj}, \href{http://dx.doi.org/10.1086/510485}{133:768--790},
  March 2007\href{http://adsabs.harvard.edu/abs/2007AJ....133..768L.
\newblock }{ (ADS)}.

\bibitem{Bohlin01}
{Bohlin}, R.~C., {Dickinson}, M.~E., \& {Calzetti}, D.
\newblock {Spectrophotometric Standards from the Far-Ultraviolet to the
  Near-Infrared: STIS and NICMOS Fluxes}.
\newblock {\em \aj}, \href{http://dx.doi.org/10.1086/323137}{122:2118--2128},
  October 2001\href{http://cdsads.u-strasbg.fr/abs/2001AJ....122.2118B.
\newblock }{ (ADS)}.

\bibitem{Bohlin04BD17}
{Bohlin}, R.~C. \& {Gilliland}, R.~L.
\newblock {Absolute Flux Distribution of the SDSS Standard BD~+17~4708}.
\newblock {\em \aj}, \href{http://dx.doi.org/10.1086/425880}{128:3053--3060},
  December 2004\href{http://adsabs.harvard.edu/abs/2004AJ....128.3053B.
\newblock }{ (ADS)}.

\bibitem{Stubbs06}
{Stubbs}, C.~W. \& {Tonry}, J.~L.
\newblock {Toward 1\% Photometry: End-to-End Calibration of Astronomical
  Telescopes and Detectors}.
\newblock {\em \apj}, \href{http://dx.doi.org/10.1086/505138}{646:1436--1444},
  August 2006,
  arXiv:astro-ph/0604285\href{http://adsabs.harvard.edu/abs/2006ApJ...646.1436S.
\newblock }{ (ADS)}.

\bibitem{Stubbs10}
{Stubbs}, C.~W., {Doherty}, P., {Cramer}, C., {et~al.}
\newblock {Precise Throughput Determination of the PanSTARRS Telescope and the
  Gigapixel Imager Using a Calibrated Silicon Photodiode and a Tunable Laser:
  Initial Results}.
\newblock {\em \apjs},
  \href{http://dx.doi.org/10.1088/0067-0049/191/2/376}{191:376--388}, December
  2010, 1003.3465\href{http://adsabs.harvard.edu/abs/2010ApJS..191..376S.
\newblock }{ (ADS)}.

\bibitem{Juramy08}
{Juramy}, C., {Barrelet}, E., {Schahmaneche}, K., {et~al.}
\newblock {SNDICE: a direct illumination calibration experiment at CFHT}.
\newblock {\em SPIE}, \href{http://dx.doi.org/10.1117/12.789051}{7014}, August
  2008\href{http://cdsads.u-strasbg.fr/abs/2008SPIE.7014E.166J.
\newblock }{ (ADS)}.

\bibitem{Barrelet08}
{Barrelet}, E. \& {Juramy}, C.
\newblock {Direct illumination LED calibration for telescope photometry}.
\newblock {\em Nuclear Instruments and Methods in Physics Research A},
  \href{http://dx.doi.org/10.1016/j.nima.2007.10.048}{585:93--101}, January
  2008\href{http://cdsads.u-strasbg.fr/abs/2008NIMPA.585...93B.
\newblock }{ (ADS)}.

\bibitem{Tonry12}
{Tonry}, J.~L., {Stubbs}, C.~W., {Lykke}, K.~R., {et~al.}
\newblock {The Pan-STARRS1 Photometric System}.
\newblock {\em \apj},
  \href{http://dx.doi.org/10.1088/0004-637X/750/2/99}{750:99}, May 2012,
  1203.0297\href{http://cdsads.u-strasbg.fr/abs/2012ApJ...750...99T.
\newblock }{ (ADS)}.

\bibitem{Burke10}
{Burke}, D.~L., {Axelrod}, T., {Blondin}, S., {et~al.}
\newblock {Precision Determination of Atmospheric Extinction at Optical and
  Near-infrared Wavelengths}.
\newblock {\em \apj},
  \href{http://dx.doi.org/10.1088/0004-637X/720/1/811}{720:811--823}, September
  2010\href{http://cdsads.u-strasbg.fr/abs/2010ApJ...720..811B.
\newblock }{ (ADS)}.

\bibitem{Massey10}
{Massey}, R., {Stoughton}, C., {Leauthaud}, A., {et~al.}
\newblock {Pixel-based correction for Charge Transfer Inefficiency in the
  Hubble Space Telescope Advanced Camera for Surveys}.
\newblock {\em \mnras},
  \href{http://dx.doi.org/10.1111/j.1365-2966.2009.15638.x}{401:371--384},
  January 2010,
  0909.0507\href{http://cdsads.u-strasbg.fr/abs/2010MNRAS.401..371M.
\newblock }{ (ADS)}.

\bibitem{Sullivan11}
{Sullivan}, M., {Guy}, J., {Conley}, A., {et~al.}
\newblock {SNLS3: Constraints on Dark Energy Combining the Supernova Legacy
  Survey Three-year Data with Other Probes}.
\newblock {\em \apj},
  \href{http://dx.doi.org/10.1088/0004-637X/737/2/102}{737:102}, August 2011,
  1104.1444\href{http://cdsads.u-strasbg.fr/abs/2011ApJ...737..102S.
\newblock }{ (ADS)}.

\bibitem{Stritzinger11}
{Stritzinger}, M.~D., {Phillips}, M.~M., {Boldt}, L.~N., {et~al.}
\newblock {The Carnegie Supernova Project: Second Photometry Data Release of
  Low-redshift Type Ia Supernovae}.
\newblock {\em \aj},
  \href{http://dx.doi.org/10.1088/0004-6256/142/5/156}{142:156}, November 2011,
  1108.3108\href{http://cdsads.u-strasbg.fr/abs/2011AJ....142..156S.
\newblock }{ (ADS)}.

\bibitem{Smith02}
{Smith}, J.~A., {Tucker}, D.~L., {Kent}, S., {et~al.}
\newblock {The u'g'r'i'z' Standard-Star System}.
\newblock {\em \aj}, \href{http://dx.doi.org/10.1086/339311}{123:2121--2144},
  April
  2002\href{http://adsabs.harvard.edu/cgi-bin/nph-bib_query?bibcode=2002AJ....123.2121S&db_key=AST.
\newblock }{ (ADS)}.

\bibitem{Albert09}
{Albert}, J., {Foster}, K., {Battat}, J., {et~al.}
\newblock {Satellite-Mounted Light Sources as Photometric Calibration
  Standards}.
\newblock {\em ArXiv e-prints}, August 2009,
  0908.3343\href{http://cdsads.u-strasbg.fr/abs/2009arXiv0908.3343A.
\newblock }{ (ADS)}.

\bibitem{Aldering00}
{Aldering}, G., {Knop}, R., \& {Nugent}, P.
\newblock {The Rise Times of High- and Low-Redshift Type IA Supernovae Are
  Consistent}.
\newblock {\em \aj}, 119:2110--2117, May
  2000\href{http://adsabs.harvard.edu/cgi-bin/nph-bib_query?bibcode=2000AJ....119.2110A&db_key=AST.
\newblock }{ (ADS)}.

\bibitem{Hook05}
{Hook}, I.~M., {Howell}, D.~A., {Aldering}, G., {et~al.}
\newblock {Spectra of High-Redshift Type Ia Supernovae and a Comparison with
  Their Low-Redshift Counterparts}.
\newblock {\em \aj}, \href{http://dx.doi.org/10.1086/497635}{130:2788--2803},
  December 2005,
  arXiv:astro-ph/0509041\href{http://cdsads.u-strasbg.fr/abs/2005AJ....130.2788H.
\newblock }{ (ADS)}.

\bibitem{blondin06}
{Blondin}, S., {Dessart}, L., {Leibundgut}, B., {et~al.}
\newblock {Using Line Profiles to Test the Fraternity of Type Ia Supernovae at
  High and Low Redshifts}.
\newblock {\em \aj}, \href{http://dx.doi.org/10.1086/498724}{131:1648--1666},
  March 2006,
  arXiv:astro-ph/0510089\href{http://cdsads.u-strasbg.fr/abs/2006AJ....131.1648B.
\newblock }{ (ADS)}.

\bibitem{Garavini07}
{Garavini}, G., {Folatelli}, G., {Nobili}, S., {et~al.}
\newblock {Quantitative comparison between type Ia supernova spectra at low and
  high redshifts: a case study}.
\newblock {\em \aap},
  \href{http://dx.doi.org/10.1051/0004-6361:20065400}{470:411--424}, August
  2007,
  arXiv:astro-ph/0703629\href{http://cdsads.u-strasbg.fr/abs/2007A%26A...470..411G.
\newblock }{ (ADS)}.

\bibitem{Foley08}
{Foley}, R.~J., {Filippenko}, A.~V., {Aguilera}, C., {et~al.}
\newblock {Constraining Cosmic Evolution of Type Ia Supernovae}.
\newblock {\em \apj}, \href{http://dx.doi.org/10.1086/589612}{684:68--87},
  September 2008,
  0710.2338\href{http://cdsads.u-strasbg.fr/abs/2008ApJ...684...68F.
\newblock }{ (ADS)}.

\bibitem{Walker11}
{Walker}, E.~S., {Hook}, I.~M., {Sullivan}, M., {et~al.}
\newblock {Supernova Legacy Survey: using spectral signatures to improve Type
  Ia supernovae as distance indicators}.
\newblock {\em \mnras},
  \href{http://dx.doi.org/10.1111/j.1365-2966.2010.17519.x}{410:1262--1282},
  January 2011,
  1008.2308\href{http://cdsads.u-strasbg.fr/abs/2011MNRAS.410.1262W.
\newblock }{ (ADS)}.

\bibitem{Hoeflich98}
{Hoeflich}, P., {Wheeler}, J.~C., \& {Thielemann}, F.~K.
\newblock {Type IA Supernovae: Influence of the Initial Composition on the
  Nucleosynthesis, Light Curves, and Spectra and Consequences for the
  Determination of Omega M and Lambda}.
\newblock {\em \apj}, \href{http://dx.doi.org/10.1086/305327}{495:617--+},
  March 1998,
  arXiv:astro-ph/9709233\href{http://cdsads.u-strasbg.fr/abs/1998ApJ...495..617H.
\newblock }{ (ADS)}.

\bibitem{Lentz00}
{Lentz}, E.~J., {Baron}, E., {Branch}, D., {Hauschildt}, P.~H., \& {Nugent},
  P.~E.
\newblock {Metallicity Effects in Non-LTE Model Atmospheres of Type IA
  Supernovae}.
\newblock {\em \apj}, \href{http://dx.doi.org/10.1086/308400}{530:966--976},
  February 2000,
  arXiv:astro-ph/9906016\href{http://cdsads.u-strasbg.fr/abs/2000ApJ...530..966L.
\newblock }{ (ADS)}.

\bibitem{Cooke11}
{Cooke}, J., {Ellis}, R.~S., {Sullivan}, M., {et~al.}
\newblock {Hubble Space Telescope Studies of Nearby Type Ia Supernovae: The
  Mean Maximum Light Ultraviolet Spectrum and its Dispersion}.
\newblock {\em \apjl},
  \href{http://dx.doi.org/10.1088/2041-8205/727/2/L35}{727:L35+}, February
  2011, 1010.2211\href{http://cdsads.u-strasbg.fr/abs/2011ApJ...727L..35C.
\newblock }{ (ADS)}.

\bibitem{Sullivan03}
{Sullivan}, M., {Ellis}, R.~S., {Aldering}, G., {et~al.}
\newblock {The Hubble diagram of type Ia supernovae as a function of host
  galaxy morphology}.
\newblock {\em \mnras}, 340:1057--1075, April
  2003\href{http://adsabs.harvard.edu/cgi-bin/nph-bib_query?bibcode=2003MNRAS.340.1057S&db_key=AST.
\newblock }{ (ADS)}.

\bibitem{Howell09}
{Howell}, D.~A., {Sullivan}, M., {Brown}, E.~F., {et~al.}
\newblock {The Effect of Progenitor Age and Metallicity on Luminosity and
  $^{56}$Ni Yield in Type Ia Supernovae}.
\newblock {\em \apj},
  \href{http://dx.doi.org/10.1088/0004-637X/691/1/661}{691:661--671}, January
  2009, 0810.0031\href{http://adsabs.harvard.edu/abs/2009ApJ...691..661H.
\newblock }{ (ADS)}.

\bibitem{Gallagher08}
{Gallagher}, J.~S., {Garnavich}, P.~M., {Caldwell}, N., {et~al.}
\newblock {Supernovae in Early-Type Galaxies: Directly Connecting Age and
  Metallicity with Type Ia Luminosity}.
\newblock {\em \apj}, \href{http://dx.doi.org/10.1086/590659}{685:752--766},
  October 2008,
  0805.4360\href{http://cdsads.u-strasbg.fr/abs/2008ApJ...685..752G.
\newblock }{ (ADS)}.

\bibitem{Kelly10}
{Kelly}, P.~L., {Hicken}, M., {Burke}, D.~L., {Mandel}, K.~S., \& {Kirshner},
  R.~P.
\newblock {Hubble Residuals of Nearby Type Ia Supernovae are Correlated with
  Host Galaxy Masses}.
\newblock {\em \apj},
  \href{http://dx.doi.org/10.1088/0004-637X/715/2/743}{715:743--756}, June
  2010, 0912.0929\href{http://cdsads.u-strasbg.fr/abs/2010ApJ...715..743K.
\newblock }{ (ADS)}.

\bibitem{Schneider92}
{Schneider}, P., {Ehlers}, J., \& {Falco}, E.~E.
\newblock {\em {Gravitational Lenses}}.
\newblock {Springer}, 1992.

\bibitem{Narayan96}
{Narayan}, R. \& {Bartelmann}, M.
\newblock {Lectures on Gravitational Lensing}.
\newblock {\em ArXiv Astrophysics e-prints}, June 1996,
  arXiv:astro-ph/9606001\href{http://cdsads.u-strasbg.fr/abs/1996astro.ph..6001N.
\newblock }{ (ADS)}.

\bibitem{Fort94}
{Fort}, B. \& {Mellier}, Y.
\newblock {Arc(let)s in clusters of galaxies}.
\newblock {\em \aapr}, \href{http://dx.doi.org/10.1007/BF00877691}{5:239--292},
  1994\href{http://cdsads.u-strasbg.fr/abs/1994A\%26ARv...5..239F.
\newblock }{ (ADS)}.

\bibitem{Peacock86}
{Peacock}, J.~A.
\newblock {Flux conservation and random gravitational lensing}.
\newblock {\em \mnras}, 223:113--128, November
  1986\href{http://cdsads.u-strasbg.fr/abs/1986MNRAS.223..113P.
\newblock }{ (ADS)}.

\bibitem{Weinberg76}
{Weinberg}, S.
\newblock {Apparent luminosities in a locally inhomogeneous universe}.
\newblock {\em \apjl}, \href{http://dx.doi.org/10.1086/182216}{208:L1--L3},
  August 1976\href{http://cdsads.u-strasbg.fr/abs/1976ApJ...208L...1W.
\newblock }{ (ADS)}.

\bibitem{Holz98b}
{Holz}, D.~E.
\newblock {Lensing and High-z Supernova Surveys}.
\newblock {\em \apjl}, \href{http://dx.doi.org/10.1086/311631}{506:L1--L5},
  October 1998,
  arXiv:astro-ph/9806124\href{http://cdsads.u-strasbg.fr/abs/1998ApJ...506L...1H.
\newblock }{ (ADS)}.

\bibitem{Bergstrom00}
{Bergstr{\" o}m}, L., {Goliath}, M., {Goobar}, A., \& {M{\" o}rtsell}, E.
\newblock {Lensing effects in an inhomogeneous universe}.
\newblock {\em \aap}, 358:13--29, June
  2000\href{http://cdsads.u-strasbg.fr/cgi-bin/nph-bib_query?bibcode=2000A%26A...358...13B&db_key=AST.
\newblock }{ (ADS)}.

\bibitem{Goobar02SNOC}
{Goobar}, A., {M{\"o}rtsell}, E., {Amanullah}, R., {et~al.}
\newblock {SNOC: A Monte-Carlo simulation package for high-z supernova
  observations}.
\newblock {\em \aap},
  \href{http://dx.doi.org/10.1051/0004-6361:20020930}{392:757--771}, September
  2002,
  arXiv:astro-ph/0206409\href{http://cdsads.u-strasbg.fr/abs/2002A\%26A...392..757G.
\newblock }{ (ADS)}.

\bibitem{ClassSurvey}
{Myers}, S.~T., {Jackson}, N.~J., {Browne}, I.~W.~A., {et~al.}
\newblock {The Cosmic Lens All-Sky Survey - I. Source selection and
  observations}.
\newblock {\em MNRAS},
  \href{http://dx.doi.org/10.1046/j.1365-8711.2003.06256.x}{341:1--12}, May
  2003\href{http://cdsads.u-strasbg.fr/cgi-bin/nph-bib_query?bibcode=2003MNRAS.341....1M&db_key=AST.
\newblock }{ (ADS)}.

\bibitem{Holz98a}
{Holz}, D.~E. \& {Wald}, R.~M.
\newblock {New method for determining cumulative gravitational lensing effects
  in inhomogeneous universes}.
\newblock {\em \prd},
  \href{http://dx.doi.org/10.1103/PhysRevD.58.063501}{58(6):063501--+},
  September 1998,
  arXiv:astro-ph/9708036\href{http://cdsads.u-strasbg.fr/abs/1998PhRvD..58f3501H.
\newblock }{ (ADS)}.

\bibitem{Amanullah03}
{Amanullah}, R., {M{\"o}rtsell}, E., \& {Goobar}, A.
\newblock {Correcting for lensing bias in the Hubble diagram}.
\newblock {\em \aap},
  \href{http://dx.doi.org/10.1051/0004-6361:20021547}{397:819--823}, January
  2003,
  arXiv:astro-ph/0204280\href{http://cdsads.u-strasbg.fr/abs/2003A\%26A...397..819A.
\newblock }{ (ADS)}.

\bibitem{HolzLinder04}
{Holz}, D.~E. \& {Linder}, E.~V.
\newblock {Safety in Numbers: Gravitational Lensing Degradation of the
  Luminosity Distance-Redshift Relation}.
\newblock {\em \apj}, \href{http://dx.doi.org/10.1086/432085}{631:678--688},
  October 2005,
  arXiv:astro-ph/0412173\href{http://cdsads.u-strasbg.fr/abs/2005ApJ...631..678H.
\newblock }{ (ADS)}.

\bibitem{Kronborg10}
{Kronborg}, T., {Hardin}, D., {Guy}, J., {et~al.}
\newblock {Gravitational lensing in the supernova legacy survey (SNLS)}.
\newblock {\em \aap},
  \href{http://dx.doi.org/10.1051/0004-6361/200913618}{514:A44+}, May 2010,
  1002.1249\href{http://cdsads.u-strasbg.fr/abs/2010A\%26A...514A..44K.
\newblock }{ (ADS)}.

\bibitem{Johnsson10}
{J{\"o}nsson}, J., {Sullivan}, M., {Hook}, I., {et~al.}
\newblock {Constraining dark matter halo properties using lensed Supernova
  Legacy Survey supernovae}.
\newblock {\em \mnras},
  \href{http://dx.doi.org/10.1111/j.1365-2966.2010.16467.x}{405:535--544}, June
  2010, 1002.1374\href{http://cdsads.u-strasbg.fr/abs/2010MNRAS.405..535J.
\newblock }{ (ADS)}.

\bibitem{Goobar09}
{Goobar}, A., {Paech}, K., {Stanishev}, V., {et~al.}
\newblock {Near-IR search for lensed supernovae behind galaxy clusters. II.
  First detection and future prospects}.
\newblock {\em \aap},
  \href{http://dx.doi.org/10.1051/0004-6361/200811254}{507:71--83}, November
  2009, 0810.4932\href{http://cdsads.u-strasbg.fr/abs/2009A%26A...507...71G.
\newblock }{ (ADS)}.

\bibitem{Goobar02}
{Goobar}, A., {M{\"o}rtsell}, E., {Amanullah}, R., \& {Nugent}, P.
\newblock {Cosmological parameters from lensed supernovae}.
\newblock {\em \aap},
  \href{http://dx.doi.org/10.1051/0004-6361:20020987}{393:25--32}, October
  2002,
  arXiv:astro-ph/0207139\href{http://cdsads.u-strasbg.fr/abs/2002A%26A...393...25G.
\newblock }{ (ADS)}.

\bibitem{Oguri03}
{Oguri}, M. \& {Kawano}, Y.
\newblock {Gravitational lens time delays for distant supernovae: breaking the
  degeneracy between radial mass profiles and the Hubble constant}.
\newblock {\em \mnras},
  \href{http://dx.doi.org/10.1046/j.1365-8711.2003.06290.x}{338:L25--L29},
  February 2003,
  arXiv:astro-ph/0211499\href{http://cdsads.u-strasbg.fr/abs/2003MNRAS.338L..25O.
\newblock }{ (ADS)}.

\bibitem{Oguri10}
{Oguri}, M. \& {Marshall}, P.~J.
\newblock {Gravitationally lensed quasars and supernovae in future wide-field
  optical imaging surveys}.
\newblock {\em \mnras},
  \href{http://dx.doi.org/10.1111/j.1365-2966.2010.16639.x}{405:2579--2593},
  July 2010, 1001.2037\href{http://cdsads.u-strasbg.fr/abs/2010MNRAS.405.2579O.
\newblock }{ (ADS)}.

\bibitem{Garnavich98}
{Garnavich}, P.~M., {Jha}, S., {Challis}, P., {et~al.}
\newblock {Supernova Limits on the Cosmic Equation of State}.
\newblock {\em \apj}, \href{http://dx.doi.org/10.1086/306495}{509:74--79},
  December 1998,
  arXiv:astro-ph/9806396\href{http://cdsads.u-strasbg.fr/abs/1998ApJ...509...74G.
\newblock }{ (ADS)}.

\bibitem{Riess07}
{Riess}, A.~G., {Strolger}, L.-G., {Casertano}, S., {et~al.}
\newblock {New Hubble Space Telescope Discoveries of Type Ia Supernovae at $z
  \geq 1$: Narrowing Constraints on the Early Behavior of Dark Energy}.
\newblock {\em \apj}, \href{http://dx.doi.org/10.1086/510378}{659:98--121},
  April 2007,
  arXiv:astro-ph/0611572\href{http://adsabs.harvard.edu/abs/2007ApJ...659...98R.
\newblock }{ (ADS)}.

\bibitem{Frieman08}
{Frieman}, J.~A., {Bassett}, B., {Becker}, A., {et~al.}
\newblock {The Sloan Digital Sky Survey-II Supernova Survey: Technical
  Summary}.
\newblock {\em \aj},
  \href{http://dx.doi.org/10.1088/0004-6256/135/1/338}{135:338--347}, January
  2008, 0708.2749\href{http://adsabs.harvard.edu/abs/2008AJ....135..338F.
\newblock }{ (ADS)}.

\bibitem{Riess99a}
{Riess}, A.~G., {Kirshner}, R.~P., {Schmidt}, B.~P., {et~al.}
\newblock {BVRI Light Curves for 22 Type IA Supernovae}.
\newblock {\em \aj}, 117:707--724, February
  1999\href{http://adsabs.harvard.edu/cgi-bin/nph-bib_query?bibcode=1999AJ....117..707R&db_key=AST.
\newblock }{ (ADS)}.

\bibitem{Jha06}
{Jha}, S., {Kirshner}, R.~P., {Challis}, P., {et~al.}
\newblock {UBVRI Light Curves of 44 Type Ia Supernovae}.
\newblock {\em \aj}, \href{http://dx.doi.org/10.1086/497989}{131:527--554},
  January 2006,
  arXiv:astro-ph/0509234\href{http://adsabs.harvard.edu/abs/2006AJ....131..527J.
\newblock }{ (ADS)}.

\bibitem{Hicken09a}
{Hicken}, M., {Challis}, P., {Jha}, S., {et~al.}
\newblock {CfA3: 185 Type Ia Supernova Light Curves from the CfA}.
\newblock {\em \apj},
  \href{http://dx.doi.org/10.1088/0004-637X/700/1/331}{700:331--357}, July
  2009, 0901.4787\href{http://cdsads.u-strasbg.fr/abs/2009ApJ...700..331H.
\newblock }{ (ADS)}.

\bibitem{LOSS-Li00}
{Li}, W.~D., {Filippenko}, A.~V., {Treffers}, R.~R., {et~al.}
\newblock {The Lick Observatory Supernova Search}.
\newblock In {S.~S.~Holt \& W.~W.~Zhang}, editor, {\em American Institute of
  Physics Conference Series}, volume 522 of {\em American Institute of Physics
  Conference Series}, pages 103--106, June 2000,
  arXiv:astro-ph/9912336\href{http://adsabs.harvard.edu/abs/2000AIPC..522..103L.
\newblock }{ (ADS)}.

\bibitem{Blondin12}
{Blondin}, S., {Matheson}, T., {Kirshner}, R.~P., {et~al.}
\newblock {The Spectroscopic Diversity of Type Ia Supernovae}.
\newblock {\em \aj},
  \href{http://dx.doi.org/10.1088/0004-6256/143/5/126}{143:126}, May 2012,
  1203.4832\href{http://cdsads.u-strasbg.fr/abs/2012AJ....143..126B.
\newblock }{ (ADS)}.

\bibitem{SNFAldering}
{Aldering}, G., {Adam}, G., {Antilogus}, P., {et~al.}
\newblock {Overview of the Nearby Supernova Factory}.
\newblock In {\em Survey and Other Telescope Technologies and Discoveries.
  Edited by Tyson, J. Anthony; Wolff, Sidney. Proceedings of the SPIE, Volume
  4836, pp. 61-72 (2002).}, pages 61--72, December
  2002\href{http://cdsads.u-strasbg.fr/cgi-bin/nph-bib_query?bibcode=2002SPIE.4836...61A&db_key=AST.
\newblock }{ (ADS)}.

\bibitem{GarnavichHST98}
{Garnavich}, P.~M., {Kirshner}, R.~P., {Challis}, P., {et~al.}
\newblock {Constraints on Cosmological Models from Hubble Space Telescope
  Observations of High-z Supernovae}.
\newblock {\em \apjl}, \href{http://dx.doi.org/10.1086/311140}{493:L53+},
  February 1998,
  arXiv:astro-ph/9710123\href{http://cdsads.u-strasbg.fr/abs/1998ApJ...493L..53G.
\newblock }{ (ADS)}.

\bibitem{PerlmutterSchmidt03}
{Perlmutter}, S. \& {Schmidt}, B.~P.
\newblock {Measuring Cosmology with Supernovae}.
\newblock In {K.~Weiler}, editor, {\em Supernovae and Gamma-Ray Bursters},
  volume 598 of {\em Lecture Notes in Physics, Berlin Springer Verlag}, pages
  195--217, 2003,
  arXiv:astro-ph/0303428\href{http://cdsads.u-strasbg.fr/abs/2003LNP...598..195P.
\newblock }{ (ADS)}.

\bibitem{Blakeslee03}
{Blakeslee}, J.~P., {Tsvetanov}, Z.~I., {Riess}, A.~G., {et~al.}
\newblock {Discovery of Two Distant Type Ia Supernovae in the Hubble Deep
  Field-North with the Advanced Camera for Surveys}.
\newblock {\em \apj}, \href{http://dx.doi.org/10.1086/374797}{589:693--703},
  June 2003,
  arXiv:astro-ph/0302402\href{http://cdsads.u-strasbg.fr/abs/2003ApJ...589..693B.
\newblock }{ (ADS)}.

\bibitem{Amanullah10}
{Amanullah}, R., {Lidman}, C., {Rubin}, D., {et~al.}
\newblock {Spectra and Hubble Space Telescope Light Curves of Six Type Ia
  Supernovae at 0.511 {\lt} z {\lt} 1.12 and the Union2 Compilation}.
\newblock {\em \apj},
  \href{http://dx.doi.org/10.1088/0004-637X/716/1/712}{716:712--738}, June
  2010, 1004.1711\href{http://cdsads.u-strasbg.fr/abs/2010ApJ...716..712A.
\newblock }{ (ADS)}.

\bibitem{Dawson09}
{Dawson}, K.~S., {Aldering}, G., {Amanullah}, R., {et~al.}
\newblock {An Intensive Hubble Space Telescope Survey for z{\gt}1 Type Ia
  Supernovae by Targeting Galaxy Clusters}.
\newblock {\em \aj},
  \href{http://dx.doi.org/10.1088/0004-6256/138/5/1271}{138:1271--1283},
  November 2009,
  0908.3928\href{http://cdsads.u-strasbg.fr/abs/2009AJ....138.1271D.
\newblock }{ (ADS)}.

\bibitem{Suzuki12}
{Suzuki}, N., {Rubin}, D., {Lidman}, C., {et~al.}
\newblock {The Hubble Space Telescope Cluster Supernova Survey. V. Improving
  the Dark-energy Constraints above z {\gt} 1 and Building an Early-type-hosted
  Supernova Sample}.
\newblock {\em \apj},
  \href{http://dx.doi.org/10.1088/0004-637X/746/1/85}{746:85}, February 2012,
  1105.3470\href{http://adsabs.harvard.edu/abs/2012ApJ...746...85S.
\newblock }{ (ADS)}.

\bibitem{Rodney12}
{Rodney}, S.~A., {Riess}, A.~G., {Dahlen}, T., {et~al.}
\newblock {A Type Ia Supernova at Redshift 1.55 in Hubble Space Telescope
  Infrared Observations from CANDELS}.
\newblock {\em \apj},
  \href{http://dx.doi.org/10.1088/0004-637X/746/1/5}{746:5}, February 2012,
  1201.2470\href{http://cdsads.u-strasbg.fr/abs/2012ApJ...746....5R.
\newblock }{ (ADS)}.

\bibitem{Miknaitis07}
{Miknaitis}, G., {Pignata}, G., {Rest}, A., {et~al.}
\newblock {The ESSENCE Supernova Survey: Survey Optimization, Observations, and
  Supernova Photometry}.
\newblock {\em \apj}, \href{http://dx.doi.org/10.1086/519986}{666:674--693},
  September 2007,
  arXiv:astro-ph/0701043\href{http://cdsads.u-strasbg.fr/abs/2007ApJ...666..674M.
\newblock }{ (ADS)}.

\bibitem{Kowalski08}
{Kowalski}, M., {Rubin}, D., {Aldering}, G., {et~al.}
\newblock {Improved Cosmological Constraints from New, Old, and Combined
  Supernova Data Sets}.
\newblock {\em \apj}, \href{http://dx.doi.org/10.1086/589937}{686:749--778},
  October 2008,
  0804.4142\href{http://cdsads.u-strasbg.fr/abs/2008ApJ...686..749K.
\newblock }{ (ADS)}.

\bibitem{KesslerChallenge10}
{Kessler}, R., {Bassett}, B., {Belov}, P., {et~al.}
\newblock {Results from the Supernova Photometric Classification Challenge}.
\newblock {\em \pasp}, \href{http://dx.doi.org/10.1086/657607}{122:1415--1431},
  December 2010,
  1008.1024\href{http://adsabs.harvard.edu/abs/2010PASP..122.1415K.
\newblock }{ (ADS)}.

\bibitem{Sako11}
{Sako}, M., {Bassett}, B., {Connolly}, B., {et~al.}
\newblock {Photometric Type Ia Supernova Candidates from the Three-year SDSS-II
  SN Survey Data}.
\newblock {\em \apj},
  \href{http://dx.doi.org/10.1088/0004-637X/738/2/162}{738:162--+}, September
  2011, 1107.5106\href{http://cdsads.u-strasbg.fr/abs/2011ApJ...738..162S.
\newblock }{ (ADS)}.

\bibitem{Bazin11}
{Bazin}, G., {Ruhlmann-Kleider}, V., {Palanque-Delabrouille}, N., {et~al.}
\newblock {Photometric selection of Type Ia supernovae in the Supernova Legacy
  Survey}.
\newblock {\em \aap},
  \href{http://dx.doi.org/10.1051/0004-6361/201116898}{534:A43+}, October 2011,
  1109.0948\href{http://cdsads.u-strasbg.fr/abs/2011A%26A...534A..43B.
\newblock }{ (ADS)}.

\bibitem{Palanque10}
{Palanque-Delabrouille}, N., {Ruhlmann-Kleider}, V., {Pascal}, S., {et~al.}
\newblock {Photometric redshifts for type Ia supernovae in the supernova legacy
  survey}.
\newblock {\em \aap},
  \href{http://dx.doi.org/10.1051/0004-6361/200913283}{514:A63+}, May 2010,
  0911.1629\href{http://cdsads.u-strasbg.fr/abs/2010A%26A...514A..63P.
\newblock }{ (ADS)}.

\bibitem{Kessler-photoz-10}
{Kessler}, R., {Cinabro}, D., {Bassett}, B., {et~al.}
\newblock {Photometric Estimates of Redshifts and Distance Moduli for Type Ia
  Supernovae}.
\newblock {\em \apj},
  \href{http://dx.doi.org/10.1088/0004-637X/717/1/40}{717:40--57}, July 2010,
  1001.0738\href{http://adsabs.harvard.edu/abs/2010ApJ...717...40K.
\newblock }{ (ADS)}.

\bibitem{DETF06}
{Albrecht}, A., {Bernstein}, G., {Cahn}, R., {et~al.}
\newblock {Report of the Dark Energy Task Force}.
\newblock September 2006,
  astro-ph/0609591\href{http://adsabs.harvard.edu/cgi-bin/nph-bib_query?bibcode=2006astro.ph..9591A&db_key=PRE.
\newblock }{ (ADS)}.

\bibitem{ESO-ESA}
{Peacock}, J.~A., {Schneider}, P., {Efstathiou}, G., {et~al.}
\newblock {ESA-ESO Working Group on ''Fundamental Cosmology''}.
\newblock Technical report, October
  2006\href{http://adsabs.harvard.edu/abs/2006ewg3.rept.....P.
\newblock }{ (ADS)}.

\bibitem{Holland02}
{Holland}, S.
\newblock {An Overview of CCD Development at Lawrence Berkeley National
  Laboratory}.
\newblock {\em Experimental Astronomy},
  \href{http://dx.doi.org/10.1023/B:EXPA.0000004341.11906.bf}{14:83--89},
  October 2002\href{http://cdsads.u-strasbg.fr/abs/2002ExA....14...83H.
\newblock }{ (ADS)}.

\bibitem{Bernstein11}
{Bernstein}, J.~P., {Kessler}, R., {Kuhlmann}, S., {et~al.}
\newblock {Supernova Simulations and Strategies For the Dark Energy Survey}.
\newblock {\em ArXiv e-prints}, November 2011,
  1111.1969\href{http://cdsads.u-strasbg.fr/abs/2011arXiv1111.1969B.
\newblock }{ (ADS)}.

\end{thebibliography}

\end{document}